\documentclass[a4paper,11pt]{article}
\pdfoutput=1
\usepackage{jheppub}
\usepackage{shuffle}
\usepackage{graphicx}
\usepackage{subcaption}
\usepackage{url}
\usepackage{xcolor}
\usepackage{easy-todo}
\usepackage{here}
\usepackage{slashed}
\usepackage{booktabs,array}

\newcommand{\dlog}{\textrm{d}\log}
\newcommand{\sodd}[2]{\Omega_1\!\left(#1, #2 \right)}
\newcommand{\dodd}[3]{\Omega_2\!\left(#1, #2, #3 \right)}
\newcommand{\ep}{\epsilon}
\newcommand{\tr}{\textrm{tr}_5}
\newcommand{\gram}[1]{\Delta_3^{(#1)}}
\newcommand{\sij}[1]{s_{#1}}
\newcommand{\vij}[1]{v_{#1}}
\newcommand{\mHsq}{q^2}
\newcommand{\mTsq}{m_t^2}
\newcommand{\cayley}[1]{\mathcal{C}_{#1}}
\newcommand{\kira}{\textsc{Kira}}
\newcommand{\fire}{\textsc{FIRE}}
\newcommand{\amflow}{\textsc{AMFlow}}
\newcommand{\diffexp}{\textsc{DiffExp}}

\newcommand{\polylogtools}{\textsc{PolyLogTools}}
\newcommand{\mathematica}{\textsc{Mathematica}}

\newcommand{\Tpb}{$T_1$}
\newcommand{\Tht}{$\tilde T_2$}
\newcommand{\ZTht}{$\mathcal{Z}(\tilde T_2)$}

\newcommand{\kite}{\textit{kite}$_7$}

\title{Two-Loop Master Integrals for Leading-Color $pp\to t\bar{t}H$ Amplitudes
with a Light-Quark Loop}

\author[1]{F.~Febres Cordero,}
\author[1]{G.~Figueiredo,}
\author[2]{M.~Kraus,}
\author[3,4]{B.~Page,}
\author[1]{and L.~Reina}

\affiliation[1]{Physics Department, Florida State University, Tallahassee,
Florida 32306-4350, USA}
\affiliation[2]{Departamento de F\'{i}sica Te\'{o}rica, Instituto de
F\'{i}sica, \\ Universidad Nacional Aut\'{o}noma de M\'{e}xico, Cd. de
M\'{e}xico C.P. 04510, M\'{e}xico}
\affiliation[3]{Theoretical Physics Department, CERN, Geneva, Switzerland}
\affiliation[4]{Department of Physics and Astronomy, Ghent University, 9000
Ghent, Belgium}

\abstract{
We compute the two-loop master integrals for leading-color QCD scattering
amplitudes including a closed light-quark loop in $t\bar{t}H$ production at
hadron colliders.
Exploiting numerical evaluations in modular arithmetic, we construct a basis of
master integrals satisfying a system of differential equations in
$\ep$-factorized form. We present the analytic form of the differential
equations in terms of a minimal set of differential one-forms.
We explore properties of the function space of analytic solutions to the
differential equations in terms of iterative integrals which can be exploited
for studying the analytic form of related scattering amplitudes. 
Finally, we solve the differential equations using generalized series
expansions to numerically evaluate the master integrals in physical phase
space.
As the first computation of a set of two-loop seven-scale master integrals, our
results provide valuable input for analytic studies of scattering amplitudes in
processes involving massive particles and a large number of kinematic scales.
}

\preprint{
CERN-TH-2023-240
\\[-5mm]
\begin{flushright}
  \includegraphics[width=2cm]{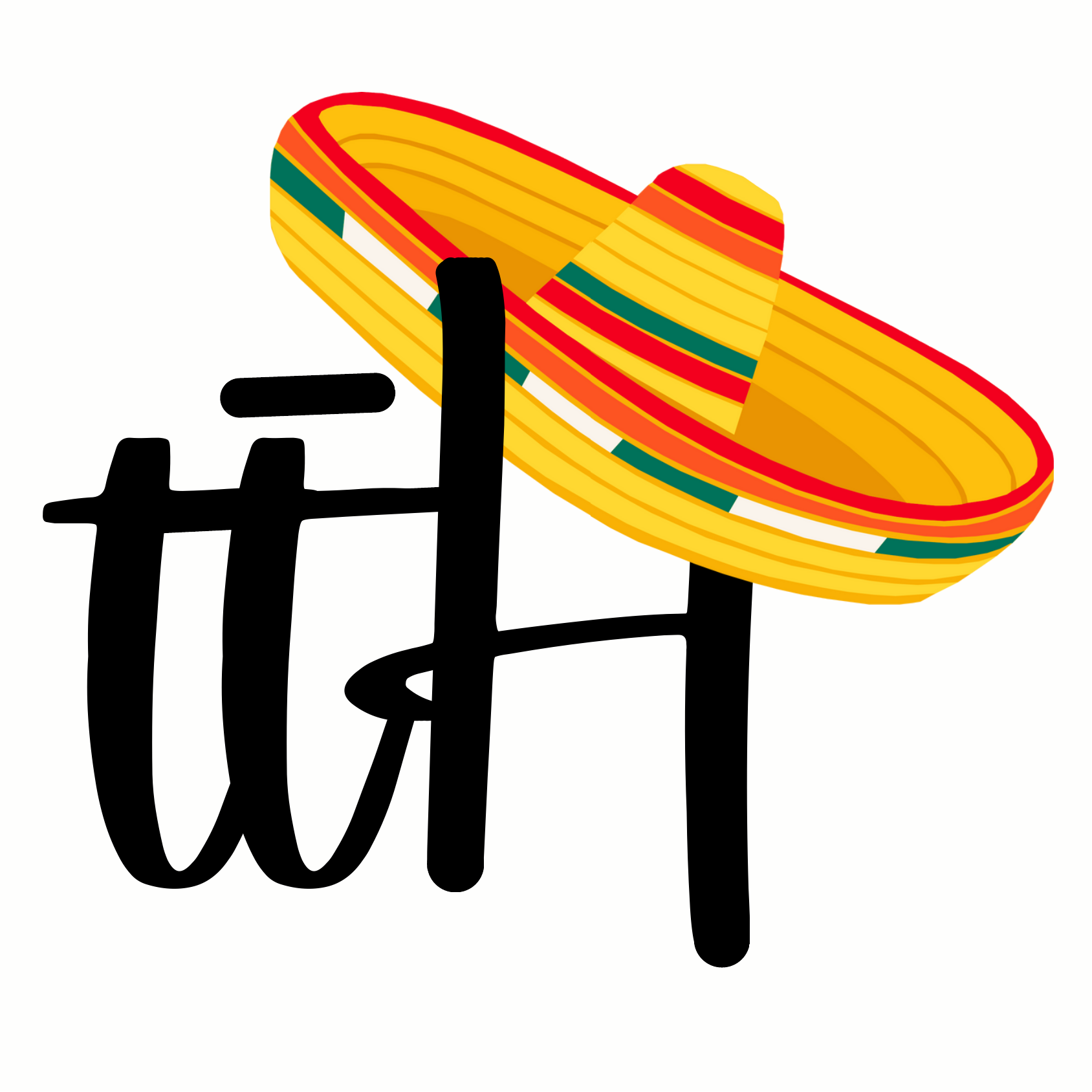}
\end{flushright}
}

\begin{document}
\maketitle
\section{Introduction}
Multi-loop Feynman integrals provide essential information about the analytic
properties of scattering amplitudes in quantum field theory. They are at the
core of making theoretical predictions for collider physics and are often the
main bottleneck for the calculation of precise predictions for scattering
processes. A particular challenge is the computation of Feynman integrals for
two-loop five-particle processes. In recent years, great effort has been
dedicated to such computations resulting in the calculation of all integrals
for fully massless
processes~\cite{Gehrmann:2015bfy,Papadopoulos:2015jft,Gehrmann:2018yef,Abreu:2018rcw,Abreu:2018aqd,Chicherin:2018old,Chicherin:2020oor},
all integrals for processes with one massive external particle and all massless
internal
particles~\cite{Abreu:2020jxa,Canko:2020ylt,Abreu:2021smk,Chicherin:2021dyp,Kardos:2022tpo,Abreu:2023rco},
and, more recently, the calculation of the first master integrals contributing
to five-particle processes involving an external massive top-quark pair and one
massive propagator~\cite{Badger:2022hno}.

A particularly important five-point process is that of $t\bar{t}H$ production
at hadron colliders, which gives a direct constraint on the top-quark Yukawa
coupling. First observed at the LHC in 2018~\cite{CMS:2018uxb,ATLAS:2018mme},
this process has by now allowed to constrain deviations from a Standard-Model-like Yukawa coupling
at the 10\% level---an impressive achievement that already challenges the
precision of existing theoretical predictions. It is expected that by the end
of the high-luminosity run at the LHC, measurements will be able to constrain
such coupling at the 3-5\% level and will be dominated by theory
uncertainties~\cite{ATL-PHYS-PUB-2014-012,CMS-PAS-FTR-21-002}. This creates a
pressing need for next-to-next-to-leading-order (NNLO) QCD
corrections~\cite{LHCHiggsCrossSectionWorkingGroup:2016ypw,Huss:2022ful,Schwienhorst:2022yqu}.

The $t\bar{t}H$ production process been studied extensively, with the
leading-order (LO) predictions known since the
mid-eighties~\cite{Ng:1983jm,Kunszt:1984ri}. Next-to-leading order (NLO) QCD
corrections were first computed in
Refs.~\cite{Beenakker:2001rj,Beenakker:2002nc,Reina:2001sf,Reina:2001bc,Dawson:2002tg,Dawson:2003zu},
and subsequently further improved by the resummation of soft-gluon
effects~\cite{Kulesza:2015vda,Broggio:2015lya,Broggio:2016lfj,Kulesza:2017ukk,Broggio:2019ewu,Ju:2019lwp,Kulesza:2020nfh},
the inclusion of first-order electroweak corrections~\cite{Frixione:2014qaa,
Zhang:2014gcy,Frixione:2015zaa}, the study of NLO off-shell
effects~\cite{Denner:2015yca,Denner:2016wet, Stremmer:2021bnk}, and the NLO QCD
matching to parton-shower event
generators~\cite{Frederix:2011zi,Garzelli:2011vp,Hartanto:2015uka,Maltoni:2015ena}.
Recently, the first NNLO QCD calculation has appeared~\cite{Catani:2022mfv},
where the two-loop amplitudes were approximated by a soft expansion in the
momentum of the Higgs boson ($p_H \to 0$). Obtaining the exact two-loop
scattering amplitudes is thus of great importance for the completion of the
NNLO QCD corrections to $t\bar{t}H$ production at hadron colliders.

\begin{figure}[ht]
\begin{center}
 \includegraphics[scale=0.7]{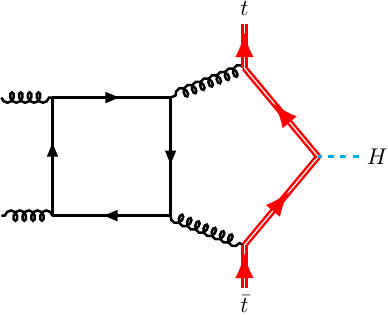}
 \hspace{5mm}
 \includegraphics[scale=0.7]{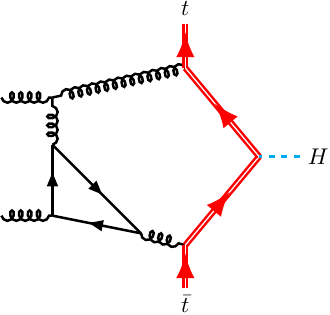}
 \hspace{5mm}
 \includegraphics[scale=0.7]{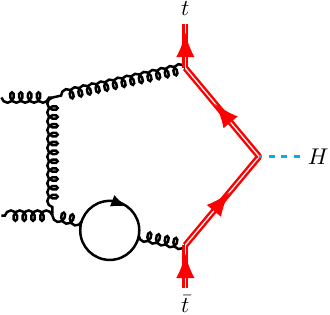}
\end{center}
 \caption{Examples of two-loop Feynman diagrams proportional to
the number of light flavors $n_f$ contributing to leading-color two-loop
scattering amplitudes for the process $gg\to t\bar{t}H$. The red double lines
represent top quarks and the external cyan dashed line the Higgs boson. Light
quarks are represented by black solid lines and gluons by black wavy lines.}
 \label{fig:feyndiags}
\end{figure}

As a first step towards this goal, in this work we compute a set of two-loop
master integrals contributing to the production of a top-quark pair in
association with a Higgs boson at hadron colliders. We focus on the Feynman
integrals arising in the calculation of the leading-color two-loop QCD
scattering
amplitudes for the parton-level processes $gg,q\bar{q}\to t\bar{t}H$
including a closed light-quark loop. Examples of related Feynman diagrams are
given in figure~\ref{fig:feyndiags} (see~\cite{Badger:2021ega,Abreu:2021asb}
for a discussion about the color decomposition of related scattering
amplitudes).
The corresponding amplitudes and Feynman integrals depend on seven different
kinematic scales, including the mass of the top quark (which also enters
internal lines of Feynman diagrams) and the mass of the Higgs boson.  
Importantly, we note that these integrals also arise in the two-loop scattering
amplitudes for processes such as $pp\to t\bar{t}Z$ and $e^+ e^- \to t \bar{t} +
2 j$.

The set of Feynman integrals that we study is organized in terms of three
integral families and contains a total of $127$ master integrals. We compute
the integrals with the method of differential
equations~\cite{Kotikov:1990kg,Kotikov:1991pm,Bern:1993kr,Remiddi:1997ny,Gehrmann:1999as,Henn:2013pwa},
constructing a basis of master integrals that satisfies a
set of differential equations in $\ep$-factorized form~\cite{Henn:2013pwa}.
Having found such a basis, we uncover a novel feature of these Feynman integrals:
their analytic description requires a nested square root function of the external
kinematics.
We then
show that the differential equations can be expressed in terms of $152$
differential one-forms, of which we are able to express all but four in $\dlog$
form. In such a compact form, our analytic differential equations clearly
manifest the singularity structure of the integrals. We then explore the
analytic properties of the master integrals by considering the iterated
integrals which arise in solutions to the differential equations. Moreover, we
solve the differential equations numerically using the generalized series
expansion method~\cite{Moriello:2019yhu} as implemented in the \diffexp{}
package~\cite{Hidding:2020ytt}. The required boundary values in the numerical
solutions are obtained with the auxiliary mass flow
method~\cite{Liu:2017jxz,Liu:2021wks,Liu:2022tji} as implemented in the
\amflow{} package~\cite{Liu:2022chg}. We provide in the ancillary files of this
article a \mathematica{} implementation based on \diffexp{} that allows to
solve the system of differential equations for points in the physical phase
space.

The rest of this article is organized as follows. In
section~\ref{sec:kinematics} we present the kinematic properties for the
process studied and define a series of relevant Lorentz invariant functions. In
section~\ref{sec:intfamilies} we define the families of Feynman integrals that
we study and describe the related master integrals. In section~\ref{sec:diffeq}
we give details of our procedure to build the basis of master integrals that
satisfy $\ep$-factorized differential equations and our method of determining
the analytic form of said differential equations. In particular we discuss the
determination of the corresponding ``alphabet'' of one-forms, and how we
construct $\dlog$ forms. In section~\ref{sec:AnalyticStructure} we consider the
analytic structure of Feynman integrals, first discussing the alphabet in
subsection~\ref{sec:list_letters} and then exploring the analytic properties of
the corresponding function space in subsection~\ref{sec:iterints}. In
section~\ref{sec:numerics} we present numerical results based on generalized
series expansions. We describe in subsection~\ref{sec:ancillary} the ancillary
files provided with this article. Finally, in section~\ref{sec:conclusions} we
give our conclusions and outlook. Appendices~\ref{app:2looppbox},
\ref{app:2looppbub}, and~\ref{app:1loop} contain a detailed description of the
integral bases we have constructed.

\section{Scattering Kinematics and Notation}
\label{sec:kinematics}
We consider the scattering process
\begin{equation}
  q_1(p_4)\, q_2(p_5)\, \to \, t(p_1)\, H(p_2) \, \bar{t}(p_3)\;,
  \label{eq:ScatteringProcess}
\end{equation}
where the initial pair of partons $(q_1,q_2)$ is either a gluon pair or a massless
quark/anti-quark pair. For convenience we work in an all-incoming convention
for the external momenta, such that momentum conservation is expressed as
$\sum_{i=1}^5 p_i = 0$. The momenta fulfill the on-shell conditions
\begin{equation}
 p_1^2=p_3^2 = \mTsq\;, \qquad p_2^2 = \mHsq\;, \qquad p_4^2=p_5^2 = 0\;,
\end{equation}
where $m_t$ is the mass of the top quark and we have kept the momentum
squared of the external massive boson in terms of the variable $\mHsq$. We
write general kinematic invariants in terms of the scalar products $\vij{ij} =
2p_i\cdot p_j$, although for convenience we sometimes also use the Mandelstam
variables $\sij{ij} = (p_i +p_j)^2$. The kinematic invariants can be expressed
in terms of $7$ independent variables which we choose to be
\begin{equation}
 \vec{s} = \{\vij{12},\vij{23},\vij{34},\vij{45},\vij{15},\mTsq,\mHsq\}\;,
\label{eq:vars}
\end{equation}
together with the parity-odd invariant
\begin{equation}
\tr = 4i\ep_{\mu\nu\alpha\beta}\,p_1^\mu p_2^\nu p_3^\alpha p_4^\beta\;,
\end{equation}
which is written in terms of the fully antisymmetric Levi-Civita symbol.
In terms of these variables we can write all remaining scalar products as
\begin{equation}
\begin{split}
 \vij{13} &= \vij{45} - \vij{12} - \vij{23} - 2\mTsq - \mHsq\;, \\
 \vij{14} &= \vij{23} - \vij{45} - \vij{15} + \mHsq\;, \\
 \vij{24} &= \vij{15} - \vij{23} - \vij{34} - \mHsq\;, \\
 \vij{25} &= \vij{34} - \vij{12} - \vij{15} - \mHsq\;, \\
 \vij{35} &= \vij{12} - \vij{34} - \vij{45} + \mHsq\;.
\end{split}
\end{equation}
When we consider the scattering process of
equation~\eqref{eq:ScatteringProcess}, the physical phase space in the
$\textrm{diag}(1,-1,-1,-1)$ Minkowski metric is a region in the space of
Mandelstam variables that is specified by the following set of inequalities
\begin{equation}
\begin{split}
  m_t^2 > 0 \ , &\qquad q^2 > 0\ ,    \\
  \vij{12} \ge 2\; m_t\; q\ , \qquad \vij{23} \ge 2\; m_t\; q\ ,  &\qquad \vij{34} \le 0 \ , \qquad \vij{15} \le 0\ ,  \\
  \vij{45} \ge (2 m_t + q)^2 \ , \qquad \det G(p_i,p_j,p_k) \ge 0\ , &\qquad \det G(p_1, p_2, p_3, p_4) \le 0\ ,
\label{eq:realps}
\end{split}
\end{equation}
where $q=\sqrt{q^2}$, the indices $i,j,k=1,\dots,5$, and we define the
Gram matrix according to $G(q_1,\dots,q_n)_{ij}=q_i\cdot q_j$.

\begin{figure}[t]
 \centering
 \begin{subfigure}{0.48\textwidth}
   \centering
   \includegraphics[scale=0.65]{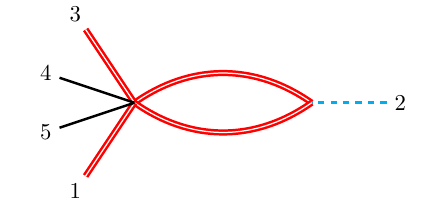}
   \vspace{6mm}
   \caption{$\cayley{1}$}
   \label{fig:cayley_1}
 \end{subfigure}
 \begin{subfigure}{0.48\textwidth}
   \centering
   \includegraphics[scale=0.6]{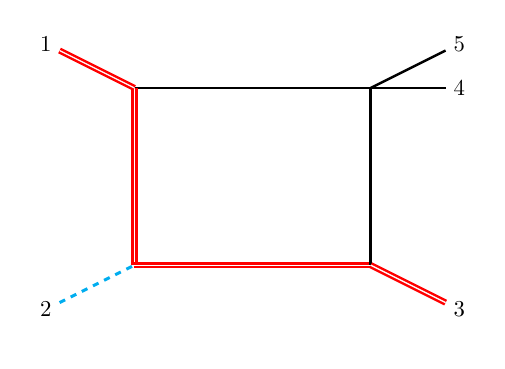}
   \caption{$\cayley{2}$}
 \end{subfigure}
 \caption{One-loop Feynman integrals whose maximal cut Baikov polynomials are
related to the $\cayley{1}$ and $\cayley{2}$ functions in
equations~\eqref{eq:c1} and \eqref{eq:c2}. The red double lines represent
massive internal and external lines (associated to top quarks in $t\bar{t}H$
production) while the dashed cyan line denotes an external off-shell line (associated to
the Higgs boson in $t\bar{t}H$ production).}
 \label{fig:cayleys}
\end{figure}

\begin{figure}[t]
 \centering
 \begin{subfigure}{0.32\textwidth}
   \centering
   \includegraphics[scale=0.54]{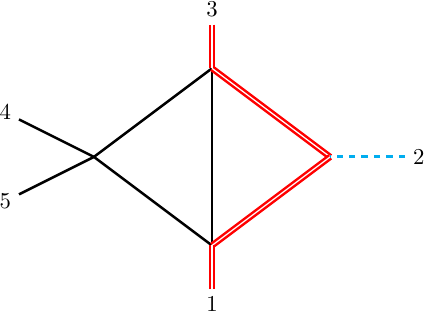}
   \caption{$r_1$}
   \label{fig:kite}
 \end{subfigure}
 \begin{subfigure}{0.32\textwidth}
   \centering
   \includegraphics[scale=0.54]{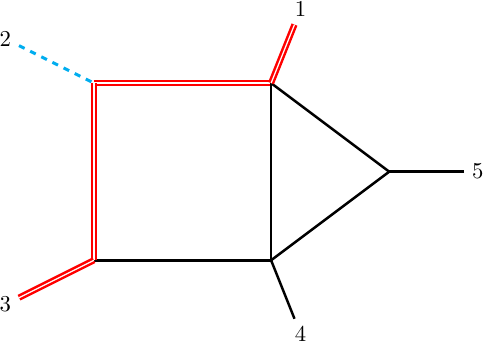}
   \caption{$r_2$}
 \end{subfigure}
 \hspace{2mm}
 \begin{subfigure}{0.32\textwidth}
   \centering
   \includegraphics[scale=0.54]{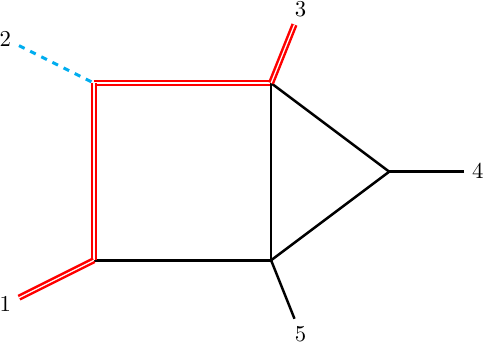}
   \caption{$r_3$}
 \end{subfigure}
 \caption{Two-loop Feynman integrals with leading singularities associated to
the $r_1$, $r_2$ and $r_3$ functions in equations~\eqref{eq:r1}--\eqref{eq:r3}.
Diagram lines are as in figure~\ref{fig:cayleys}.}
 \label{fig:maxcut_roots}
\end{figure}

The integrals considered in this paper can be expressed in terms of a basis of
special functions. One finds that these functions possess algebraic branch
points on various surfaces. Some are given by the zero
sets of the following Gram determinants
\begin{align}
 \gram{1} &= -4\det G(p_1,p_2) = \vij{12}^2 - 4\mTsq\mHsq\;, \\
 \gram{2} &= -4\det G(p_2,p_3) = \vij{23}^2 - 4\mTsq\mHsq\;, \\
 \gram{3} &= -4\det G(p_1,p_2+p_3) = (\mHsq+\vij{23}-\vij{45})^2 - 4\mTsq\vij{45}\;,\\ 
 \gram{4} &= -4\det G(p_1+p_2,p_3) = (\mHsq+\vij{12}-\vij{45})^2 - 4\mTsq\vij{45}\;, \\
 \gram{5} &= -4\det G(p_2,p_3+p_4) = (\mHsq+\vij{34}-\vij{15})^2 - 4\mHsq(\mTsq+\vij{34})\;, \\
 \Delta_5\ \!\ &= 16\det G(p_1,p_2,p_3,p_4) = \tr^2\;.
\end{align}
An important subtlety here is that one cannot identify $\tr$ with
$\sqrt{\Delta_5}$ as the first picks up a sign under parity, while the second
is invariant. For simplicity, when handling the algebraic branch points, we
will use $\sqrt{\Delta_5}$ instead of $\tr$ throughout this paper, analogous to
the conventions of Ref.~\cite{Abreu:2021smk}. Two further 
surfaces are given by the zero sets of of the functions
\begin{align}
 \cayley{1} &= \mHsq(\mHsq-4\mTsq)\;, \label{eq:c1} \\
 \cayley{2} &= \Big[(\mHsq+\vij{12})(\mHsq+\vij{23})-\mHsq\vij{45}\Big]\Big[(\mHsq+\vij{12})(\mHsq+\vij{23})-(\mHsq-4\mTsq)\vij{45}\Big] \label{eq:c2} \;.
\end{align}
These functions can be associated to the maximal cut of one-loop Baikov
polynomials~\cite{Baikov:1996iu} of the Feynman integrals in
figure~\ref{fig:cayleys}. Alternatively, they can be understood as modified
Cayley determinants (see e.g. Ref.~\cite{Abreu:2017mtm}).
Three additional functions associated to leading singularities of the two-loop
Feynman integrals shown in figure~\ref{fig:maxcut_roots} will also be needed,
and we define them according to
\begin{align}
  r_1 &= (v_{24} + v_{25})^2 - 4\mHsq\vij{45}\;, \label{eq:r1} \\
  r_2 &= [\mHsq{} \vij{35} + \vij{23} (\vij{35} + \vij{45})]^2 - 4 \mTsq{} \vij{45} [\vij{23} \vij{25} - \mHsq{} (\vij{15} + \vij{35})] \;, \label{eq:r2}  \\
  r_3 &= [\mHsq{} \vij{14} + \vij{12} (\vij{14} + \vij{45})]^2 - 4 \mTsq{} \vij{45} [\vij{12} \vij{24} - \mHsq{} (\vij{14} + \vij{34})] \label{eq:r3} \;. 
\end{align}
In contrast to previous two-loop five-point master integral computations, the
algebraic branch point structure is richer and involves nested square roots.
Indeed, we will need to employ square roots of the quantities 
\begin{equation}
 N_{\pm} = \mHsq{} \left(N_b \pm \sqrt{N_b^2 - N_c}\right) \;,
\label{eq:Npm}
\end{equation}
where
\begin{align}
  N_b &= \mHsq{}\left[(\vij{14}+\vij{15})^2 + (\vij{34}+\vij{35})^2\right] - 2 \mTsq{} (\vij{24} + \vij{25})^2\;, \\
  N_c &= \cayley{1} (\vij{12}-\vij{23})^2 (\vij{24} + \vij{25} + 2 \vij{45})^2\;.
\end{align}
Note that there is a subtlety when considering square roots of $N_{\pm}$, as
such roots are not algebraically independent from $\sqrt{\cayley{1}}$. This
follows as the $N_{\pm}$ fulfill
\begin{equation}
  N_+\cdot N_- = (\mHsq)^2~N_c\;. 
\end{equation}
In practice, throughout this manuscript, we choose to use this relation to
\textit{define} the symbol $\sqrt{N_-}$ in terms of the functions $\sqrt{N_+}$
and $\sqrt{\cayley{1}}$ according to
\begin{equation}
 \sqrt{N_-} \equiv \sqrt{\cayley{1}}\, \frac{\mHsq(\vij{12}-\vij{23})(2\mHsq+\vij{12}+\vij{23}-2\vij{45})}{\sqrt{N_+}}\;.
 \label{eq:NminusDefinition}
\end{equation}
When considering algebraic branch points an important associated algebraic
object is the Galois group. The Galois group is composed of all automorphisms
of the field extension that is implicitly defined by the functions
with such branch point singularities.
In previous two-loop five-point Feynman integral computations, the
automorphisms were given by transformations that flipped the signs of square
roots in the algebraic functions. Naturally, the more complicated square root
structure that we find by considering square roots of $N_{\pm}$ results in
elements of the Galois group that are more intricate. Clearly, we have the
standard sign flip associated to $\sqrt{N_+}$. However, we can see from
equation~\eqref{eq:NminusDefinition} that the sign flip of $\sqrt{N_-}$ is
achieved simultaneously with the sign flip of $\sqrt{\cayley{1}}$. Further, we
also have the automorphism
\begin{equation}
  \alpha \quad : \quad \sqrt{N_+} \leftrightarrow \sqrt{N_-}\;, \quad \sqrt{N_b^2 - N_c} \leftrightarrow -\sqrt{N_b^2 - N_c}\;,
\label{eq:galois}
\end{equation}
which simultaneously flips the sign of the ``inner'' square root, and swaps
$\sqrt{N_+}$ with $\sqrt{N_-}$.

Finally, we also note that there is an interesting relevant kinematic
map. Specifically, the set of integrals maps into itself under
\begin{equation}
  \mathcal{Z} \quad : \quad p_1 \leftrightarrow p_3, \quad p_4 \leftrightarrow p_5\;.
  \label{eq:Zmap}
\end{equation}
with $p_2$ left unchanged. Under this map our set of independent kinematic
variables in equation~\eqref{eq:vars} transforms as:
\begin{equation}
\{\vij{12},\vij{23},\vij{34},\vij{45},\vij{15},\mTsq,\mHsq\} \ \
\xrightarrow{\mathcal{Z}} \ \
\{\vij{23},\vij{12},\vij{15},\vij{45},\vij{34},\mTsq,\mHsq\}\;,
\end{equation}
and then all functions defined above transform under $\mathcal{Z}$ according to
\begin{equation}
  \gram{1} \leftrightarrow \gram{2}\;, \quad \gram{3} \leftrightarrow \gram{4}\;, \quad r_2 \leftrightarrow r_3\;, \quad \sqrt{N_{-}} \rightarrow - \sqrt{N_{-}}\;,
\end{equation}
while $\gram{5}$, $\Delta_5$, $\cayley{1}$, $\cayley{2}$, $r_1$ and $N_+$
remain invariant. Naturally, $\alpha$ and $\mathcal{Z}$ can be composed and we
denote the composition as $[\alpha \circ \mathcal{Z}]$, which acts on a
function $f$ as
\begin{equation}
 [\alpha \circ \mathcal{Z}](f) \equiv \alpha\left(\mathcal{Z}(f)\right)\;.
\end{equation}
%

\section{Feynman Integral Families}
\label{sec:intfamilies}
There are six types of eleven-propagator Feynman integral families, namely one
penta-box, two hexa-triangle, and three hepta-bubble families that contribute
to the considered scattering amplitudes. We start by analyzing the number of
master integrals that are associated to them. To find them we construct
integration-by-parts (IBP)
identities~\cite{Tkachov:1981wb,Chetyrkin:1981qh,Laporta:2000dsw} employing
numerical evaluations, independently obtained with the software
packages~\kira{}~\cite{Maierhofer:2017gsa, Klappert:2020nbg} and
\fire{}~\cite{Smirnov:2019qkx}. We find the following three main structures for
the master integrals.
\begin{description}
  \item[Penta-box integral family:] This family is associated to the propagator
structure of the left diagram of figure~\ref{fig:feyndiags}. We find that this
family has $111$ master integrals, including integrals with $3$ to $8$
propagators. We label this family of integrals as \Tpb{}, and we specify it
below in detail. We notice that this integral family maps into itself under the
$\mathcal{Z}$ transformation introduced in the previous section.

  \item[Hexa-triangle integral family:] This family is associated to the
propagator structure of the central diagram in figure~\ref{fig:feyndiags}. We
find that this family contains $46$ master integrals, of which $38$ are already
contained in the \Tpb{} family. This leaves $8$ distinct master integrals all
of which are contained in a penta-bubble subsystem with $19$ master integrals.
We label this subsystem as \Tht{}\footnote{The \Tht{} notation is
  chosen to remind that this (sub)family is part of a bigger 8-propagator
  family (to be denoted as $T_2$) that appears in the full calculation of $t\bar{t}H$
  amplitudes and of which only the \Tht{} portion is relevant for
  this paper.} and specify it in detail below. The
hexa-triangle integral family does not transform into itself under the
$\mathcal{Z}$ map, therefore the $8$ distinct master integrals above map under
$\mathcal{Z}$ to $8$ additional independent master integrals. We denote the
corresponding integral family by \ZTht{}.

\item[Hepta-bubble families:] These families are associated to the propagator
structure of the right diagram in figure~\ref{fig:feyndiags}. They can be
described in terms of the propagator structures that result from a massless
bubble insertion in each of the internal massless lines of the corresponding
one-loop diagram. We find that each of these families contain $19$ master
integrals, all of which are contained in the \Tpb{}, \Tht{}, and \ZTht{}
families described above.
\end{description}

In summary we have $127$ two-loop master integrals, decomposed into three
distinct Feynman integral families that we denote as \Tpb{} (with $111$ master
integrals), \Tht{} (with $8$ independent master integrals), and \ZTht{} (with
$8$ independent master integrals). In the following subsections, as
well as in Appendices~\ref{app:2looppbox} and \ref{app:2looppbub}, we give further
details about them. We notice that for completeness, in
Appendix~\ref{app:1loop} we also provide details of our basis of one-loop
master integrals for the propagator structure of the diagram shown in
figure~\ref{fig:1loopdiag}. In the following, we will refer to the one-loop
topology as $T_0$.

\subsection{The Penta-Box Family \Tpb{}}
\begin{figure}[ht]
 \centering
 \includegraphics[scale=0.8]{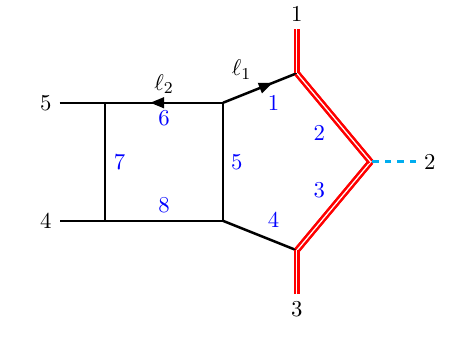}
 \caption{The propagator structure associated to the \Tpb{} integral family,
with the routing of loop momenta $\ell_i$ ($i=1,2$) chosen as in
equation~\eqref{eq:T1props}. The red double lines represent massive propagators
or external on-shell momenta (with mass $m_t$), the black solid lines represent
massless propagators or external on-shell momenta, and the cyan dashed line
represents an external off-shell momentum. The integers labeling the external
lines refer to the corresponding momenta $p_i$ ($i=1,\cdots,5$) as defined in
section~\ref{sec:kinematics}, while the blue inner integers correspond to the
inverse propagators $\rho_j$ ($j=1,\cdots,8$) as defined in
equation~\eqref{eq:T1props}.}
 \label{fig:T1}
\end{figure}

The \Tpb{} family is an eleven-propagator Feynman integral family, where three
propagators are introduced to make the family \textit{complete}, such that all
scalar products between loop momenta and external momenta can be expressed in
terms of inverse propagators. It is defined according to
\begin{equation}
 T_1[\vec{\nu}] = \int \frac{d^d\ell_1}{i\pi^{d/2}}\frac{d^d\ell_2}{i\pi^{d/2}}
 \frac{\rho_9^{-\nu_9}\rho_{10}^{-\nu_{10}}\rho_{11}^{-\nu_{11}}}
 {\rho_1^{\nu_1}\rho_2^{\nu_2}\rho_3^{\nu_3}\rho_4^{\nu_4}\rho_5^{\nu_5}\rho_6^{\nu_6}
 \rho_7^{\nu_7}\rho_8^{\nu_8}}\;,
 \label{eq:family}
\end{equation}
where $\vec\nu=(\nu_1,\cdots,\nu_{11})$ is a vector of integers (the propagator
powers) and $\nu_i \le 0$ for $i=9, 10$ and $11$. We work in dimensional
regularization with $d=4-2\ep$ and the inverse propagators are defined
according to 
\begin{align}
 &\rho_1 = \ell_1^2\;, &&\rho_2 = (\ell_1+p_1)^2 - \mTsq\;, &&&\rho_3 = (\ell_1+p_{12})^2 - \mTsq\;,\nonumber \\
 &\rho_4 = (\ell_1+p_{123})^2\;, &&\rho_5 = (\ell_1+\ell_2)^2\;, &&&\rho_6 =\ell_2^2\;,\label{eq:T1props} \\
 &\rho_7 = (\ell_2+p_5)^2\;, &&\rho_8 = (\ell_2+p_{45})^2\;, \nonumber
\end{align}
where $p_{i\cdots j}=p_i+\cdots+p_j$. These propagators correspond to the
diagram shown in figure~\ref{fig:T1}, to which we add the three irreducible
scalar products
\begin{align}
 &\rho_9 = (\ell_1-p_5)^2\;, &&\rho_{10} = (\ell_2-p_{12})^2 - \mTsq\;, 
 &&&\rho_{11} = (\ell_2-p_1)^2 - \mTsq\;.
\label{eq:T1isps}
\end{align}
The integral family \Tpb{} defines a vector space of integrals. Each element of
this vector space can be systematically expressed in terms of basis elements
with the help of IBP identities. As described above, the dimension of this
vector space is $\textrm{dim}\left(T_1\right) = 111$. There is a lot of freedom
in the choice of a basis of integrals, the so-called master integrals, and in
section~\ref{sec:diffeq} we describe how we construct a basis that satisfies a
set of differential equations in $\ep$-factorized form. In such way we make
explicit the singularity structure of all master integrals. We notice that $50$
master integrals of \Tpb{} have never been studied in the literature, with the
rest being one-loop squared integrals, $3$-propagator integrals, integrals with
all massless propagators, or integrals studied in Ref.~\cite{Badger:2022hno}.
In Appendix~\ref{app:2looppbox} we include the definition of all $111$
integrals in our basis.

\subsection{The Penta-Bubble Family \Tht{}}
\begin{figure}[h!]
 \centering
 \includegraphics[scale=0.7]{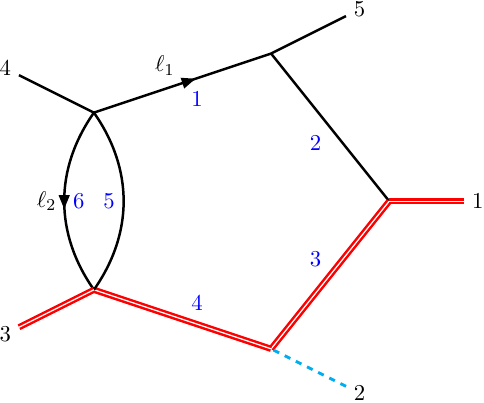}
 \caption{The propagator structure associated to the \Tht{} integral family,
with the routing of loop momenta $\ell_i$ ($i=1,2$) chosen in
equation~\eqref{eq:T2tprops}. See caption of figure~\ref{fig:T1} for details on
the notation.}
 \label{fig:T2t}
\end{figure}

The penta-bubble family \Tht{} is defined according to:
\begin{equation}
 \tilde T_2[\vec{\nu}] = \int \frac{d^d\ell_1}{i\pi^{d/2}}\frac{d^d\ell_2}{i\pi^{d/2}}
 \frac{\rho_7^{-\nu_7}\rho_{8}^{-\nu_{8}}\rho_{9}^{-\nu_{9}}\rho_{10}^{-\nu_{10}}\rho_{11}^{-\nu_{11}}}
 {\rho_1^{\nu_1}\rho_2^{\nu_2}\rho_3^{\nu_3}\rho_4^{\nu_4}\rho_5^{\nu_5}\rho_6^{\nu_6}}\;,
 \label{eq:T2tfamily}
\end{equation}
where $\nu_i \le 0$ if $i=7,\ldots,11$. The inverse propagators are defined
according to 
\begin{align}
 &\rho_1 = \ell_1^2\;, &&\rho_2 = (\ell_1+p_5)^2\;, &&&\rho_3 = (\ell_1+p_{15})^2 - \mTsq\;,\nonumber \\
 &\rho_4 = (\ell_1+p_{125})^2 - \mTsq\;, &&\rho_5 = (\ell_1+\ell_2-p_4)^2\;, &&&\rho_6 =\ell_2^2\;,
\label{eq:T2tprops} 
\end{align}
which correspond to the diagram shown in figure~\ref{fig:T2t}. We add the
following five irreducible scalar products to complete the family
\begin{align}
 &\rho_7 = (\ell_1-p_3)^2\;, &&\rho_{8} = (\ell_2-p_1)^2\;,  &&&\rho_{9} = (\ell_2-p_2)^2\;, \nonumber \\
 &\rho_{10} = (\ell_2-p_3)^2\;, &&\rho_{11} = (\ell_2-p_4)^2\;.
\label{eq:T2tisps}
\end{align}
The integral family \Tht{} defines a vector space with $\textrm{dim}(\tilde
T_2) = 19$. As described above, only $8$ of those integrals are not included in
the \Tpb{} family, and $4$ of them have never been studied in the literature.
Considering also the \ZTht{} family introduces $16$ more master integrals in
our analysis. In Appendix~\ref{app:2looppbub} we include the definition of
all \Tht{} master integrals not included in \Tpb{}.

\section{Differential Equations in $\ep$-Factorized Form}
\label{sec:diffeq}
The method of differential equations~\cite{Kotikov:1990kg,Kotikov:1991pm,
Bern:1993kr,Remiddi:1997ny,Gehrmann:1999as,Henn:2013pwa} has become one of the
most used approaches for computing master integrals, especially in the case of
integrals involving multiple scales. In practice, it turns out that obtaining
the analytic form of the differential equation can be challenging.
Nevertheless, a major simplification is achieved when the master integrals are
chosen to satisfy a differential equation in $\ep$-factorized
form~~\cite{Henn:2013pwa}. Despite great theoretical progress (see
e.g.~\cite{Lee:2014ioa,Prausa:2017ltv,
Gituliar:2017vzm,Dlapa:2020cwj,Henn:2020lye,Lee:2020zfb,Dlapa:2021qsl,
Chen:2022lzr,Dlapa:2022wdu,Gorges:2023zgv}), finding such sets of master
integrals for multi-scale problems remains a major problem. In this section we
give details of our approach to construct a basis of master integrals
satisfying $\ep$-factorized differential equations, which builds upon the
approaches presented in Refs.~\cite{Gehrmann:2014bfa,Meyer:2017joq} and
in Refs.~\cite{Abreu:2020jxa,Abreu:2021smk,Tschernow:2022kit}.

Let us denote a basis of master integrals as $\vec{J}$ for a family of Feynman
integrals $T$. The integrals $\vec{J}$ are functions of the kinematic
invariants $\vec{s}$ and the dimensional regulator $\ep = (4-d)/2$. A basis of
master integrals satisfies a set of first-order partial differential equations
\begin{equation}
 \frac{\partial \vec{J}}{\partial s_i} = B_i(\vec{s},\ep)~\vec{J}\;,
 \label{eq:deqgeneral}
\end{equation}
where the $B_i(\vec{s},\ep)$ are $N\times N$ matrices, with
$N=\textrm{dim}(T)$ and entries which are rational functions of $\ep$. If the
choice of basis $\vec{J}$ does not include algebraic functions of $\vec{s}$,
such as square roots, then the differential equation matrices
$B_i(\vec{s},\ep)$ are also rational functions of $\vec{s}$. Differential
equations of the form~(\ref{eq:deqgeneral}) are neither easy to construct, nor
easy to solve. To remedy these difficulties, we change to a basis of master
integrals $\vec{I}$ via $\vec{J} = U \vec{I}$, which satisfy differential
equations of the form
\begin{equation}
 \frac{\partial \vec{I}}{\partial s_i} = \ep\, A_i(\vec{s}\!\;)\vec{I}\;.
 \label{eq:deqcanonical}
\end{equation}
The differential equation matrices $A_i(\vec{s}\,)$ are related to those of the
original differential equation by
\begin{equation} 
 \ep\, A_i(\vec{s}\!\;) =  U ^{-1} B_i(\vec{s},\ep) U -U^{-1} \frac{\partial U}{\partial s_i} \;.
 \label{eq:changeofbasis}
\end{equation}
The key feature of equation~\eqref{eq:deqcanonical} is that it is in
``$\ep$-factorized'' form. In general, finding a basis of master integrals
$\vec{I}$ that satisfies such differential equations is a non-trivial task. We
describe our approach in the next section. In practice, we find that we are
able to achieve such an $\ep$-factorized form with a change of basis matrix $U$
that is algebraic in the kinematic invariants and rational in $\ep$. Hence, the
$A_i(\vec{s}\,)$ are algebraic in the kinematics invariants.

In practice, it is useful to unify the $7$ differential equations into a single
one using the language of differential forms. In such a language, we write that
\begin{equation}
 \mathrm{d} \vec{I} = \ep \sum_{\alpha = 1}^{\kappa} M_\alpha\, \omega_\alpha \vec{I}\;,
 \label{eq:diffFormDEQ}
\end{equation}
where we express the differential equation in terms of a set of $\kappa$
linearly independent differential one-forms $\omega_\alpha$. We call such a
one-form a ``letter'', and the full collection of all $\kappa$ letters, the
``alphabet''. The coefficient matrices $M_{\alpha}$ are $N\times N$ matrices
with rational number entries. 

In the following subsections we describe the procedure to construct our basis
of master integrals $\vec{I}$ that satisfy equation~\ref{eq:diffFormDEQ},
alongside the associated set of letters $\omega_\alpha$, and rational number
matrices $M_{\alpha}$.

\subsection{Construction of Master Integral Basis}
\label{sec:MasterIntegralBasisConstruction}
In order to construct a basis of master integrals that satisfy an
$\ep$-factorized differential equation, we follow the strategies employed in
Refs.~\cite{Abreu:2020jxa,Abreu:2021smk,Tschernow:2022kit}. Starting from a
choice of master integrals, for example by following the Laporta
algorithm~\cite{Laporta:2000dsw}, we construct the differential equations
\eqref{eq:deqgeneral} on a fixed kinematic point while keeping the full $\ep$
dependence. This allows to explore the analytic form of the differential
equations as a function of the dimensional regulator for multiple choices of
integral bases in an efficient way. We employ the \kira{}
program~\cite{Maierhofer:2017gsa,Klappert:2020nbg} for these reductions using
finite fields $\mathbb{F}_p$ where $p$ is a large prime
number~\cite{vonManteuffel:2014ixa,Peraro:2016wsq}. We then follow a number of
approaches to refine the basis choice, building on experience from the
literature and using a variety of techniques that we summarize here. 

As a first step, we search for a collection of master integrals where the
differential equations are linear in $\ep$. That is, we search for a basis such
that the matrix $B_i(\vec{s},\ep)$ in equation~\eqref{eq:deqgeneral} takes the
form
\begin{equation}
 B_i(\vec{s},\ep) = B_i^{(0)}(\vec{s}\!\;) + \ep B_i^{(1)}(\vec{s}\!\;)\;,
\label{eq:Blinear}
\end{equation}
We proceed in a bottom-up fashion, starting with master integrals
with the fewest number of propagators. For integrals with a low number of
propagators, we search through a collection of basis integrals with raised
propagator powers, until the differential equations take a form that is linear
in $\ep$. Often, such bases must be normalized by various $\ep$-dependent
functions, which we read from the differential equations evaluated on a
numerical kinematic point. A number of such basis choices can be interpreted as
dimension shift relations~\cite{Tarasov:1996br,Lee:2009dh} of subloops, such as
considering tadpole and bubble subloops into $2-2\ep$ dimensions. This procedure
gives rise to a large number of integrals in our basis, for example those given
in equations~\eqref{eq:T1_N15}, \eqref{eq:T1_N73} and
\eqref{eq:T1_N91}. For many integrals with a higher number of propagators, we
instead consider simple tensor insertions in order to arrive at a differential
equation that is linear in $\ep$.

From the refined starting point of equation~\eqref{eq:Blinear}, we apply a
number of techniques to obtain $\ep$-factorized differential equations. For
integrals with box or pentagon subloops, we follow techniques introduced
elsewhere in the literature~\cite{Abreu:2020jxa,Abreu:2021smk,Abreu:2023rco}.
Some examples were constructed by considering a four-dimensional $\dlog$-form
integrand analysis (see e.g.~\cite{Henn:2020lye}). Others make use of
numerators built from $\ep$-dimensional scalar products~\cite{Abreu:2018rcw}
\begin{equation}
 \mu_{ij} = \ell_i^{[d-4]}\cdot\ell_j^{[d-4]}\;, \qquad i,j=1,2\;,
\end{equation}
where we write the loop momenta as $\ell_i=(\ell_i^{[4]},\ell_i^{[d-4]})$, i.e.
decomposing them in terms of their 4- and $(d-4)$-dimensional parts. Examples
of integrands obtained through this procedure for \Tpb{} can be found in
equations~\eqref{eq:T1_N78}, \eqref{eq:T1_N90} and \eqref{eq:T1_N110}, and for
\Tht{} in equations~\eqref{eq:T2_N12}, \eqref{eq:T2_N18} and \eqref{eq:T2_N19}.

For many other integrals, it was fruitful to employ an approach based on the
structure of the $\ep \to 0$ limit of the differential equation matrix. One
advantage of the linear-in-$\ep$ form is that a change of basis matrix that
satisfies
\begin{equation}
 \frac{\partial}{\partial s_i} U =  B^{(0)}_i(\vec{s}\,) U\;,
 \label{eq:MagnusDE}
\end{equation}
will result in a differential equation in $\ep$-factorized form. This allows us
to use techniques based on the Magnus exponential~\cite{Argeri:2014qva}, which
we combine with analytic reconstruction techniques. In practice, we work sector
by sector\footnote{A sector of a Feynman integral family refers to integrals
that share the same set of inverse propagators with positive powers.}, or
equivalently block by block of $B^{(0)}_i$, and make a series of partial basis
changes to sequentially improve the basis. In most sectors we find that the
$B^{(0)}_i$ are triangular, and proceed in two stages. In the first stage, we
restrict our analysis of equation~\eqref{eq:MagnusDE} to diagonal entries,
which reduces equation~\eqref{eq:MagnusDE} to a collection of $1 \times 1$
systems. In practice, we find that these systems take the form
\begin{equation}
 \frac{\partial}{\partial s_i} u = b_i u\;, \qquad \text{where} \qquad b_i = \frac{\partial}{\partial s_i} \log(\tilde{b})\;,
\end{equation}
where $u$ is a diagonal entry of $U$ and $b_i$  are diagonal entries of
$B_i^{(0)}$. By analytically reconstructing the $b_i$ and integrating, we find an associated normalization of the
corresponding integral. We note that it may be the case that the $b_i$ are
rational, while $\tilde{b}$ is algebraic. In practice, we find that this
procedure is often easier to automate than a leading singularity calculation in
momentum space. In the second stage, we can assume that the relevant block of
each $B^{(0)}_i$ is strictly lower triangular. As an example, let us assume
that the relevant block is $2 \times 2$. Larger cases can be similarly handled.
One then has that
\begin{equation}
  \frac{\partial}{\partial s_i} U =
  \left( 
  \begin{array}{cc}
      0 & 0 \\
      b_{10, i} & 0 \\
  \end{array}
  \right)  U\;,
  \qquad \text{where} \qquad b_{10, i} = \frac{\partial}{\partial s_i} \tilde{b}_{10}\;,
\end{equation}
and in practice we find that $\tilde{b}_{10}$ is an algebraic function of the
kinematics. This differential equation is then solved by
\begin{equation}
  U =
    \left( 
      \begin{array}{cc}
        1 & 0 \\
        \tilde{b}_{10} & 1 \\
      \end{array}
    \right),
\end{equation}
which can be read as an instruction to redefine the second integral in the
block by subtracting the first with a factor of $\tilde{b}_{10}$. When
encountering situations like this, we analytically reconstruct $b_{10}$ to find
the associated basis change. Example of integrals obtained through this
procedure are in equations~\eqref{eq:T1_N93}, \eqref{eq:T1_N97} and
\eqref{eq:T1_N107} for \Tpb{}.

\begin{figure}
  \centering
  \includegraphics[scale=0.7]{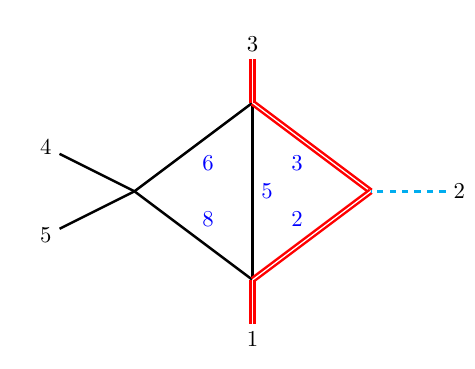}
  \caption{The ``\kite{}'' sector with seven master integrals. When studying
the differential equations for these integrals using Magnus exponential
techniques, the associated matrix is not triangular. The result of the Magnus
exponential for this sector introduces the nested square roots $\sqrt{N_{\pm}}$
(see section~\ref{sec:kinematics}).}
  \label{fig:kiteMagnus}
\end{figure}
A particular five-propagator sector, displayed in figure~\ref{fig:kiteMagnus},
involves $7$ master integrals and requires special attention, as the relevant
block structure of $B_i^{(0)}$ is not triangular. We refer to this sector as
the \kite{}. Four such integrals arise in a lower triangular block and so an
$\ep$-factorizing basis can be constructed with the procedures described
before. However, it was necessary to study a non-triangular $3 \times 3$ block.
Here, we again start by focusing on the diagonal entries, leading to
normalizations for the involved integrals which eliminate the diagonal elements
of the $B_i^{(0)}$. This leads to studying the differential equation for the
integrals with the numerators
\begin{equation} 
\begin{split}
 \widetilde{\mathcal{N}}_{64}^{(1)} &= \ep^3\sqrt{\mHsq}\sqrt[4]{N_b^2-N_c}\left(\frac{1}{\rho_3} - \frac{1}{\rho_2}\right)\;, \\
 \widetilde{\mathcal{N}}_{65}^{(1)} &= \ep^3\sqrt{\mHsq-4\mTsq}\sqrt[4]{N_b^2-N_c}\left(\frac{1}{\rho_3} + \frac{1}{\rho_2}\right)\;, \\
 \widetilde{\mathcal{N}}_{66}^{(1)} &= \ep^2\frac{\mTsq \vij{45}(\mHsq+\vij{12})(\mHsq+\vij{23})}{2\mHsq+\vij{12}+\vij{23}}\left(\frac{1}{\rho_2\rho_6} + \frac{1}{\rho_3\rho_8}\right)\;.
\label{eq:64-66ints1}
\end{split}
\end{equation} 
where the tilded notation highlights that this is an intermediate step
in the production of the final $\epsilon$-factorized basis.
The corresponding differential equations take the form
\begin{equation}
  \frac{\partial}{\partial s_i}\begin{pmatrix}
    \widetilde{I}_{64} \\ \widetilde{I}_{65} \\ \widetilde{I}_{66}
  \end{pmatrix} = 
  \begin{pmatrix}
    0 & a_i & 0
    \\
    a_i & 0 & 0
    \\ b_i & c_i &  0
  \end{pmatrix}
  \begin{pmatrix}
    \widetilde{I}_{64} \\ \widetilde{I}_{65} \\\widetilde{I}_{66}
  \end{pmatrix}
  + \mathcal{O}(\ep), \qquad \text{where} \qquad a_i = \frac{1}{4} \frac{\partial}{\partial s_i} \log(\tilde{a})\;,
  \label{eq:kiteMagnusBlock}
\end{equation}
and $a_i, b_i, c_i$ and $\tilde{a}$ are algebraic functions of the kinematics.
We then work to construct a change of basis which renders the matrix lower
triangular, and therefore amenable to the previous techniques. This leads us to
focus only on the upper $2 \times 2$ block. To remove this block, we solve the
associated differential equation for the change of basis matrix of
equation~\eqref{eq:MagnusDE} using the Magnus exponential. As the two
off-diagonal entries involve the same functions $a_i$, the Magnus exponential
truncates at its first order. Nevertheless, the factor of $\frac{1}{4}$ in
equation~\eqref{eq:kiteMagnusBlock} leads to a complicated algebraic procedure
which eventually results in removing the quartic roots in
equation~\eqref{eq:64-66ints1}, but introducing the nested roots $\sqrt{N_\pm}$
of equation~\eqref{eq:Npm}. After this step, the resulting block is now lower
triangular, and can be handled as discussed above. To apply this procedure in
practice, we reconstruct the analytic form of the $a_i, b_i$ and $c_i$ from
numerical samples. The integrals obtained through this procedure are presented
in equations~\eqref{eq:T1_N64}, \eqref{eq:T1_N65} and \eqref{eq:T1_N66}. 

Altogether, using this suite of approaches, we were able to construct a basis
of master integrals that satisfy a set of differential equations in
$\ep$-factorized form. We present the full set of basis integrals in the
appendices~\ref{app:2looppbox} and~\ref{app:2looppbub} for the \Tpb{} and
\Tht{} families respectively.

\subsection{Analytic Reconstruction of the Differential Equations}
Given that the bases of integrals obtained in the last section satisfy a set of
differential equations in $\ep$-factorized form, we are now in a position to
compute the analytic form of the differential equations. We follow the general
procedure of Ref.~\cite{Abreu:2020jxa}. We begin by computing $\kappa$, the
number of linearly independent letters $\omega_{\alpha}$ that arise in
equation~\eqref{eq:diffFormDEQ}. As described in Ref.~\cite{Abreu:2020jxa},
this can be computed from repeated numerical evaluations of the differential
equations. In practice, we find that the number of linearly independent
letters, or the dimension of the alphabet, is 
\begin{equation}
 \kappa = 152\;.
\end{equation}
Furthermore, we observe that by itself the \Tpb{} family contains all $152$
independent letters, and that the families \Tht{}, $\mathcal{Z}$(\Tht) and
$T_0$ can be expressed in terms of a subset of the same letters. 

Next, we focus on reconstructing a basis of letters, onto which we will later
fit the full differential equations. This approach avoids reconstructing the
functional form of each entry of the differential equations with numerical
evaluations, which can become computationally prohibitive. Following
Refs.~\cite{Abreu:2018rcw,Abreu:2018aqd,Abreu:2020jxa}, we choose a basis of
linearly independent letters by prioritizing entries of the differential
equation that lie on the block-diagonal. In practice, we find that $80$ of the
letters can be obtained from maximal-cut\footnote{For a given integral sector
we call the maximal-cut differential equations those obtained when working
modulo subsectors. Correspondingly, (next-to)$^k$-maximal-cut differential
equations are those that only keep subsectors with $k$ fewer propagators.},
$70$ from next-to-maximal-cut, and $2$ are found on next-to-next-to-maximal-cut
differential equations.

Our approach to reconstructing the basis of letters is based upon expectations
for their analytic form. Given that we have found a basis with an
$\ep$-factorized differential equation using only algebraic functions, we
naively expect that the letters can be expressed in $\dlog$ form, i.e.
\begin{equation}
 \omega_{\alpha} = \dlog(W_\alpha)\;.
 \label{eq:dlogAssumption}
\end{equation}
which gives a strong constraint on their analytic structure. We will return to the validity
of this assumption later. Further constraints on the analytic structure of the
letters follow from their properties under Galois transformations. We start
with letters that are independent of the square roots and therefore have
trivial Galois transformations. We call these \textit{even} letters. If a given
letter $\omega_\alpha$ is even
\begin{equation}
 \omega_\alpha  = \sum_{i} \omega_{\alpha,i}\; \textrm{d}s_i\;,
\end{equation}
it is clear that the $\omega_{\alpha,i}$ are rational functions. Given the
$\dlog$-form expectation, we should have that
\begin{equation}
 \omega_{\alpha, i} = \frac{\frac{\partial}{\partial s_i} W_\alpha}{W_\alpha}\;.
 \label{eq:denoms}
\end{equation}
Thus, irreducible polynomial factors of denominators of the $\omega_{\alpha,
i}$ are natural candidates for even letters. We therefore proceed to compute
the rational functions $\omega_{\alpha, i}$ through functional reconstruction
techniques (see e.g. Refs.~\cite{Peraro:2016wsq,Klappert:2019emp}), and take
the irreducible factors as letter candidates. In practice, this procedure
allowed us to compute a set of $33$ $W_\alpha$, which generate the full
subspace of even letters $\omega_\alpha$.

Next, we consider letters with non-trivial properties under Galois
transformations. We first consider those which do not depend on the nested
roots $\sqrt{N_{\pm}}$ but may depend on any other square root. In contrast to
the even letters, finding the corresponding $W_\alpha$ of
equation~\eqref{eq:dlogAssumption} is considerably more challenging and we
dedicate most of the rest of this section to their determination. Due to the
way in which the algebraic functions are introduced into our choice of basis,
the corresponding entries of the differential equation matrices (and therefore
letters) all pick up a sign under sign-flip Galois transformations. We refer to
these as \textit{odd} letters. An important feature observed in all odd letters
in two-loop five-point computations to date is that the possible denominator
factors of the associated differential form correspond to the even letters.
Given these two features, denoting the relevant square root as $\sqrt{R}$, we
are motivated to consider an initial ansatz for odd $\omega_{\alpha}$ of the
form
\begin{equation}
 \omega_{\alpha} = \frac{1}{\sqrt{R}} \frac{N_{\alpha}}{\prod_{\beta}(W_\beta^{\text{even}})^{q_{\alpha \beta}}}\;,
 \label{eq:OddDifferentialFormStructure}
\end{equation}
where $N_{\alpha}$ is a polynomially valued differential form and the
$q_{\alpha \beta}$ take values in $\{0, 1\}$ and therefore select which even
letters arise in the denominators. In order to determine these exponents, given
that the set of even letters has already been determined, we apply univariate
reconstruction approaches~\cite{Abreu:2018zmy} to each of the individual
$\omega_{\alpha, i}$, and take the lowest common multiple of the set in $i$ of
all denominators in the $\omega_{\alpha,i}$.

Naturally, for each such letter, two steps remain: first we must determine the
numerators $N_{\alpha}$ and then we must perform the integration to rewrite
$\omega_{\alpha}$ as a $\dlog$ form. In practice each of these steps are
computationally and theoretically demanding. Instead, we perform both
operations together, using an ansatz approach. Specifically, we build ans\"atze
for the arguments of our $\dlog$ forms $W_{\beta}$ using the function
\begin{equation}
 \Omega_1(w, R)=\frac{w-\sqrt{R}}{w + \sqrt{R}}\;.
 \label{eq:O1}
\end{equation}
For the purposes of the ansatz procedure, we will consider
$w$ as an unknown rational function of the Mandelstam invariants. In order to
constrain $w$, let us consider the structure of a
$\dlog$ form arising from $\Omega_1$,
\begin{equation}
 \dlog(\Omega_1(w, R)) = \frac{-w (\mathrm{d} R) + 2 R (\mathrm{d} w) }{\sqrt{R}(w^2 - R)}\;.
 \label{eq:ExplicitOddDLog}
\end{equation}
Here, we see that the denominator of the $\dlog$ form is given by $w^2- R$. If
we write $w = w_N/w_D$, where $w_X$ (with $X$ either $N$ or $D$) are
polynomials and compare the denominators of equation~\eqref{eq:ExplicitOddDLog}
and equation~\eqref{eq:OddDifferentialFormStructure} we find
\begin{equation}
 w_N^2 - R w_D^2 \sim \prod_{\beta}(W_\beta^{\text{even}})^{q_{\alpha \beta}}\;,
 \label{eq:O1fact}
\end{equation}
where we use $\sim$ to state that the left-hand side and right-hand side are
(polynomially) proportional. We therefore reduce the problem to finding
polynomials $w_N$ and $w_D$ that satisfy equation~\eqref{eq:O1fact}. This
constraint is similar to that proposed in Ref.~\cite{Heller:2019gkq}, however
here the product of even letters is known a priori.

In order to solve equation~\eqref{eq:O1fact}, we use two methods. First, we
consider an ansatz for $w_N$ and $w_D$ where they are taken to be as
multivariate polynomials with rational numbers as coefficients. That is, we
write
\begin{equation}
 w_{X} = \sum_{|\vec{\gamma}| = \text{deg}(w_X)} \tau_{{X}, \vec{\gamma}} \left( \prod_{i=1}^7 s_i^{\gamma_i} \right)\;,
 \label{eq:MultivariateRationalFunctionAnsatz}
\end{equation}
where the $\tau_{{X},\vec{\gamma}}$ are unknown rational numbers,
$|\vec{\gamma}| = \sum_i \gamma_i$ and $\text{deg}(f)$ is the total degree of
the polynomial $f$. The sum in
equation~\eqref{eq:MultivariateRationalFunctionAnsatz} is over all exponents
$\vec{\gamma}$ that have the same degree as the $w_X$. By
equation~\eqref{eq:O1fact}, the degrees of $w_{N}$ and $w_{D}$ are constrained
such that their difference is the mass dimension of $w$. The degree of $w_{D}$
is therefore another unknown, and in practice we vary this in the ansatz
procedure. With this parametrization, we can rewrite equation~\eqref{eq:O1fact}
as
\begin{equation}
 w_N^2 - R w_D^2 \quad \text{mod} \quad \prod_{\beta}(W_\beta^{\text{even}})^{q_{\alpha \beta}} = 0\;.
 \label{eq:O1Modulo}
\end{equation}
The modulo operation can then be implemented with polynomial reduction
techniques that are commonly implemented in computer algebra systems (CAS). In
this way, equation~\eqref{eq:O1Modulo} becomes a quadratic set of constraints
that the $\tau_{X, \vec{\gamma}}$ must satisfy.

Let us consider how to solve equation~\eqref{eq:O1Modulo} given the
multivariate polynomial ans\"atze for $w_N$ and $w_D$. First, note that any
rescaling of equation~\eqref{eq:MultivariateRationalFunctionAnsatz} of the form
\begin{equation}
 (\tau_{N, \vec{\gamma}}, \tau_{D, \vec{\gamma}}) \rightarrow (\lambda \tau_{N, \vec{\gamma}}, \lambda \tau_{D, \vec{\gamma}})\;,
 \label{eq:ScalingInvariance}
\end{equation}
for any non-zero $\lambda$, will leave $w$ invariant. This implies that
solutions to equation~\eqref{eq:O1Modulo} are not unique and come in families.
This non-uniqueness can be avoided in any case where we know that some
$\tau_{X, \vec{\gamma}}$ is non-zero as we can use the rescaling to set it to
1. For example, if we consider a case where the degree of $w_{D}$ is 0, i.e.
$w$ is simply a polynomial, then it is natural to use the rescaling to set $w_D
= 1$. In practice, we find that this results in a quadratic system of equations
for the $(\tau_{N, \vec{\gamma}}, \tau_{D, \vec{\gamma}})$ that have a finite
number of solutions. Algorithms for enumerating the solutions of such systems
are commonly implemented in computer algebra systems. In practice, we find it
helpful to further organize the system by (repeatedly) solving all equations of
the form $x^2 = 0$. Nevertheless, if $w_{D}$ has non-zero degree, then it is a
priori unclear which term in $w_{D}$ is non-zero and we must walk through all
possibilities before we find a term which we can choose to have unit
coefficient. 

In practice, as the number of terms in a multivariate polynomial grows rapidly
as a function of the polynomial degree, the enumeration through all possible
non-zero terms can be computationally prohibitive. To address this problem, we
consider a second approach. Here, we single out one of the Mandelstam
invariants, denoting it as $s_i$, which is chosen in an ad-hoc manner. We then
consider an ansatz for $w_N$ and $w_D$ where they are univariate rational
function in the variable $s_i$, that is
\begin{equation}
 w_X = \sum_{k=0}^{\text{deg}(w_X)} \tau_{X, k}^{(\hat s_i)} (s_i)^k\;,
 \label{eq:UnivariateAnsatz}
\end{equation}
where the $\tau_{X,k}^{(\hat s_i)}$ are rational functions of all invariants
other than $s_i$. Here, we fix the degrees in $s_i$ of the $w_X$ to be the
largest possible, without the left hand side of equation~\eqref{eq:O1fact}
having greater degree in $s_i$ than the right hand side of
equation~\eqref{eq:O1fact}. With the ansatz in
equation~\eqref{eq:UnivariateAnsatz}, we again consider
equation~\eqref{eq:O1Modulo}, this time implementing the modulo operation using
polynomial reduction with respect to only $s_i$. Once again, this leads to a
quadratic system of equations that the $\tau_{X,k}^{(\hat s_i)}$ solve. While
one must also determine which $\tau_{X,k}^{(\hat s_i)}$ is non-zero by
enumeration, the number of terms in a univariate rational function is much
smaller, and is hence more tractable. Nevertheless, the $\tau_{X,k}^{(\hat
s_i)}$ are rational functions, and solving for them is a non-trivial exercise
which we confront using an in-house implementation of ``companion matrix''
techniques (see, for example, Ref.~\cite{Jiang:2017phk,cox2006using}). In
practice, we find this univariate ansatz method is able to handle the suite of
complicated cases we study in this work.

In summary, with this ansatz procedure, we are able to construct a collection
of $\dlog$ forms whose denominators must correspond to the provided
differential form. We then use these even $\dlog$ forms as an ansatz for the
odd forms (under a single square root) of the differential equations. The
validity of this ansatz is confirmed by a numerical fitting procedure.
Importantly, we did not directly reconstruct the numerator of
equation~\eqref{eq:OddDifferentialFormStructure}. Instead, we have implicitly
constructed it by using the expectation that
equation~\eqref{eq:OddDifferentialFormStructure} corresponds to a $\dlog$ form.

Beyond the letters which are odd under a single Galois transformation, there
are also letters which are odd under two such transformations. To handle these
cases, we follow an analogous approach to the one just described. Specifically,
denoting the two corresponding roots as $\sqrt{R_1}$ and $\sqrt{R_2}$, we start
from an ansatz for the argument of the $\dlog$ form using the function
\begin{equation}
\Omega_2(w, R_1, R_2) = \frac{w - \sqrt{R_1} \sqrt{R_2}}{w + \sqrt{R_1} \sqrt{R_2}}\;,
\label{eq:O2}
\end{equation}
where $w$ is again an unknown rational function. Finally, it is sometimes
useful to consider an ansatz for the argument of the $\dlog$ form using the
function
\begin{equation}
 \tilde\Omega(w_0, w_1\sqrt{R_1}, \sqrt{R_2}) = \frac{(w_0 + w_1 \sqrt{R_1} + \sqrt{R_2})(w_0 - w_1\sqrt{R_1} - \sqrt{R_2})}{(w_0 - w_1\sqrt{R_1} + \sqrt{R_2})(w_0 + w_1\sqrt{R_1} - \sqrt{R_2})}\;.
 \label{eq:Otilde}
\end{equation}
Here, $w_0$ and $w_1$ are unknown rational functions. Such forms of letters
have been previously found in multi-scale two-loop five-point amplitude
calculations~\cite{Abreu:2020jxa, Abreu:2021smk, Badger:2022hno}.
Indeed, early iterations of the multivariate rational function ansatz procedure
were used to determine the most complicated letters in~\cite{Abreu:2021smk}.
In practice,
this can be a fruitful approach to simplify the result as the mass dimension of
the unknown functions in equation~\eqref{eq:Otilde} is lower than those of
equation~\eqref{eq:O2}.

We now return to the collection of letters that depend on the nested square
roots $\sqrt{N_{\pm}}$. In these cases, we construct the full differential form
using functional reconstruction techniques. In practice, this was the most
computationally demanding reconstruction procedure. In all but four cases,
these were then integrated to $\dlog$ forms using ad-hoc techniques.
The remaining four letters proved resilient to being cast in $\mathrm{d}\log$
form. In these cases, one can show that the geometry associated to the square
roots in the integrand is that of an elliptic curve. While there is a wealth of
literature where such cases can be cast in $\mathrm{d}\log$ form, it is
important to emphasize that this is a property of the form itself. Indeed, a
priori, algebraic forms involving an elliptic curve could correspond to
differentials of first, second or third kind. Given that the differential
equation is in $\epsilon$-factorized form, and the change of basis matrix is
only algebraic, this is perhaps a surprising statement. Nevertheless, such cases
have arisen~\cite{VitaPrivateCommunication}. Currently, we are not aware of any
technique that could definitively classify the nature of such algebraic
forms and we suggest that this would be an interesting avenue for future
investigation.

Finally, with this complete alphabet in hand, we determine the rational number
matrices $M_{\alpha}$ of equation~\eqref{eq:diffFormDEQ} using the procedure
described in Ref.~\cite{Abreu:2020jxa}. Specifically, we sample the
differential equation $\kappa$ times, and use this information to fit the
differential equations onto the basis of letters, yielding the $M_{\alpha}$ and
therefore the analytic form of equation~\eqref{eq:diffFormDEQ}. Note that for
this procedure to work it is enough to know the partial derivatives of the
letters $\omega_\alpha$ for which we could not find a $\dlog$ form. We provide
the results of this procedure in the next section, as well as in a series of
ancillary files that we describe in section \ref{sec:ancillary}.

\section{Analytic Structure of the Feynman Integrals}
\label{sec:AnalyticStructure}

Given our analytic determination of an $\ep$-factorized differential equation
for the Feynman integrals studied in this paper, we are now in the position to
investigate the analytic structure of these integrals. In the following we
first discuss the organization of the letters in the alphabet of
the differential equation and afterwards the analytic structures of the
solution of the Feynman integrals.

\subsection{The Alphabet}
\label{sec:list_letters}

In section~\ref{sec:diffeq}, we described a procedure to determine a set of $148$ out of
$152$ letters as $\dlog$ forms and that can be used to express the differential equation matrices. Here
we present the results of that procedure, that is the analytic forms for the letters. Almost
all letters are so-called $\dlog$-forms, and take the form
\begin{equation}
    \omega_\alpha = \dlog(W_\alpha)\;, 
\end{equation}
for some expression $W_\alpha$ that is algebraic in the Mandelstam invariants, depending on a
limited series of square roots defined in section~\ref{sec:kinematics}.
In a slight abuse of language, in situations where there is no ambiguity, for
letters $\omega_\alpha$ that are in $\mathrm{d}\log$ form we will also call the
associated $W_\alpha$ a letter.
As a first organizational criteria for the alphabet, we note that a number of
letters do not arise in iterated integral solutions before
$\mathcal{O}(\ep^5)$ and hence do not contribute to the NNLO QCD corrections to
associated physical observables. We will return to this discussion in
section~\ref{sec:iterints}.
We denote the set of letters that do contribute as \textit{relevant letters} and
consider these first.

The first set of relevant letters that we consider are those that appear in
the first entry of iterated integral solutions,
\begin{equation}
  \{ W_1, \ldots, W_7 \} \!=\! \left\{ \!\sij{12} - \mTsq{}, \sij{23} - \mTsq{}, \sij{34} - \mTsq{}, \sij{15} - \mTsq{}, \vij{45}, \mTsq{}, \frac{\mHsq \!+\! \sqrt{\cayley{1}}}{\mHsq \!-\! \sqrt{\cayley{1}}} \right\}\;.
 \label{eq:first_entries}
\end{equation}
Note that one such letter is algebraic, and is odd under the Galois
transformation associated to $\sqrt{\cayley{1}}$. This letter can readily be
associated to the one-loop bubble shown in figure~\ref{fig:cayley_1}.

Beyond the first entries, the letters can be organized in terms of the square
roots that arise. The first class does not depend on any square root and are
denoted by \textit{even} letters. There are $33$ such relevant letters, and we
arrange them by mass dimension. Besides the $6$ even letters already accounted for in the first entries, we find $13$ additional letters which are linear in
Mandelstam invariants
\begin{multline}
  \{W_8, \ldots, W_{20}\} = \{ \mHsq{}, \sij{12}, \sij{23}, \sij{34}, \sij{15}, \vij{24}, \vij{25}, \vij{14},\vij{35}, \\ \vij{34}+\vij{45},\vij{15}+\vij{45}, \sij{12} - \sij{34}, \sij{23} - \sij{15} \}\;.
\end{multline}
We make use next of the object $\mathrm{tr}_{\pm}$, which we define
as\footnote{Let us comment on a subtlety in representing 
  letters in terms of $\mathrm{tr}_{\pm}$. If one re-expresses such objects in
  terms of Mandelstam invariants, one may find an explicit dependence on
  $\mathrm{tr}_5$ and not $\sqrt{\Delta_5}$. In our alphabet, in such a case we
  make the replacement $\mathrm{tr}_5 \rightarrow \sqrt{\Delta_5}$, such that all
  of our letters are parity invariant. This is the same convention as used in
  Ref~\cite{Abreu:2021smk}.
}
\begin{equation}
  \mathrm{tr}_{\pm}(i_1, \ldots, i_n) \equiv \mathrm{tr}\left( \left[ \frac{1 \pm \gamma_5}{2} \right]\slashed{p}_{i_1} \cdots \slashed{p}_{i_n} \right)\;.
\end{equation}
This object is multilinear in the momenta, and allows us to rewrite many
expressions in a form that manifestly vanishes in limits where involved momenta
become soft.

We obtained $4$ more even letters which are quadratic in Mandelstam invariants
\begin{equation}
  \{W_{21}, \ldots, W_{24}\} = \{\mathrm{tr}{}_+(4151), \mathrm{tr}{}_+(4353), \mathrm{tr}{}_+(15[2+3][4+5]), \mathrm{tr}{}_+(34[1+2][4+5])\}\;,
\end{equation}
and $8$ additional even letters which are cubic
\begin{equation}
  \begin{split}
    \{ W_{25}, \ldots, W_{32} \} = \{&
    \mathrm{tr}{}_+(125215), \mathrm{tr}{}_+(324234), \mathrm{tr}{}_+(124214), \mathrm{tr}{}_+(235325), \\
    &\mTsq{}(\vij{15} \!-\! \vij{34})^2 + \mHsq{} \vij{15} \vij{34}, \mTsq{} \vij{25}^2 + \mHsq{} \vij{15} (\vij{15} \!+\! \vij{25}), \\
    &\mTsq{} \vij{24}^2 + \mHsq{} \vij{34} (\vij{24} \!+\! \vij{34}), \\
    &\mTsq{} ([\vij{24} \!+\! \vij{25}]^2 - 4 \mHsq{} \vij{45}) - \sij{13} (\vij{12} \vij{23} - \mHsq{} \sij{13})
    \}\;.
  \end{split}
\end{equation}
Finally, there is a single even letter that is quartic in Mandelstam invariants
\begin{equation}
  W_{33} = \mathrm{tr}_+[4 3 (4+5) 1 5 1 (4+5) 3]\;,
\end{equation}
and a single even letter that is sextic
\begin{multline}
  W_{34} = (\mTsq{} \vij{24} \vij{25})^2 + (\mHsq{})^2 \vij{14} \vij{15} \vij{34} \vij{35} \\ + \mHsq{} \mTsq{} \left(  [\vij{14} \vij{15} \!-\! \vij{34} \vij{35}]^2 - \vij{24} \vij{25} [\vij{14} \vij{15} \!+\! \vij{34} \vij{35}]\right)\;.
\end{multline}

Beyond these letters, there are letters which
transform under the subgroup of the Galois group associated to each square
root. We first consider the letters which are invariant under the Galois
operation $\alpha$ associated to the nested roots $\sqrt{N_{\pm}}$ (see
equation~\eqref{eq:galois}), but do transform under the
Galois operation associated to a sign flip of the other square roots.
As it is standard, we organize these letters so that under the action of the
associated Galois group element, they pick up a sign.
We begin with those that only transform non-trivially under a single Galois
transformation, organizing all letters making use of $\sodd{w}{R_1}$ from equation~\eqref{eq:O1}.
Note that, in principle, $w$ is a rational function, and we organize our
letters by the mass dimension of the numerator of $w$ when considered in common
denominator form. Beginning with cases which are linear in Mandelstam variables
we have the following collection that depend on the roots of three-mass Gram determinants
$\Delta_3^{(n)}$ and $r_1$
\begin{equation}
  \begin{split}
  \{W_{35}, \ldots, W_{47} \} = 
  \Bigg\{
  & \sodd{\vij{12}}{{\Delta_3^{(1)}}}, \, \sodd{\vij{23}}{{\Delta_3^{(2)}}}, \, \sodd{2 \mTsq{} + \vij{12}}{{\Delta_3^{(1)}}}, \, \\
  & \sodd{2 \mTsq{} + \vij{23}}{{\Delta_3^{(2)}}}, \, \sodd{\mHsq{} + \vij{23} + \vij{45}}{{\Delta_3^{(3)}}}, \, \\
  & \sodd{\mHsq{} + \vij{12} + \vij{45}}{{\Delta_3^{(4)}}}, \, \sodd{\vij{13} + \vij{12}}{{\Delta_3^{(3)}}}, \, \\
  & \sodd{\vij{13} + \vij{23}}{{\Delta_3^{(4)}}}, \, \sodd{\vij{15}-\vij{14}}{{\Delta_3^{(3)}}}, \, \\
  & \sodd{\vij{35} - \vij{34}}{{\Delta_3^{(4)}}}, \, \sodd{\vij{12} + \vij{25}}{{\Delta_3^{(5)}}}, \, \\
  & \sodd{\vij{23} + \vij{24}}{{\Delta_3^{(5)}}}, \sodd{\vij{24}+\vij{25}}{{r_1}}
  \Bigg\}\;.
\end{split}
\end{equation}
The next set of letters is quadratic in the Mandelstam variables
\begin{equation}
  \begin{split}
    \{W_{48}, \ldots, W_{59}\} =
    \Bigg\{
        &\sodd{\frac{\vij{12}\vij{15} - 2 \mTsq{} \vij{25}}{\vij{15}}}{\Delta_3^{(1)}}, \, \sodd{\frac{\vij{23}\vij{34} - 2 \mTsq{} \vij{24}}{\vij{34}}}{\Delta_3^{(2)}}, \, \\
        &\sodd{\frac{\vij{12}\vij{14} - 2 \mTsq{} \vij{24}}{\vij{14}}}{\Delta_3^{(1)}}, \, \sodd{\frac{\vij{23}\vij{35} - 2 \mTsq{} \vij{25}}{\vij{35}}}{\Delta_3^{(2)}}, \, \\
        &\sodd{\frac{\mHsq{} \vij{15} - (2 \mTsq{} +\vij{15})(\vij{15}-\vij{34})}{\vij{15}}}{\Delta_3^{(5)}}, \, \\
        &\sodd{\frac{\vij{25}(\vij{13}+\vij{14}) - \vij{12}(\vij{35}+\vij{45})}{\vij{15}}}{\Delta_3^{(5)}}, \, \\
        &\sodd{\frac{\vij{24}(\vij{13}+\vij{35}) - \vij{23}(\vij{14}+\vij{45})}{\vij{34}}}{\Delta_3^{(5)}}, \, \\
        &\sodd{\mHsq{} - 2 \mTsq{} + 2\frac{\mTsq{}\vij{34}}{\vij{15}}}{\mathcal{C}_1}, \, \sodd{\frac{\mHsq{}(\vij{25}+2\vij{15})}{\vij{25}}}{\mathcal{C}_1}, \, \\
        &\sodd{\frac{\mHsq{}(\vij{24}+2\vij{34})}{\vij{24}}}{\mathcal{C}_1}, \, \\
        &\sodd{(\mHsq{} + \vij{12})(\mHsq{} + \vij{23}) - \mHsq{} \vij{45}}{\mathcal{C}_2}, \, \frac{\mathrm{tr}_{+}(1 4 3 5)}{\mathrm{tr}_{-}(1 4 3 5)}
    \Bigg\}\;,
  \end{split}
\end{equation}
where we observe that only the last letter depends on $\sqrt{\Delta_5}$
and can be written compactly by making use of $\mathrm{tr}_{\pm}$.  Next, we
collect the letters whose numerator is degree $3$ in common denominator form,
\begin{equation}
  \begin{split}
    \{W_{60}, &\ldots, W_{69}\} = \\
    \bigg\{ 
    &\sodd{
      \frac{2 \mTsq{} \vij{24} \vij{25} - \mHsq{} (\vij{14} \vij{15} + \vij{34} \vij{35})}{\vij{34} \vij{35}-\vij{14} \vij{15}}
    }{\mathcal{C}_1}, \\
    &
    \sodd{
      2 \mTsq{} \vij{45} + \vij{12} \vij{23} - \mHsq{} \sij{13} - 2 \frac{\mTsq{} \vij{34} \vij{45}}{\vij{15}}
    }{\mathcal{C}_2}, \\
    &\sodd{\vij{14} \vij{35} - \vij{13}\vij{45} - \vij{34}\vij{45} + \vij{15}\vij{35} - \frac{2 \mTsq{} \vij{15} \vij{45}}{\vij{34}} }{r_2},
    \mathcal{Z}(W_{62}), 
    \\
    &\sodd{ \vij{14} \vij{15} + (\mTsq{} + \sij{14}) \vij{25}  + \sij{25} \vij{45} - \frac{(\mHsq{} + \sij{25}) \vij{15} \vij{35}}{\vij{25}} }{r_2},
    \mathcal{Z}(W_{64}),
    \\
    &\sodd{
      \frac{(\mTsq - \sij{23}) \vij{15} \vij{35} + \mHsq{} \sij{14} \vij{45} - (\sij{23} \sij{25} + \mTsq{} \vij{25}) \vij{45}}{\vij{15} + \vij{45}}
    }{r_2},
    \mathcal{Z}(W_{66}),
    \\
    &\sodd{
      \mTsq{} \vij{35} - \sij{23} \vij{35} - \vij{23} \vij{45} + \frac{2 \mTsq{} \vij{25} \vij{45}}{\vij{35}}
    }{r_2},
    \mathcal{Z}(W_{68})
    \bigg\}\;,
  \end{split}
\end{equation}
where we make use of the $\mathcal{Z}$ operation defined in equation \eqref{eq:Zmap}.
Furthermore, we find that $5$ letters can also be written compactly in terms of
$\mathrm{tr}_{\pm}$.
\begin{equation}
  \begin{split}
    \{W_{70}, \ldots, W_{74}\} = \bigg\{ &
    \frac{\mathrm{tr}_+[1 2 4 1 5 4]}{ \mathrm{tr}_-[1 2 4 1 5 4] },
    \frac{\mathrm{tr}_+[1 2 4 2 3 4]}{ \mathrm{tr}_-[1 2 4 2 3 4] },
    \frac{\mathrm{tr}_+[1 2 5 1 4 5]}{ \mathrm{tr}_-[1 2 5 1 4 5] },
    \frac{\mathrm{tr}_{+}[1 2 5 2 3 5]    }{ \mathrm{tr}_{-}[1 2 5 2 3 5]    },
    \\
    &
    \frac{\mathrm{tr}_{+}[1 4 3 (4+5) 1 5]}{ \mathrm{tr}_{-}[1 4 3 (4+5) 1 5]}
    \bigg\}\;.
  \end{split}
\end{equation}
Finally, there is a single sextic case
\begin{equation}
  W_{75} = \sodd{
    \vij{14} \vij{25} \!-\! \vij{15} \vij{24} \!-\! \vij{12} \vij{45} + 
    \frac{2}{\vij{25}} \left[ \mHsq{} \vij{15} \vij{45} - W_{25} \tilde{w}_{75}
    \right]
  }{\Delta_5},
\end{equation}
where we write the letter in a way that emphasizes its simplifications on the
$W_{25} = 0$ surface and make use of the auxiliary function
\begin{equation}
  \tilde{w}_{75} = \frac{(\mTsq{} \vij{24}^2 \vij{25} + \mHsq{} \vij{34} [\vij{14}\vij{15} + \vij{14}\vij{35} + \vij{35} \vij{45}])}
  {(\mTsq{} \vij{24} \vij{25} [\vij{15} - \vij{34}] - \mHsq{} [\sij{24} + \vij{12}] \vij{15} \vij{34})}\;.
\end{equation}

Beyond this, a number of letters depend on two square roots, and are odd under
the sign flip of each of them. We find that we can cast these in one of two
forms. First, there are 21 letters which make use of $\dodd{w}{R_1}{R_2}$ of equation~\eqref{eq:O2}.
While $w$ could, in principle, be a rational function, we find that it is always
polynomial. Organizing again by the mass dimension of $w$, there are $10$ quadratic cases
\begin{equation}
  \begin{split}
    \{W_{76}, &\ldots, W_{85}\} = \\
    \bigg\{
  &\dodd{\vij{12} [\vij{14} + \vij{15}] - 2 \mTsq{} [\vij{24} + \vij{25}]}{\Delta_3^{(1)}}{\Delta_3^{(3)}}, \mathcal{Z}(W_{76}), \\
  &\dodd{ \vij{13} [\vij{24} + \vij{25}] - \vij{23}[\vij{14} - \vij{15}]}{\Delta_3^{(1)}}{\Delta_3^{(4)}}, \mathcal{Z}(W_{78}), \\
  &\dodd{\vij{12} \vij{23} - 2 \mHsq{} \vij{13}}{\Delta_{3}^{(1)}}{\Delta_{3}^{(2)}}, \\ 
  &\dodd{\vij{12} \vij{23} + \vij{45}[2 \mTsq{} - \vij{13}] -\mHsq{} \sij{13}}{\Delta_3^{(3)}}{\Delta_3^{(4)}}, \\
  &\dodd{2 \mHsq{} \sij{13} - \vij{12} [\vij{12} + \vij{23}]}{\Delta_3^{(1)}}{r_1}, \mathcal{Z}(W_{82}), \\
  &\dodd{\vij{12} \vij{45} - \mHsq{} [2 \sij{13} + \vij{12}] - \sij{13} \vij{23}}{\Delta_3^{(3)}}{r_1}, \mathcal{Z}(W_{84})
  \bigg\}\;.
\end{split}
\end{equation}
We also obtained the following $10$ cubic cases which are more complicated
\begin{equation}
\begin{split}
\{W_{86}, &\ldots, W_{95}\} = \\
\bigg\{
 &\dodd{\tilde{w}_{86}}{\Delta_3^{(1)}}{\mathcal{C}_2} , \mathcal{Z}(W_{86}) , \dodd{\tilde{w}_{88}}{\Delta_3^{(3)}}{\mathcal{C}_2} , \mathcal{Z}(W_{88}), \\
 &\dodd{\mHsq{} (\vij{12} \vij{23} + 4 \mTsq{} \vij{45} - \mHsq{} \sij{13})}{\mathcal{C}_1}{\mathcal{C}_2}, \\
 &\dodd{\tilde{w}_{91}}{\Delta_3^{(1)}}{\Delta_5} , \mathcal{Z}(W_{91}), \dodd{\tilde{w}_{93}}{\Delta_3^{(3)}}{\Delta_5} , \mathcal{Z}(W_{93}), \\
 &\dodd{ \mHsq{} [2 \mTsq{} \sij{13} + \vij{15} \vij{34} + \vij{14} \vij{35} - \vij{13} \vij{45}] - 2 \mTsq{} \sij{24} \sij{25}}{\mathcal{C}_1}{\Delta_5} 
\bigg\}\;,
\end{split}
\end{equation}
where we have defined the following polynomials
\begin{equation}
\begin{split}
 \tilde{w}_{86} =&\ \mHsq{} \mTsq{} (\vij{34} + \vij{35} - 3 \sij{13} - \vij{23}) \\ 
      &\ - \vij{12} [\sij{13} (\sij{14} + \sij{15}) + \vij{14} \vij{23} + \vij{15} \vij{23} +\vij{13} \vij{45}]\;, \\
 \tilde{w}_{88} =&\ \mHsq{} \sij{13} (\sij{13} + \vij{12}) + 2 \mTsq{} \vij{23} \vij{45} + \vij{12} (\vij{14} \vij{23} + \vij{15} \vij{23} - \sij{13} \vij{45} + \vij{13} \vij{45})\;, \\
 \tilde{w}_{91} =&\ \vij{12} (\vij{15} \vij{23} + \sij{13} \vij{25} - \vij{12} \vij{35}) \\ 
 &\ + 2 \mTsq{} [\sij{24} \sij{25} + \vij{25}^2 - \mHsq{} (\sij{13} + \vij{45})] - 2 \mHsq{} \sij{13} \vij{15}\;, \\
 \tilde{w}_{93} =&\ \vij{12} (\vij{15} \vij{34} - \vij{14} \vij{35}) - \sij{13} (\vij{25} \vij{34} - \vij{24} \vij{35}) +  \vij{13} \vij{45} (\vij{24} - \vij{25})\;.
\end{split}
\end{equation}
At last, there is a single letter that is quartic in Mandelstam variables
\begin{equation}
  W_{96} = \dodd{\tilde{w}_{96}}{\mathcal{C}_2}{\Delta_5}\;,
\end{equation}
where
\begin{equation}
 \tilde{w}_{96} = [\mHsq{} \sij{13} - \vij{12} \vij{23}] [\vij{15} \vij{34} - \vij{14} \vij{35} + \sij{13} \vij{45}] - 2 \mTsq{} [2 \mHsq{} \sij{13} + \sij{24} \vij{12} + \sij{25} \vij{23}] \vij{45}\;.
\end{equation}

Next, we have a number of letters which can be expressed more compactly in terms
of $\tilde{\Omega}(w_0, w_1\sqrt{R_1}, \sqrt{R}_2)$ of equation~\eqref{eq:Otilde}.
Organizing by the mass dimension of $w_0$, there are five linear cases
\begin{equation}
  \begin{split}
  \{W_{97}, \ldots, W_{101}\} = \Bigg\{
       &\tilde{\Omega}\left(\vij{25}, \sqrt{\Delta_3^{(1)}} , \sqrt{\Delta_3^{(5)}}\right), \mathcal{Z}(W_{97}), \\
       &\tilde{\Omega}\left(\mHsq{}+\vij{12}, \sqrt{\Delta_3^{(1)}}, \sqrt{\mathcal{C}_1}\right), \mathcal{Z}(W_{99}), \\
       &\tilde{\Omega}\left( \vij{34} - \vij{15}, \sqrt{\Delta_{3}^{(5)}}, \sqrt{\mathcal{C}_1} \right)
        \Bigg\}\;,
  \end{split}
\end{equation}
and $11$ quadratic cases
\begin{equation}
  \begin{split}
  \{W_{102}, \ldots, W_{112}\} = \Bigg\{
       &\tilde{\Omega}\left(\vij{14} (\sij{34} \!-\! \sij{12}) - \vij{15} \vij{35} + \vij{25} \vij{45}, \vij{45} \sqrt{\mathcal{C}_1}, \sqrt{r_2} \right), \mathcal{Z}(W_{102}), \\
       &\tilde{\Omega}\left((\mHsq{}+\vij{12})\vij{14}, \vij{45} \sqrt{\Delta_3^{(1)}}, \sqrt{r_3} \right), \mathcal{Z}(W_{104}), \\
       &\tilde{\Omega}\left(\vij{15}\vij{24} - \vij{25}\vij{14}, \vij{45} \sqrt{\Delta_3^{(1)}}, \sqrt{\Delta_5} \right), \mathcal{Z}(W_{106}), \\
       &\tilde{\Omega}\left(\vij{15}\vij{24} - \vij{25}\vij{14} - \vij{25}\vij{45}, \vij{45} \sqrt{\Delta_3^{(5)}}, \sqrt{\Delta_5} \right),  \\
       &\tilde{\Omega}\left(  (\mHsq{} + \vij{23}) \vij{34}, \sqrt{\mathcal{C}_2}, \sqrt{r_2} \right), \mathcal{Z}(W_{109}),  \\
       &\tilde{\Omega}\left(  \vij{15}\vij{34} + \vij{15}\vij{35} + \vij{34} \vij{45}, \sqrt{\Delta_5}, \sqrt{r_2}\right), \mathcal{Z}(W_{111})
       \Bigg\}\;.
  \end{split}
\end{equation}

A remaining set of letters depend on the nested root, with non-trivial Galois
properties. Firstly, we have a set of $6$ letters, which are odd with respect to the sign-flip
of $\sqrt{N_+}$, given by 
\begin{equation}
  \begin{split}
    \{W_{113}, &\ldots, W_{118}\}
    = \\
    \Bigg\{
    &\sodd{\mHsq{}[\vij{45} + \sij{13} - \mHsq{}]}{N_+}, \alpha(W_{113}),  \\
    &\sodd{\mHsq{}[\vij{12} \!-\! \vij{23}] }{N_+} \sodd{\frac{f_1}{\vij{24} \!+\! \vij{25}}}{N_+}, \alpha(W_{115}), \\
    &\sodd{\frac{N_+ ([\vij{14} \!-\! \vij{15}] - [\vij{34} \!-\! \vij{35}]) + (\mHsq{} - 4 \mTsq{}) (\vij{24} \!-\! \vij{25}) f_1 }
      {2 ( \mHsq{} [\vij{14} \!-\! \vij{15}] [\vij{34} \!-\! \vij{35}] -\mTsq{} [\vij{24} \!-\! \vij{25}]^2) - \sqrt{N_b^2 \!-\! N_c}}}{N_+}, \\
    &\alpha(W_{117})\Bigg\}\;,
    \end{split}
\end{equation}
where we use the $\alpha$ operation defined in equation \eqref{eq:galois}, 
$W_{115}$ and $W_{116}$ are given as products of two $\Omega_1$ functions and
we also introduced the following polynomial
\begin{equation}
  f_1 = \mHsq{} (\vij{12} \!-\! \vij{23})(\vij{45} + \sij{13} \!-\! \mHsq{})\;.
\end{equation}
We find a single letter that is odd under the $\alpha$
transformation~\eqref{eq:galois}
\begin{equation}
  W_{119} = \sodd{N_b + \mHsq{} (4 [2 \mTsq{} - \vij{13}] \vij{45} - [\vij{12} - \vij{23}]^2)}{N_b^2 - N_c}.
\end{equation}
A further set of $10$ letters have non-trivial Galois transformations with respect
to the sign flips of both $\sqrt{N_+}$ and another square root
\begin{equation}
  \begin{split}
    \{W_{120}, &\ldots, W_{129}\} = \\
    \Bigg\{
    &\dodd{f_1}{r_1}{N_+}, \alpha(W_{120}),  \\
    &\dodd{\frac{1}{2}f_1 + \frac{N_+}{2\mHsq{}}}{\Delta_3^{(3)}}{ N_+}, \mathcal{Z}(W_{122}), \alpha(W_{122}), [\alpha \circ \mathcal{Z}](W_{122}),  \\
    &\dodd{-\mHsq{} [\vij{12} \!-\! \vij{23}] ([\mHsq{} \!+\! \vij{12}] [\mHsq{} \!+\! \vij{23}] - [\mHsq{} \!-\! 4 \mTsq{}] \vij{45})}{\cayley{2}}{N_+}, \alpha(W_{126}),  \\
    &\dodd{\frac{(N_+ r_1 - f_1^2) f_2 -4 \mHsq{} N_+ \Delta_5 }{N_+ (\vij{25} \!-\! \vij{24}) + \mHsq{} f_1 ([\vij{15}-\vij{14}] -[\vij{34}-\vij{34}]) }}{N_+}{\Delta_5}, \alpha(W_{128})
    \Bigg\}\;,
  \end{split}
\end{equation}
where we defined the auxiliary function
\begin{equation}
  f_2 = \mHsq{} (\vij{15} \vij{34} + \vij{14} \vij{35} - \vij{13} \vij{45}) - 2 \mTsq{} (\vij{24} \vij{25} - \mHsq{} \vij{45})\;.
\end{equation}

A remaining class of letters we were unable to express in terms of
$\dlog$ forms. All such letters can be found in the maximal-cut
differential equation of the \kite{} integrals in figure \ref{fig:kiteMagnus}.
Notably, the four letters are generated by a single letter,
$\omega^{E}$. Specifically, this set of letters is given by
\begin{equation}
  \{\omega_{130}, \ldots, \omega_{133}\} = \Big\{\omega^{E}, \mathcal{Z}(\omega^{E}), \alpha(\omega^{E}), [\alpha \circ \mathcal{Z}](\omega^{E})\Big\}\;,
\end{equation}
where one can write $\omega^{E}$ in the form
\begin{equation}
  \omega^{E} = \frac{\Omega^E}{\mTsq{} (\mHsq{} - \vij{23}) W_{32} \sqrt{\Delta^{(2)}_3}\sqrt{N_+}\sqrt{N_b^2 - N_c}},
\end{equation}
where $\Omega^E$ is a polynomially-valued differential form.  Importantly, one
can understand the singularities of $\omega^E$ by expanding around the zeros of
the denominator, which correspond to a number of surfaces.  We find that around
each such surface $\omega^E$ has at worst single poles.
Analytic expressions for these one-forms are provided in the ancillary files
(see section~\ref{sec:ancillary}).

The final set of relevant letters are all square roots,
\begin{multline}
    \{ W_{134}, \ldots, W_{143} \} = \Big\{\sqrt{\mathcal{C}_1}, \sqrt{\Delta_3^{(1)}}, \sqrt{\Delta_3^{(2)}}, \sqrt{\Delta_3^{(3)}}, \\ \sqrt{\Delta_3^{(4)}}, 
                                     \sqrt{\Delta_3^{(5)}}, \sqrt{\mathcal{C}_2}, \sqrt{\Delta_5}, \sqrt{r_2}, \sqrt{r_3} \Big\}\;.
\end{multline}
Let us stress that $\{\omega_{134}, \ldots, \omega_{143}\}$ are Galois invariant
as $\mathrm{d}\log(\sqrt{f}) = \frac{1}{2}\mathrm{d}\log(f)$. It is interesting
to note that the root $\sqrt{N_b^2 - N_c}$ does not appear among the list of letters.

Beyond this, we finally have a set of $9$ \textit{irrelevant} letters, which do
not arise in solutions to the differential equation before
$\mathcal{O}(\ep^5)$, and hence are not expected to contribute to the NNLO QCD corrections of
associated physical observables. Firstly, we have two Galois invariant letters
\begin{equation}
  \{W_{144}, \ldots, W_{145}\} = \{\mathrm{tr}{}_+(4252), \mathrm{tr}_+[5 3 (4+5) 1 4 1 (4+5) 3]\}\;.
\end{equation}
These are followed by a set of Galois non-trivial letters,
\begin{equation}
  \begin{split}
    \{W_{146}, &\ldots, W_{151}\} = \\
    \bigg\{
    &\sodd{\vij{25} - \vij{24}}{{r_1}}, \frac{\mathrm{tr}_{+}[1 4 2 5]}{\mathrm{tr}_{-}[1 4 2 5]}, \frac{\mathrm{tr}_{+}[1 4 1 5 3 (4+5)]}{ \mathrm{tr}_{-}[1 4 1 5 3 (4+5)]}, \\
    &\sodd{\frac{(\vij{12} \vij{23} \!-\! \mHsq{} \sij{13}) (\vij{14} \vij{34}
\!+\! \vij{15} \vij{35}) + 2 \mTsq{} (\vij{15} \!-\! \vij{34}) (\vij{35} \!-\!
\vij{14} ) \vij{45}}{\vij{14} \vij{34} \!-\! \vij{15} \vij{35}}}{\cayley{2}}, \\
    &\dodd{\sij{13} [\vij{15} \vij{24} - \vij{14} \vij{25}] + \mHsq{} [\vij{15} \vij{34} - \vij{14} \vij{35}] - \vij{45}[\sij{24} \vij{34}  - \sij{25} \vij{35}]}{r_1}{\Delta_5}, \\
    &\tilde{\Omega}\left(\vij{14}\vij{34} - \vij{15}\vij{35}, \sqrt{\mathcal{C}_2}, \sqrt{\Delta_5} \right)
    \bigg\}\;.
  \end{split}
\end{equation}
Finally, one of the square roots itself is an irrelevant letter:
\begin{equation}
  W_{152} = \sqrt{r_1}\;.
\end{equation}

\subsection{Analytic Structures of the Function Space}
\label{sec:iterints}

In this section, we explore properties of the space of functions that arises in
the Feynman integrals under consideration. Given the $\ep$-factorized differential
equation~\eqref{eq:diffFormDEQ}, one can find solutions for the 
differential equation order by order in the dimensional regulator $\ep$ in terms
of Chen's iterated integrals~\cite{Chen:1977oja,Brown:2013qva}.
These special functions have proven to be a powerful tool for exploring
analytic and numerical properties of multi-scale integrals (see e.g.
Refs.~\cite{Chicherin:2020oor,Chicherin:2021dyp,Abreu:2023rco}).
In the following, we discuss the classes of iterated integrals that can arise in
the solution to our differential equation and leave construction of dedicated
solutions to future work.

We denote by $\vec{I}_i$ the vector of pure integrals, where $i=\;$0, 1 or 2
refers to the families $T_0$, \Tpb{}, and \Tht{} respectively (see
section~\ref{sec:intfamilies} and appendix~\ref{app:1loop} for details). We expand the integrals in $\ep$
and define
\begin{equation}
 \vec{I}_i(\ep,\vec{s}\!\;) = \sum_{n=0}^\infty \ep^n\vec{I}_i^{\;(n)}(\vec{s}\!\;)\;,
\label{eq:Iexpansion}
\end{equation}
where by construction the expansions start at $\mathcal{O}(\ep^0)$.
By equation~\eqref{eq:diffFormDEQ}, each term in the $\ep$ expansion can be
constructed iteratively as
\begin{equation}
 \vec{I}^{\;(n)}_i(\vec{s}\!\;) = \vec{b}^{\;(n)}_i(\vec{s}_0) + \sum_{\alpha=1}^{152}M_\alpha\int_\gamma \omega_\alpha\,
 \vec{I}_i^{\;(n-1)}(\gamma)\;, 
\label{eq:Iiter}
\end{equation}
where $\gamma$ is a path that connects the points $\vec{s}_0$ and $\vec{s}$ and
$\vec{b}^{\;(n)}_i(\vec{s}_0)$ are the vectors of boundary values.
In order to study the classes of iterated integrals that arise, in this section
we work modulo boundary values, except for the leading term
$\vec{I}_i^{\;(0)}=\vec{b}^{\;(0)}_i(\vec{s}_0)$, i.e. we set
$\vec{b}^{\;(n)}_i(\vec{s}_0) = 0$ for $n\ge 1$.

The integrals in equation~\eqref{eq:Iiter} can be expressed in terms of the iterated
integrals, which we define recursively according to
\begin{equation}
\begin{split}
 \left[\omega_{j_1},\ldots,\omega_{j_m}\right]_{\vec{s}_0}(\vec{s}\!\;) &=
\int_\gamma \omega_{j_m}\left[\omega_{j_1},\ldots,\omega_{j_{m-1}}\right]_{\vec{s}_0}(\gamma)\;, \\
 \left[~~\right]_{\vec{s}_0}(\vec{s}\!\;) &= 1\;.
\end{split}
\end{equation}
These functions form a graded algebra with their weight defined by
the depth $m$ of nested integrations. They also fulfill shuffle algebra relations~\cite{Brown:2013qva}
\begin{equation}
\left[\omega_{a_1},\ldots,\omega_{a_m}\right]_{\vec{s}_0}(\vec{s}\!\;)\left[\omega_{b_1},\ldots,\omega_{b_n}\right]_{\vec{s}_0}(\vec{s}\!\;)
= \sum_{\vec{c}\in
\vec{a}\shuffle\vec{b}}\left[\omega_{c_1},\ldots,\omega_{c_{m+n}}\right]_{\vec{s}_0}(\vec{s}\!\;)\;,
\label{eq:shuffle}
\end{equation}
where the \textit{shuffle} operator $\shuffle$ combines in all possible ways the components of the
vectors $\vec{a}$ and $\vec{b}$ but keeping always the relative order of the
components of both of them.
We use these iterated integrals to express the master integral coefficients
$\vec{I}_i^{\;(n)}$ as a combination of weight $n$ functions
\begin{equation}
  \vec{I}_i^{\;(n)}(\vec{s}\!\;) \sim \vec{J}_i^{\;(n)}(\vec{s}\!\;) \qquad \text{where} \qquad \vec{J}_i^{\;(n)}(\vec{s}\!\;) = \sum_{j_1,\ldots,j_n=1}^{152} \vec{c}^{\;(n)}_{i;j_1,\ldots,j_n} \left[\omega_{j_1},\ldots,\omega_{j_n}\right]_{\vec{s}_0}(\vec{s}\!\;)\;,
\label{eq:Isols}
\end{equation}
the coefficients $\vec{c}^{\;(n)}_{i;j_1,\ldots,j_n}$ are vectors of
rational numbers and $\sim$ is an equivalence relation working modulo boundary
terms $\vec{b}^{\;(n)}_i(\vec{s}_0)$ for $n\ge 1$.
According to equation~\eqref{eq:Iexpansion}, if we assign a
weight of $-1$ to $\ep$, all master integrals have a uniform weight equal to 0
at all orders in $\ep$. 

As one can see from equation~\eqref{eq:Iiter}, to construct these solutions we
need the constants $\vec{I}_i^{\;(0)}$. The boundary constants are obtained
from numerical evaluations using \amflow{} as described in the upcoming
section~\ref{sec:numerics}.  We also provide for convenience the explicit
weight-$0$ boundary terms in Appendices~\ref{app:2looppbox}, \ref{app:2looppbub}
and \ref{app:1loop}.
Using these boundary constants, we construct the iterative solutions
$\vec{J}_i^{\;(n)}(\vec{s}\!\;)$.

We will now discuss a number of
properties of these solutions.
First, as commented at the end of the previous subsection, we find that nine
letters do not arise in
iterated integral solutions up to weight $4$:
\begin{equation}
 \Big\{ \omega_{144}, \ldots, \omega_{152} \Big\}\;,
\end{equation}
that is, they do not enter in any of the corresponding
$\left[\omega_{j_1},\dots,\omega_{j_n}\right]_{\vec{s}_0}$ functions appearing in
the solutions $\{\vec{J}_0^{\;(n)}, \vec{J}_1^{\;(n)}, \vec{J}_2^{\;(n)},
\mathcal{Z}(\vec{J}_2^{\;(n)}) \}$ for $n\le 4$.
Next, we analyze the space of linear combinations of iterated integrals that
arise in the combined solutions to all two-loop Feynman integrals considered
in this work, in order to give an idea of the complexity of the function space.
First, we compute the number of linearly independent functions
at each order in $\ep$, finding $121$ linearly independent functions at
weight $4$.
Furthermore, we have explicitly checked that the linear relations that one finds
between the $\vec{J}_i^{\;(n)}$ are also linear relations of the
$\vec{J}_i^{\;(n')}$ for $n' < n \le 4$. That is, relations at higher weight also
hold at lower weight. It would be interesting to find an explanation for this
phenomenon.
Due to the shuffle algebra of equation~\eqref{eq:shuffle}, some of these functions are actually
products of lower weight functions.
Therefore, we also compute the number of linearly independent functions modulo
such shuffle relations, which we obtain with the help of the \polylogtools{}
package~\cite{Duhr:2019tlz}. This tells us the number of linearly independent
\textit{irreducible} functions and we find $114$ such irreducible functions at
weight $4$.
We summarize the results of this analysis in table~\ref{tab:funcspace}. 
\begin{table}[t]
\centering
\begin{tabular}{ccc}
 \toprule
 $n$ & Linearly independent & Irreducible \\
 \midrule
 1 & 7 & 7\\
 2 & 31 &16 \\
 3 & 85 & 69\\
 4 & 121 & 114\\
 \bottomrule
\end{tabular}
\caption{Working modulo boundary constants up to weight $4$, the number of
	linearly independent and irreducible functions at each weight $n$ for
	the combined set of two-loop integral solutions $\{\vec{J}_1^{\;(n)},
	\vec{J}_2^{\;(n)}, \mathcal{Z}(\vec{J}_2^{\;(n)})\}$.}
\label{tab:funcspace}
\end{table}
The total number of independent master integrals in the families \Tpb{}, \Tht{}
and \ZTht{} is $127$. Given that we encounter at weight four $121$ linearly
independent functions, it means, that modulo boundary constants, there are six
non-trivial relations between the master integrals arising at weight 4.

\begin{table}[t]
  \centering
  \begin{tabular}{ccccccccccccc}
    \toprule
    $n$ & \!$\sqrt{\gram{1}}\!$ & \!$\sqrt{\gram{2}}$\! & \!$\sqrt{\gram{3}}$\! & \!$\sqrt{\gram{4}}$\! & \!$\sqrt{\gram{5}}$\! & \!$\sqrt{\Delta_5}$\! & \!$\sqrt{\cayley{1}}$\! & \!$\sqrt{\cayley{2}}$\! & \!$\sqrt{r_1}$\! & \!$\sqrt{r_2}$\! & \!$\sqrt{r_3}$\! & \!$\sqrt{N_+}$\! \\
    \midrule
    1 & $0$ & $0$ & $0$ & $0$ & $0$ & $0$ & $1$ & $0$ & $0$ & $0$ & $0$ & $0$\\
    2 & $1$ & $1$ & $1$ & $1$ & $1$ & $0$ & $5$ & $1$ & $0$ & $0$ & $0$ & $0$\\
    3 & $4$ & $4$ & $4$ & $4$ & $4$ & $1$ & $6$ & $4$ & $0$ & $1$ & $1$ & $2$\\
    4 & $6$ & $6$ & $9$ & $9$ & $4$ & $7$ & $6$ & $6$ & $1$ & $1$ & $1$ & $2$\\
    \bottomrule
  \end{tabular}
  \caption{Working modulo boundary constants up to weight $4$, the number of
	linearly independent functions at each weight $n$ that are odd under
	the operation $\sqrt{X} \to -\sqrt{X}$, which arise in the combined set
	of two-loop integral solutions $\{\vec{J}_1^{\;(n)}, \vec{J}_2^{\;(n)},
	\mathcal{Z}(\vec{J}_2^{\;(n)})\}$.}
  \label{tab:funcsflip}
\end{table}
In order to further understand the properties of the functions which arise as
solutions to our differential equations, we also explore the behavior of the
space of special functions under Galois transformations. 
More precisely, in table~\ref{tab:funcsflip} we show the number of 
linearly independent functions in the solutions to the two-loop integrals, $\{\vec{J}_1^{\;(n)}, \vec{J}_2^{\;(n)}, \mathcal{Z}(\vec{J}_2^{\;(n)})\}$, for $k = 1, \ldots, 4$
that are odd under the transformation $\sqrt{X} \to -\sqrt{X}$. The only Galois
transformation acting non-trivially at weight $1$ is that of the sign flip of
$\sqrt{\cayley{1}}$. Galois transformations associated to sign flips of five-point
square roots, i.e $\sqrt{\Delta_5}, \sqrt{r_2}, \sqrt{r_3}$, do not enter until
weight 3.
The Galois transformation associated to the sign flip of the nested root
$\sqrt{N_+}$ first acts non-trivially at weight 3.
Interestingly, the Galois transformation associated to the sign flip
of $\sqrt{r_1}$ first acts non-trivially at weight 4. 
Finally, we also studied the number of linearly independent functions arising in 
the two-loop integrals, $\{\vec{J}_1^{\;(n)}, \vec{J}_2^{\;(n)}, \mathcal{Z}(\vec{J}_2^{\;(n)})\}$, which are
odd under the action of $\alpha$. At weights 1 and 2 there are no such functions,
while at weights 3 and 4 there is a single such function.

\section{Numerical Evaluations of the Master Integrals}
\label{sec:numerics}

In this section, we present numerical results obtained for the master integrals.
The compact structure of their analytic differential
equations~\eqref{eq:diffFormDEQ} makes it naturally suitable for efficient numerical
evaluations. While a more detailed implementation ready for phenomenological
studies is left to future work, here we provide tools for their evaluation and
present benchmark values up to order $\mathcal{O}(\ep^4)$ for all of the
integrals in the physical phase space for the scattering process
in~\eqref{eq:ScatteringProcess}. In terms of Mandelstam invariants, this
space is defined by the relations in~\eqref{eq:realps}. Our numerical evaluations make use of the
public packages \amflow{}~\cite{Liu:2022chg} and
\diffexp{}~\cite{Hidding:2020ytt}.

\subsection{Boundary Values}
Solving linear differential equations requires a single set of boundary values, the
$\vec{b}_i^{\;(n)}(\vec{s}_0)$ vectors in equation~\eqref{eq:Iiter}. For
numerical solutions these boundary values can be computed to very high
precision, in generic regions of parameter space and up to high orders in $\ep$
with the auxiliary mass flow method~\cite{Liu:2017jxz,Liu:2021wks,Liu:2022tji}.
We use this method to extract boundary values, with $100$ decimal
digit precision, in the physical region employing the corresponding
implementation provided in the \amflow{} package~\cite{Liu:2022chg}. First, we
use \amflow{} to numerically compute a set of scalar master integrals in
the following phase space point in the physical region~\eqref{eq:realps}
\begin{equation}
\vec{s}_0 = \left\{\frac{562}{11}\;, \frac{89}{11}\;, -\frac{36}{13}\;, \frac{305}{3}\;, -\frac{52}{21}\;, \frac{9}{56}\;, \frac{360}{197}\right\}\;,
\label{eq:boundary}
\end{equation}
with $\vec{s}$ as in equation~\eqref{eq:vars} and where the figures are in units
of the regularization scale.
Afterwards, we perform basis transformations into the bases constructed in
section~\ref{sec:MasterIntegralBasisConstruction}, $\vec{I}_i(\vec{s}_0)$
($i=0,1,2$), keeping terms up to $\mathcal{O}(\ep^4)$. 
We deliver the boundary values in the ancillary files (see
section~\ref{sec:ancillary}).

\subsection{Numerical Results and Validation}
\label{sec:numresults}

In this section, we provide numerical benchmark results for the following
phase space points in the physical region~\eqref{eq:realps}
\begin{equation}
\begin{split}
\vec{s}_1 &= \left\{\frac{19}{3}\;, \frac{46}{3}\;, -\frac{24}{7}\;, \frac{383}{5}\;, -\frac{61}{28}\;, \frac{25}{118}\;, \frac{97}{896}\right\}\;, \\ 
\vec{s}_2 &= \left\{\frac{124}{3}\;, \frac{34}{3}\;, -\frac{100}{13}\;, \frac{518}{5}\;, -\frac{36}{5}\;, \frac{176}{255}\;, \frac{37}{9}\right\}\;, \\ 
\vec{s}_3 &= \left\{47\;, 5\;, -\frac{25}{12}\;, 96\;, -\frac{23}{49}\;, \frac{149}{593}\;, \frac{62}{61}\right\}\;, \\
\vec{s}_4 &= \left\{73781\;, 74098\;, -82315\;, 307009\;, -76978\;, (173)^2\;, (125)^2\right\}\;,
\label{eq:PSPs}
\end{split}
\end{equation}
with $\vec{s}$ as in equation~\eqref{eq:vars} and were figures are in units of
the regularization scale. 
These points have been chosen
randomly in the physical phase space, except for $\vec s_4$ where we have
forced the invariants $m_t^2$ and $q^2$ to have values associated to
corresponding parameters of the Standard Model of particle physics. We use the differential
equations~\eqref{eq:diffFormDEQ} to transport the master integrals $\vec{I}_i$ from
the boundary~\eqref{eq:boundary} to
the points $\vec{s}_j$ for $j= 1,\dots,4$.  To this end, we employ the method of
generalized series expansions (see e.g.
Refs.~\cite{Moriello:2019yhu,Abreu:2020jxa}) using the implementation provided
by the package \diffexp{}~\cite{Hidding:2020ytt}.

This procedure works directly for obtaining numerical results for the master
integrals $\vec{I}_0$ and $\vec{I}_2$ 
associated to the families $T_0$ and \Tht{} respectively. However, to evaluate
$\vec{I}_1(\vec{s}_j)$ we required an additional step. This is due to a feature
of the \diffexp{} package. \diffexp{} numerically solves the differential
equation by moving on a straight segment between the
initial and final points. As it does this, it requires to perform
analytic continuations when crossing singularities, which might be endpoints of
branch cuts. In particular \diffexp{} has implementations to handle the analytic
continuation of logarithmic functions and those of square roots of polynomials.
However, our $\ep$-factorizing basis of master integrals for the \Tpb{} family, specifically in
the \kite{} integrals shown in equations~\eqref{eq:T1_N64} and
\eqref{eq:T1_N65}, include the nested square roots $\sqrt{N_\pm}$ of
equation~\eqref{eq:Npm}.

Due to this feature, and for practical reasons, we choose to resolve this by 
constructing an \textit{auxiliary} integral basis where
the integrals in~\eqref{eq:T1_N64} and \eqref{eq:T1_N65} are replaced by
\begin{align}
&\mathcal{N}^{aux}_{64} = \ep^3(\mHsq)^2\left(\frac{1}{\rho_3} - \frac{1}{\rho_2}\right)\;,
\label{eq:N64decan}
\\
&\mathcal{N}^{aux}_{65} = \ep^3(\mHsq)^2\left(\frac{1}{\rho_3} + \frac{1}{\rho_2}\right)\;,
\label{eq:N65decan}
\end{align}
while all other integrals match our $\ep$-factorizing basis.
Using equation~\eqref{eq:changeofbasis} we transform the differential equations
of the $\ep$-factorizing basis to the auxiliary basis, which results in
differential equations matrices $B_i(\vec{s},\ep)$ that are explicitly
linear in $\ep$, as in equation~\eqref{eq:Blinear}. We explicitly checked for the
auxiliary basis the integrability condition
\begin{equation}
  \Big[B_i(\vec{s},\ep),B_j(\vec{s},\ep)\Big] =
  \frac{\partial B_j(\vec{s},\ep)}{\partial s_i} - \frac{\partial B_i(\vec{s},\ep)}{\partial s_j}\;.
\end{equation}
In this auxiliary basis, the basis integrals do not involve nested square roots,
and therefore neither do the matrices $B_i$. In this way, we construct a form of
the differential equation suitable for use with the \diffexp{} package.
We provide these expressions in the ancillary files (see
subsection~\ref{sec:ancillary}). 
After the transport is done we make a basis change from the auxiliary basis to
the $\ep$-factorizing basis. 
We note that the file size of the differential equations in the auxiliary basis
is considerably larger than the $\epsilon$-factorized basis. Indeed, this is a
general feature of working with an $\epsilon$-factorized basis, which renders
compact analytic expressions for the differential equations.

In tables \ref{tab:box_triangle_benchmark}, \ref{tab:pentabox_benchmark}, and
\ref{tab:pentabubble_benchmark} we
show numerical results up to $\mathcal{O}(\ep^4)$ for the point $\vec{s}_1$  of
equation~\eqref{eq:PSPs}. We include a selection of our master integrals in the
families \Tpb{} and \Tht{}, but in the ancillary files we provide high-precision
numerical results for all master integrals for the integral families $T_0$, \Tpb{}, \Tht{} and \ZTht{}. 
Tables
\ref{tab:box_triangle_benchmark} and \ref{tab:pentabox_benchmark} display some
of the most complex integrals in $T_1$: the box-triangle
integrals~\eqref{eq:T1_N88}--\eqref{eq:T1_N93} and the penta-box
integrals~\eqref{eq:T1_N109}--\eqref{eq:T1_N111}. We choose to display the
integrals~\eqref{eq:T1_N88}--\eqref{eq:T1_N93} since they involve five-point
kinematics, and mix with the \kite{} integrals via the differential equation
during integration. In addition we provide in table~\ref{tab:pentabubble_benchmark} the
penta-bubble integrals \eqref{eq:T2_N18} and \eqref{eq:T2_N19} of
\Tht{}.

\begin{table}
\centering
\begin{tabular}{cp{1.5cm}p{2.4cm}p{2.4cm}p{2.4cm}p{2.4cm}}
 \toprule
 & $\mathcal{O}(\ep^0)$& $\mathcal{O}(\ep^1)$&$\mathcal{O}(\ep^2)$ & $\mathcal{O}(\ep^3)$&$\mathcal{O}(\ep^4)$ \\
 \midrule
 $(\vec{I}_1)_{88}$& $0$ &  $0$& $-1.697405869$&$8.990085874$ $+ 2.959793778\ i$& $-23.70912261$ $ +  12.35416236\ i$ \\
 \midrule
 $(\vec{I}_1)_{89}$& $0$ &  $0$& $0$&$-3.703380133$ $ + 5.885655074 \ i $&$ -15.40231055$ $ - 6.37555295 \ i$ \\
 \midrule
 $(\vec{I}_1)_{90}$& $0$ &  $0$& $0$&$3.703380133$ $- 5.885655074 \ i $& $13.15415510$ $ + 20.45624479 \ i$ \\
 \midrule
 $(\vec{I}_1)_{91}$& $0$ &  $0$& $5.811380795$ $ - 2.687806077 \ i$&$-14.63593742$ $ + 31.14397715 \ i $&$ -66.82494671$ $ - 70.56864014 \ i $\\
 \midrule
 $(\vec{I}_1)_{92}$& $0$ &$ - 1.461994703 \ i$&$ -4.592991817$ $ + 4.774264642 \ i$&$-2.99771383$ $ - 17.32856509 \ i$ & $31.78963784$ $ - 7.30297630 \ i$ \\
 \midrule
 $(\vec{I}_1)_{93}$& $0$ &  $0.4534743273$& $ -2.546669141$ $-1.424631615 \ i$&$10.954658459$ $ + 0.602688704 \ i $&$ -12.24416802$ $ + 16.57486204 \ i$ \\
 \bottomrule
\end{tabular}
 \caption{Numerical results up to $\mathcal{O}(\ep^4)$ for the integrals
$(\vec{I}_1)_{88}$ through $(\vec{I}_1)_{93}$ evaluated at the phase space
point $\vec{s}_1$ of equation~\eqref{eq:PSPs}. See
equations~\eqref{eq:T1_N88}--\eqref{eq:T1_N93} for the definition of the integrals.}
 \label{tab:box_triangle_benchmark}
\end{table}
\begin{table}
\centering
\begin{tabular}{cp{1cm}p{2.6cm}p{2.6cm}p{2.6cm}p{2.6cm}}
 \toprule
 & $\mathcal{O}(\ep^0)$& $\mathcal{O}(\ep^1)$&$\mathcal{O}(\ep^2)$ & $\mathcal{O}(\ep^3)$&$\mathcal{O}(\ep^4)$ \\
 \midrule
 $(\vec{I}_1)_{109}$& $0$ &  $0$& $0$&$ -3.703380133$ $ + 5.885655074\ i$& $2.149576969$ $- 10.432322830\ i$ \\
 \midrule
 $(\vec{I}_1)_{110}$& $0$ &  $0$& $0$&$0$&$0$ \\
 \midrule
 $(\vec{I}_1)_{111}$& $0$ &  $0$& $-1.306045093$ $ - 12.647039669\ i $& $2.05552771$ $ + 25.35139955\ i$&$-85.55528965$ $ - 75.93834102\ i$ \\
 \bottomrule
\end{tabular}
 \caption{Numerical results up to $\mathcal{O}(\ep^4)$ for the integrals
$(\vec{I}_1)_{109}$ through $(\vec{I}_1)_{111}$ evaluated at the phase space
point $\vec{s}_1$ of equation~\eqref{eq:PSPs}. See
equations~\eqref{eq:T1_N109}--\eqref{eq:T1_N111} for the definition of the integrals.
Consistent with the fact that it is an evanescent integral (see e.g.~\cite{Gambuti:2023eqh})
we find that the value of $(\vec{I}_1)_{110}$ is 0 through $\mathcal{O}(\ep^4)$.
}
 \label{tab:pentabox_benchmark}
\end{table}

\begin{table}
\centering
\begin{tabular}{cp{1cm}p{2.6cm}p{2.6cm}p{2.6cm}p{2.6cm}}
 \toprule
 & $\mathcal{O}(\ep^0)$& $\mathcal{O}(\ep^1)$&$\mathcal{O}(\ep^2)$ & $\mathcal{O}(\ep^3)$&$\mathcal{O}(\ep^4)$ \\
 \midrule
 $(\vec{I}_2)_{18}$ & $0.5$ & $-4.931720031$ $ + 4.712388980\ i$ & $6.90383844$ $ - 36.51486280 \ i$ & $63.72515614$ $ + 86.40251641\ i$ & $-188.2874920$ $ - 14.5546057 \ i$ \\
 \midrule
  $(\vec{I}_2)_{19}$ & $0$ & $0$ & $0$ & $3.703380133$ $- 5.885655074 \ i$ & $11.33274441$ $+ 26.71395384 \ i$ \\
 \bottomrule
\end{tabular}
 \caption{Numerical results up to $\mathcal{O}(\ep^4)$ for the integrals
$(\vec{I}_2)_{18}$ and $(\vec{I}_2)_{19}$ evaluated at the phase space point
$\vec{s}_1$ of equation~\eqref{eq:PSPs}. See
equations~\eqref{eq:T2_N18} and \eqref{eq:T2_N19} for the definition of the integrals.
}
 \label{tab:pentabubble_benchmark}
\end{table}

We observe that high-precision evaluation can be achieved with our
differential equations and leave more detailed analysis of numerical features
to future work. Although we consider these numerical explorations
to be preliminary, we note that when compared to evaluations using the
\amflow{} package, the evaluations based on our differential equations took
more than two orders of magnitude less computation time to achieve results with
comparable precision, using comparable computing resources.

The high-precision numerical evaluations that we have obtained provide a highly
non-trivial validation of the analytic form of the differential equation that
we have computed. Indeed we compared the results obtained with independent
evaluations using \amflow{} in all the points $\vec{s}_j$ ($j\ge 1$). We find
excellent agreement, that is, agreement to $95$ or more decimal digits.
We have also performed comparisons with fully numerical integrations for a
handful of integrals via sector decomposition~\cite{Heinrich:2008si}
and tropical Feynman integration~\cite{Borinsky:2020rqs}, employing the
corresponding implementations in the packages \textsc{pySecDec}
\cite{Borowka:2017idc} and \textsc{feyntrop} \cite{Borinsky:2023jdv}. Agreement
is observed, though restricted to only the few decimal digits that the numerical
integration errors allow.

\subsection{Ancillary Files}
\label{sec:ancillary}
We provide a series of ancillary files containing our analytic results,
numerical benchmarks, and the computer script we use for numerical solutions to
the differential equations. Here, we describe each of the files included.
\begin{description}
 \item[\texttt{README.md}:] Instructions to run the computer script
\texttt{transport.wl} and a description of all ancillary files.
\item[\texttt{transport.wl}:] A script which performs the transport of all
  integral families to one of the phase space points in equation~\eqref{eq:PSPs},
  as specified by a command-line argument.
 \item[\texttt{roots.m}:] Contains the definition of all square roots (see
Section~\ref{sec:kinematics}) appearing in the differential equations.
 \item[\texttt{oneForms.m}] A list of all $152$ one-forms as described
in Section~\ref{sec:list_letters}.
 \item[\texttt{X/muijs.m}:] Replacement rules for the expressions for
$\mu_{ij}$ insertions in terms of inverse propagators $\rho_i$ for the integral
families 
\begin{center}
\texttt{X}
$\in \{$ \texttt{T0}, \texttt{T1}, \texttt{T2}, \texttt{ZT2} $\}\!\;$.
\end{center}
\item[\texttt{X/basis.m}:] Machine-readable files containing the definitions
  of our integral bases as written in the Appendices~~\ref{app:2looppbox},
  \ref{app:2looppbub}, and~\ref{app:1loop}.
 \item[\texttt{X/M\_alpha.m}:] The rational coefficient matrices $M_{\alpha}$ 
of the corresponding one-forms for the integral family \texttt{X}.
 \item[\texttt{T1/deq/d\_1.m}:] The entries of the differential equation matrix in the
auxiliary integral basis of \Tpb{} which can be written in d log form.
 \item[\texttt{T1/deq/d\{v12,v23,v34,v45,v15,mTsq,qsq\}\_\{0,1\}.m:}] These 14
files provide the extra differential equation matrices in the \Tpb{} auxiliary
basis, to complete the information included in \texttt{T1/deq/d\_1.m}.
 \item[\texttt{X/boundaries.m}:] Contain values of the $\ep$-factorizing 
integral basis (see appendix~\ref{app:2looppbox}-\ref{app:1loop}) at the point
$\vec{s}_0$ in equation~\eqref{eq:boundary} with $100$ digit accuracy for all
integral families.
 \item [\texttt{X/benchmarks/sn.m}:] Benchmark numerical results
with 30-digit accuracy for all master integrals at the points $\vec{s}_n$
specified in equation~\eqref{eq:PSPs}.
\item[\texttt{points.m}:] Machine-readable version of the physical phase space
  points in equation~\eqref{eq:PSPs}.
\end{description}
%

\section{Conclusions}
\label{sec:conclusions}

In this paper we have presented the first set of two-loop master integrals
needed for the NNLO QCD corrections to $t\bar{t}H$
production at hadron colliders. These seven-scale master integrals are some of
the most complex ones computed to date.
We have provided the master integrals needed for the calculation
of the two-loop leading-color QCD scattering amplitudes that are
proportional to the number of light flavors ($n_f$) for the processes $gg,q\bar{q}\to t\bar{t}H$.
We have constructed a basis of master integrals that satisfy
$\ep$-factorized differential equations and have computed the analytic form
of the differential
equations that they fulfill in a compact manner, by writing them in terms of $152$
differential one-forms. Some of these differential one-forms involve complicated
algebraic functions of invariants, including nested square root functions.
Using Chen's iterated integrals we also studied the properties of the functions
that arise in solutions of the master integrals. Furthermore, we have provided
high-precision numerical evaluations employing generalized series expansions
and boundary values obtained with the auxiliary mass flow method. These numerical
evaluations provide a highly non-trivial validation of our results.

Given the phenomenological relevance of the associated production of a
top-quark pair and a Higgs boson, our results will have an important impact in
the physics programs at the LHC and the high-luminosity LHC.
We expect to continue refining the numerical implementations of the integrals
presented here to allow fast and precise evaluations of the associated
scattering amplitudes.
We also anticipate to continue studying a larger set of master integrals as
needed for a complete set of scattering amplitudes for this process at leading
color, beyond the light-quark loop case, where preliminary investigations show
the presence of Feynman integrals with elliptic maximal cuts.

\section*{Acknowledgements}
%
We thank Samuel Abreu and Vasily Sotnikov for helpful conversations. 
The work of F.F.C., G.F.\ and L.R.\ is supported in part by the U.S. Department
of Energy under grant DE-SC0010102. 
This work has been made possible in part through the support of the FSU Council
on Research and Creativity (``Black Holes Under the Microscope"; SEED
Grant, 2023).
M.K. is supported by the DGAPA-PAPIIT grant IA102224 and the PIIF grant at
UNAM.
L.R. acknowledges the Aspen Center for Physics, supported by National Science
Foundation grant PHY-2210452, and its kind hospitality while she was
working on this project.
The authors acknowledge the Instituto de F\'{i}sica (UNAM) for providing
computing infrastructure and Carlos Ernesto L\'{o}pez Natar\'{e}n for his HPC
support.  
The computing for this project was partly performed on the HPC cluster at the
Research Computing Center at the Florida State University (FSU).

\appendix
\section{Master Integral Basis for the \Tpb{} Feynman Integral Family}
\label{app:2looppbox}

In this appendix we provide the definition of all master integrals that we have
computed for the Feynman integral family \Tpb{} which is shown in
equations~\eqref{eq:family}, \eqref{eq:T1props}, and \eqref{eq:T1isps}. We
organize them in subsections from the integrals with the least (3) to the
integrals with the most (8) propagators.

For each integral we provide information which exactly specifies it and can be
used to reproduce our results in any common software for computing Feynman
integrals. This includes:
\begin{description}
 \item[\textbf{IBP sector:}] A binary code computed for each integral sector, i.e.
for each group of integrals that share the same set of inverse propagators with
positive powers. A sector is defined by the non-negative propagator powers $\vec{\nu} =\{\nu_1,\ldots,\nu_{11}\}$
according to 
 \begin{equation*}
  \textrm{sector} \equiv \sum_{n=1}^{11} 2^{n-1}\Theta(\nu_n-1/2)\;.
 \end{equation*}
 \item[\textbf{Figure:}] Each integral sector is shown with a figure that
contains the associated propagators in accordance with the full family presented
in figure~\ref{fig:T1}.
 \item[\textbf{Numerator insertions:}] For each integral we present its
corresponding numerator insertion $\mathcal{N}^{(1)}_j$ where the superscript
indicates that this integrand belongs to the \Tpb{} family and where
the index $j$ is an integer between 1 and 111. That is, the integral is defined as
\begin{equation*}
 \left(\vec I_{1}\right)_j = \int \frac{d^d\ell_1}{i\pi^{d/2}}\frac{d^d\ell_2}{i\pi^{d/2}}
 \frac{\mathcal{N}^{(1)}_j}{\rho_1^{\nu_1}\rho_2^{\nu_2}\rho_3^{\nu_3}\rho_4^{\nu_4}\rho_5^{\nu_5}\rho_6^{\nu_6}\rho_7^{\nu_7}\rho_8^{\nu_8}}\;,
\end{equation*}
where $\nu_i \in \{0,1\}$.  Notice
that we employ kinematic invariants and functions defined in
section~\ref{sec:kinematics}.
\end{description}

For completeness, we also print the values of the integrals at weight 0 as used
in the discussion of iterated integral solutions of section~\ref{sec:iterints},
\begin{equation}
  \begin{split}
    \vec{I}_1^{\;(0)} = \Big\{&1, 1, 0, -\frac{1}{2}, 0, -1, 0, -1, 0, -1, 1, 0, 0, 0, 0, 0, 0, 0, 0, \frac{1}{2}, 0, 0, 0, 0, 0, 0, \\ 
    &0, 0, 1, 0, 0, 0, -\frac{1}{4}, 0, 0, -\frac{1}{4}, 0, 0, 0, -\frac{1}{6}, 0, -\frac{5}{12}, 0, -\frac{5}{12}, 0, -\frac{1}{6}, 0, -\frac{1}{4}, \\
    &0, -\frac{3}{4}, 0, -\frac{5}{6}, 0, 0, 0, 0, 0, 0, 0, 0, 0, 0, 0, 0, 0, -\frac{1}{4}, 0, 0, -\frac{5}{6}, 0, -\frac{3}{4}, 0, -\frac{1}{4}, \\
    &0, -\frac{5}{4}, 0, -\frac{5}{4}, 0, 0, 0, 0, 0, 0, 0, 0, 0, 0, 0, 0, 0, 0, 0, 0, 0, \frac{5}{24}, 0, \frac{1}{6}, 0, \frac{11}{24}, 0, \frac{5}{12}, \\
    &\frac{11}{24}, 0, \frac{5}{12}, \frac{5}{24}, 0, \frac{1}{6}, 0, 0, 0, 0 \Big\}\;.
  \end{split}
\end{equation}

\subsection*{$\vec I_1$: 3 Propagator Integrals}
$\qquad$

\begin{minipage}{0.3\textwidth}
{\flushleft Sector: 50}\\
\begin{figure}[H]
 \centering
 \includegraphics[width=4.5cm]{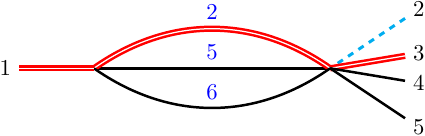}
\end{figure}
\end{minipage}
\begin{minipage}{0.65\textwidth}
\begin{align}
 \mathcal{N}^{(1)}_{1} &= \frac{\ep^2(1-2\ep)(1-3\ep)(2-3\ep)}{(1-4\ep)\mTsq}\;. \label{eq:T1_N1}
\end{align}
\end{minipage}

\begin{minipage}{0.3\textwidth}
{\flushleft Sector: 52}\\
\begin{figure}[H]
 \centering
 \includegraphics[width=4.5cm]{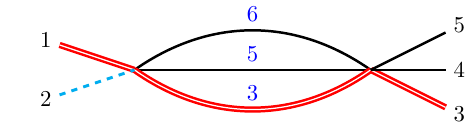}
\end{figure}
\end{minipage}
\begin{minipage}{0.65\textwidth}
\begin{align}
 \mathcal{N}^{(1)}_{2} &= \ep^2\left( \frac{\mHsq+\vij{12}}{\rho_6} - \frac{2\mTsq}{\rho_3}\right)\frac{1}{\rho_5}\;, \label{eq:T1_N2} \\
 \mathcal{N}^{(1)}_{3} &= \ep^2(\mHsq+\mTsq+\vij{12})\frac{1}{\rho_3\rho_5}\;. \label{eq:T1_N3}
\end{align}
\end{minipage}

\begin{minipage}{0.3\textwidth}
{\flushleft Sector: 56}\\
\begin{figure}[H]
 \centering
 \includegraphics[width=4.5cm]{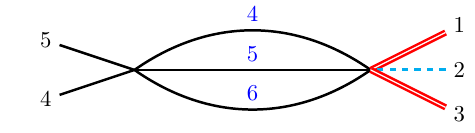}
\end{figure}
\end{minipage}
\begin{minipage}{0.65\textwidth}
\begin{align}
 \mathcal{N}^{(1)}_{4} &= \frac{\ep(1-2\ep)(1-3\ep)(2-3\ep)}{\vij{45}}\;. \label{eq:T1_N4}
\end{align}
\end{minipage}

\begin{minipage}{0.3\textwidth}
{\flushleft Sector: 82}\\
\begin{figure}[H]
 \centering
 \includegraphics[width=4.5cm]{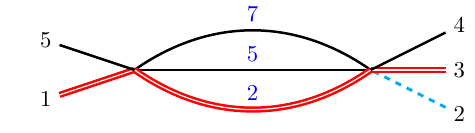}
\end{figure}
\end{minipage}
\begin{minipage}{0.65\textwidth}
\begin{align}
 \mathcal{N}^{(1)}_{5} &= \ep^2(\mTsq+\vij{15})\frac{1}{\rho_2\rho_7}\;, \label{eq:T1_N5} \\
 \mathcal{N}^{(1)}_{6} &= \ep^2\left( \frac{2\mTsq}{\rho_2} - \frac{\vij{15}}{\rho_5}\right)\frac{1}{\rho_7}\;. \label{eq:T1_N6} 
\end{align}
\end{minipage}

\begin{minipage}{0.3\textwidth}
{\flushleft Sector: 84}\\
\begin{figure}[H]
 \centering
 \includegraphics[width=4.5cm]{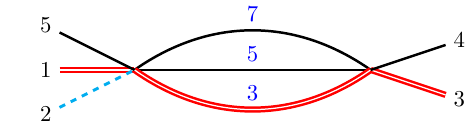}
\end{figure}
\end{minipage}
\begin{minipage}{0.65\textwidth}
\begin{align}
 \mathcal{N}^{(1)}_{7} &= \ep^2(\mTsq+\vij{34})\frac{1}{\rho_3\rho_7}\;, \label{eq:T1_N7} \\
 \mathcal{N}^{(1)}_{8} &= \ep^2\left( \frac{2\mTsq}{\rho_3} - \frac{\vij{34}}{\rho_5}\right)\frac{1}{\rho_7}\;. \label{eq:T1_N8} 
\end{align}
\end{minipage}

\begin{minipage}{0.3\textwidth}
{\flushleft Sector: 146}\\
\begin{figure}[H]
 \centering
 \includegraphics[width=4.5cm]{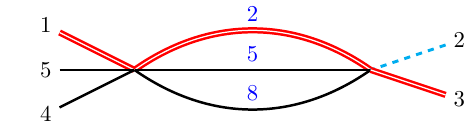}
\end{figure}
\end{minipage}
\begin{minipage}{0.65\textwidth}
\begin{align}
 \mathcal{N}^{(1)}_{9} &= \ep^2(\mHsq+\mTsq+\vij{23})\frac{1}{\rho_2\rho_8}\;, \label{eq:T1_N9} \\
 \mathcal{N}^{(1)}_{10} &= \ep^2\left( \frac{2\mTsq}{\rho_2} - \frac{\mHsq+\vij{23}}{\rho_5}\right)\frac{1}{\rho_8}\;. \label{eq:T1_N10} 
\end{align}
\end{minipage}

\begin{minipage}{0.3\textwidth}
{\flushleft Sector: 162}\\
\begin{figure}[H]
 \centering
 \includegraphics[width=4.5cm]{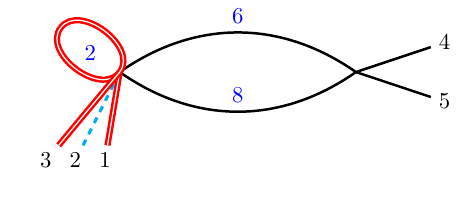}
\end{figure}
\end{minipage}
\begin{minipage}{0.65\textwidth}
\begin{align}
 \mathcal{N}^{(1)}_{11} &= \frac{\ep^2(1-2\ep)(1-\ep)}{\mTsq}\;. \label{eq:T1_N11} 
\end{align}
\end{minipage}

\subsection*{$\vec I_1$: 4 Propagator Integrals}
$\qquad$
\begin{minipage}{0.3\textwidth}
{\flushleft Sector: 54}\\
\begin{figure}[H]
 \centering
 \includegraphics[width=4.5cm]{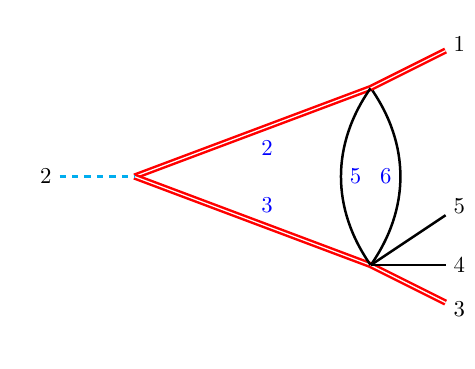}
\end{figure}
\end{minipage}
\begin{minipage}{0.65\textwidth}
\begin{align}
 \mathcal{N}^{(1)}_{12} &= \ep^3\sqrt{\gram{1}}\frac{1}{\rho_5}\;, \label{eq:T1_N12} \\
 \mathcal{N}^{(1)}_{13} &= \ep^2\sqrt{\cayley{1}}\Bigg[ (1-2\ep)\left(\frac{1}{\rho_2}+\frac{1}{\rho_3}\right) - \frac{\ep}{\rho_5}\Bigg]\;. \label{eq:T1_N13}
\end{align}
\end{minipage}

\begin{minipage}{0.3\textwidth}
{\flushleft Sector: 58}\\
\begin{figure}[H]
 \centering
 \includegraphics[width=4.5cm]{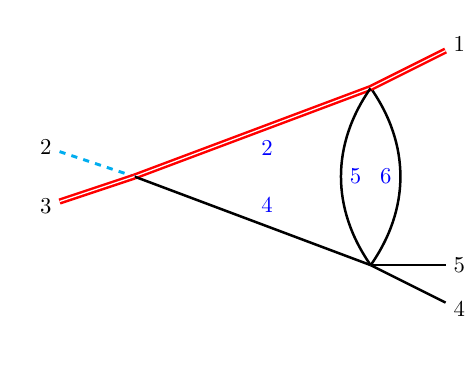}
\end{figure}
\end{minipage}
\begin{minipage}{0.65\textwidth}
\begin{align}
 \mathcal{N}^{(1)}_{14} &= \ep^3\sqrt{\gram{3}}\frac{1}{\rho_5}\;, \label{eq:T1_N14} \\
 \mathcal{N}^{(1)}_{15} &= \ep^2\Bigg[(1-2\ep)(\mHsq+\mTsq+\vij{23})\frac{1}{\rho_2} \nonumber \\
 & \qquad\qquad -\ep (\mHsq+\vij{23}+\vij{45})\frac{1}{2\rho_5}\Bigg]\; \label{eq:T1_N15} .
\end{align}
\end{minipage}

\begin{minipage}{0.3\textwidth}
{\flushleft Sector: 60}\\
\begin{figure}[H]
 \centering
 \includegraphics[width=4.5cm]{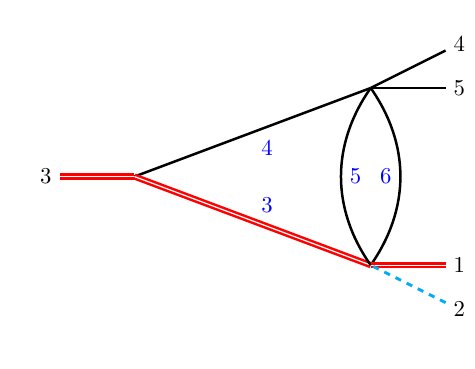}
\end{figure}
\end{minipage}
\begin{minipage}{0.65\textwidth}
\begin{align}
 \mathcal{N}^{(1)}_{16} &= \ep^3 \sqrt{\gram{4}}\frac{1}{\rho_5}\;. \label{eq:T1_N16}
\end{align}
\end{minipage}

\begin{minipage}{0.3\textwidth}
{\flushleft Sector: 85}\\
\begin{figure}[H]
 \centering
 \includegraphics[width=4.5cm]{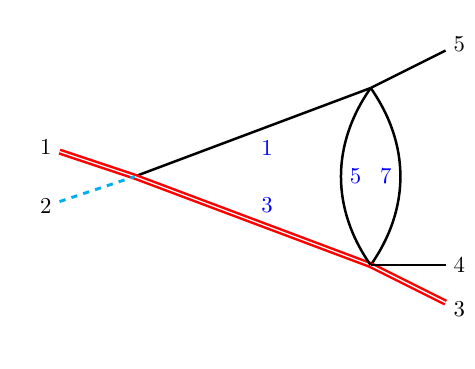}
\end{figure}
\end{minipage}
\begin{minipage}{0.65\textwidth}
\begin{align}
 \mathcal{N}^{(1)}_{17} &= \ep^3(\mHsq+\vij{12}-\vij{34})\frac{1}{\rho_5}\;. \label{eq:T1_N17}
\end{align}
\end{minipage}

\begin{minipage}{0.3\textwidth}
{\flushleft Sector: 86}\\
\begin{figure}[H]
 \centering
 \includegraphics[width=4.5cm]{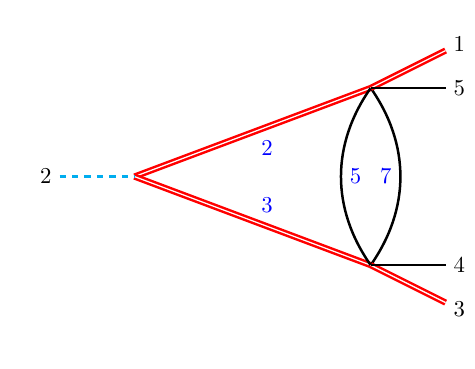}
\end{figure}
\end{minipage}
\begin{minipage}{0.65\textwidth}
\begin{align}
 \mathcal{N}^{(1)}_{18} &= \ep^3 \sqrt{\gram{5}}\frac{1}{\rho_5}\;, \label{eq:T1_N18} \\
 \mathcal{N}^{(1)}_{19} &= \ep^2 \sqrt{\cayley{1}}\Bigg[ (1-2\ep)\left(\frac{1}{\rho_2}+\frac{1}{\rho_3}\right) - \frac{\ep}{\rho_5}\Bigg]\;. \label{eq:T1_N19}
\end{align}
\end{minipage}

\begin{minipage}{0.3\textwidth}
{\flushleft Sector: 89}\\
\begin{figure}[H]
 \centering
 \includegraphics[width=4.5cm]{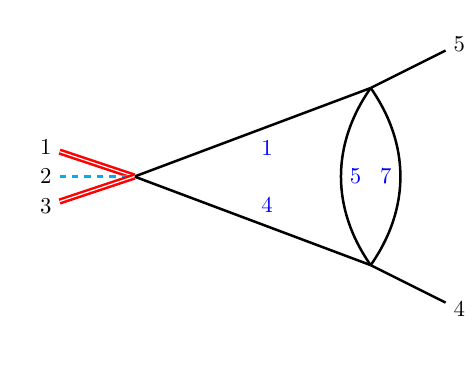}
\end{figure}
\end{minipage}
\begin{minipage}{0.65\textwidth}
\begin{align}
 \mathcal{N}^{(1)}_{20} &= \ep^2(1-2\ep)(1-3\ep)\;. \label{eq:T1_N20}
\end{align}
\end{minipage}

\begin{minipage}{0.3\textwidth}
{\flushleft Sector: 90}\\
\begin{figure}[H]
 \centering
 \includegraphics[width=4.5cm]{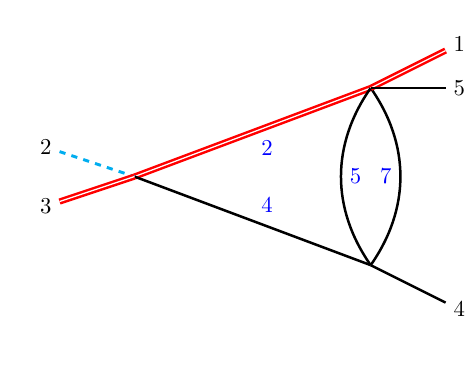}
\end{figure}
\end{minipage}
\begin{minipage}{0.65\textwidth}
\begin{align}
 \mathcal{N}^{(1)}_{21} &=\ep^3 (\mHsq+\vij{23}-\vij{15})\frac{1}{\rho_5}\;. \label{eq:T1_N21}
\end{align}
\end{minipage}

\begin{minipage}{0.3\textwidth}
{\flushleft Sector: 147}\\
\begin{figure}[H]
 \centering
 \includegraphics[width=4.5cm]{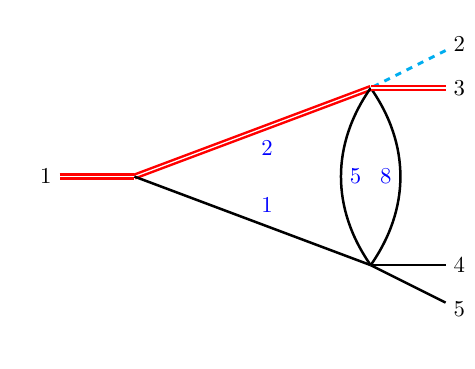}
\end{figure}
\end{minipage}
\begin{minipage}{0.65\textwidth}
\begin{align}
 \mathcal{N}^{(1)}_{22} &= \ep^3 \sqrt{\gram{3}}\frac{1}{\rho_5}\;. \label{eq:T1_N22}
\end{align}
\end{minipage}

\begin{minipage}{0.3\textwidth}
{\flushleft Sector: 149}\\
\begin{figure}[H]
 \centering
 \includegraphics[width=4.5cm]{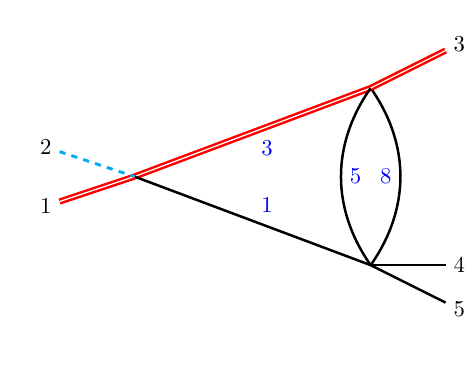}
\end{figure}
\end{minipage}
\begin{minipage}{0.65\textwidth}
\begin{align}
 \mathcal{N}^{(1)}_{23} &= \ep^3\sqrt{\gram{4}}\frac{1}{\rho_5}\;, \label{eq:T1_N23} \\
 \mathcal{N}^{(1)}_{24} &= \ep^2\Bigg[(1-2\ep)(\mHsq+\mTsq+\vij{12})\frac{1}{\rho_3} \nonumber \\
 &\qquad\qquad - \ep (\mHsq+\vij{12}+\vij{45})\frac{1}{2\rho_5}\Bigg]\; \label{eq:T1_N24} .
\end{align}
\end{minipage}

\begin{minipage}{0.3\textwidth}
{\flushleft Sector: 150}\\
\begin{figure}[H]
 \centering
 \includegraphics[width=4.5cm]{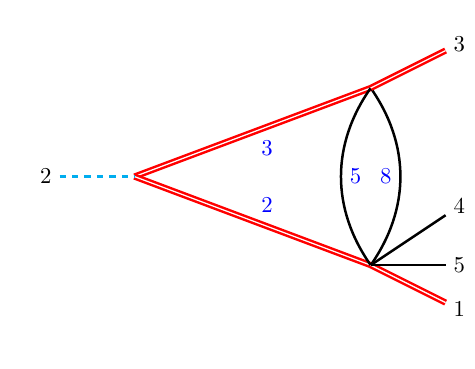}
\end{figure}
\end{minipage}
\begin{minipage}{0.65\textwidth}
\begin{align}
 \mathcal{N}^{(1)}_{25} &= \ep^3\sqrt{\gram{2}}\frac{1}{\rho_5}\;, \label{eq:T1_N25} \\
 \mathcal{N}^{(1)}_{26} &= \ep^2\sqrt{\cayley{1}}\Bigg[ (1-2\ep)\left(\frac{1}{\rho_2}+\frac{1}{\rho_3}\right) - \frac{\ep}{\rho_5}\Bigg]\;. \label{eq:T1_N26}
\end{align}
\end{minipage}

\begin{minipage}{0.3\textwidth}
{\flushleft Sector: 165}\\
\begin{figure}[H]
 \centering
 \includegraphics[width=4.5cm]{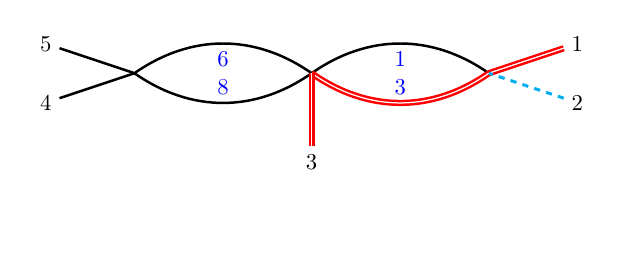}
\end{figure}
\end{minipage}
\begin{minipage}{0.65\textwidth}
\begin{align}
 \mathcal{N}^{(1)}_{27} &= \ep^2(1-2\ep)(\mHsq+\mTsq+\vij{12})\frac{1}{\rho_3}\;. \label{eq:T1_N27}
\end{align}
\end{minipage}

\begin{minipage}{0.3\textwidth}
{\flushleft Sector: 166}\\
\begin{figure}[H]
 \centering
 \includegraphics[width=4.5cm]{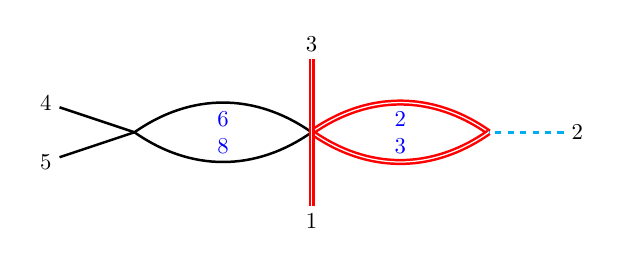}
\end{figure}
\end{minipage}
\begin{minipage}{0.65\textwidth}
\begin{align}
 \mathcal{N}^{(1)}_{28} &= \ep^2(1-2\ep)\sqrt{\cayley{1}}\frac{1}{\rho_3}\;. \label{eq:T1_N28}
\end{align}
\end{minipage}

\begin{minipage}{0.3\textwidth}
{\flushleft Sector: 169}\\
\begin{figure}[H]
 \centering
 \includegraphics[width=4.5cm]{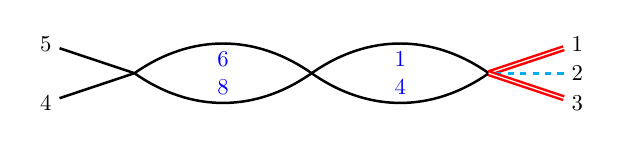}
\end{figure}
\end{minipage}
\begin{minipage}{0.65\textwidth}
\begin{align}
 \mathcal{N}^{(1)}_{29} &= \ep^2(1-2\ep)^2\;. \label{eq:T1_N29}
\end{align}
\end{minipage}

\begin{minipage}{0.3\textwidth}
{\flushleft Sector: 170}\\
\begin{figure}[H]
 \centering
 \includegraphics[width=4.5cm]{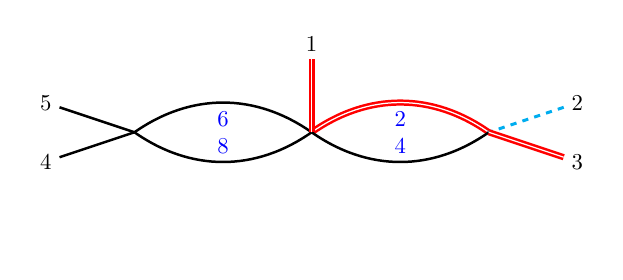}
\end{figure}
\end{minipage}
\begin{minipage}{0.65\textwidth}
\begin{align}
 \mathcal{N}^{(1)}_{30} &= \ep^2(1-2\ep)(\mHsq+\mTsq+\vij{23})\frac{1}{\rho_2}\;. \label{eq:T1_N30}
\end{align}
\end{minipage}

\begin{minipage}{0.3\textwidth}
{\flushleft Sector: 178}\\
\begin{figure}[H]
 \centering
 \includegraphics[width=4.5cm]{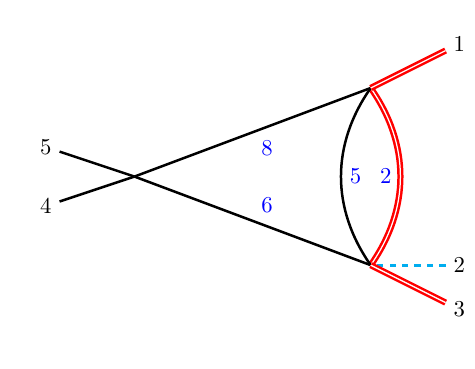}
\end{figure}
\end{minipage}
\begin{minipage}{0.65\textwidth}
\begin{align}
 \mathcal{N}^{(1)}_{31} &= \ep^3 \sqrt{\gram{3}}\frac{1}{\rho_2}\;, \label{eq:T1_N31} \\
 \mathcal{N}^{(1)}_{32} &= \ep^3 \sqrt{\gram{3}}\frac{1}{\rho_5}\;, \label{eq:T1_N32} \\
 \mathcal{N}^{(1)}_{33} &= \ep^2\Bigg[ \frac{\mTsq\vij{45}}{\rho_2\rho_6} + \ep(\mHsq+\vij{23}-\vij{45})\left(\frac{1}{\rho_2} + \frac{1}{2\rho_5}\right)\Bigg]\;. \label{eq:T1_N33}
\end{align}
\end{minipage}

\begin{minipage}{0.3\textwidth}
{\flushleft Sector: 180}\\
\begin{figure}[H]
 \centering
 \includegraphics[width=4.5cm]{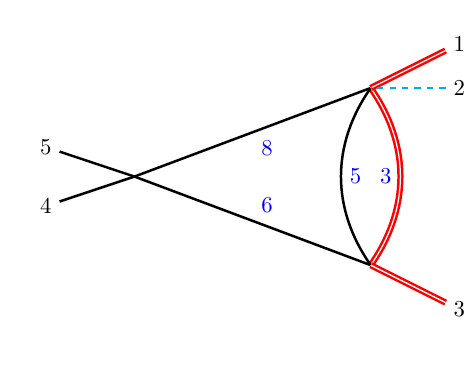}
\end{figure}
\end{minipage}
\begin{minipage}{0.65\textwidth}
\begin{align}
 \mathcal{N}^{(1)}_{34} &= \ep^3 \sqrt{\gram{4}}\frac{1}{\rho_3}\;, \label{eq:T1_N34} \\
 \mathcal{N}^{(1)}_{35} &= \ep^3 \sqrt{\gram{4}}\frac{1}{\rho_5}\;, \label{eq:T1_N35} \\
 \mathcal{N}^{(1)}_{36} &= \ep^2\Bigg[ \frac{\mTsq\vij{45}}{\rho_3\rho_8} + \ep(\mHsq+\vij{12}-\vij{45})\left(\frac{1}{\rho_3} + \frac{1}{2\rho_5}\right)\Bigg]\;. \label{eq:T1_N36}
\end{align}
\end{minipage}
\subsection*{$\vec I_1$: 5 Propagator Integrals}
$\qquad$
\begin{minipage}{0.3\textwidth}
{\flushleft Sector: 62}\\
\begin{figure}[H]
 \centering
 \includegraphics[width=4.5cm]{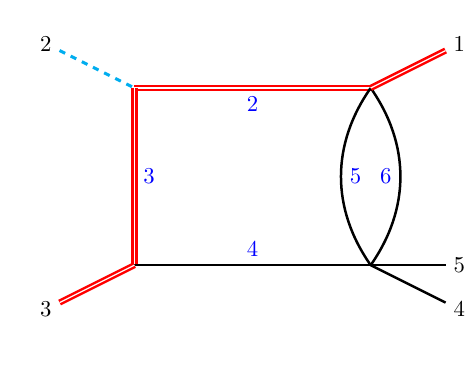}
\end{figure}
\end{minipage}
\begin{minipage}{0.65\textwidth}
\begin{align}
 \mathcal{N}^{(1)}_{37} &= \ep^3(1-2\ep)\sqrt{\gram{2}}\;, \label{eq:T1_N37} \\
 \mathcal{N}^{(1)}_{38} &= \ep^3 \sqrt{\cayley{2}}\frac{1}{\rho_5}\;. \label{eq:T1_N38}
\end{align}
\end{minipage}

\begin{minipage}{0.3\textwidth}
{\flushleft Sector: 87}\\
\begin{figure}[H]
 \centering
 \includegraphics[width=4.5cm]{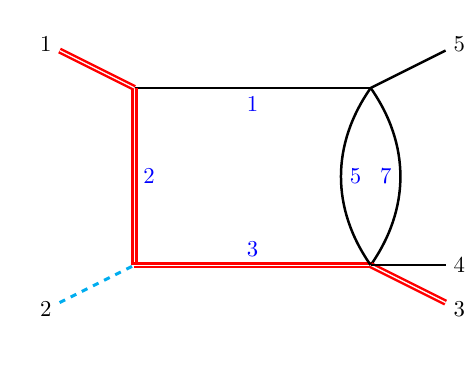}
\end{figure}
\end{minipage}
\begin{minipage}{0.65\textwidth}
\begin{align}
 \mathcal{N}^{(1)}_{39} &= \ep^3(1-2\ep)\sqrt{\gram{1}}\;, \label{eq:T1_N39} \\
 \mathcal{N}^{(1)}_{40} &= \ep^3 (\mHsq+\vij{12})\vij{15}\frac{1}{\rho_5}\;. \label{eq:T1_N40}
\end{align}
\end{minipage}

\begin{minipage}{0.3\textwidth}
{\flushleft Sector: 91}\\
\begin{figure}[H]
 \centering
 \includegraphics[width=4.5cm]{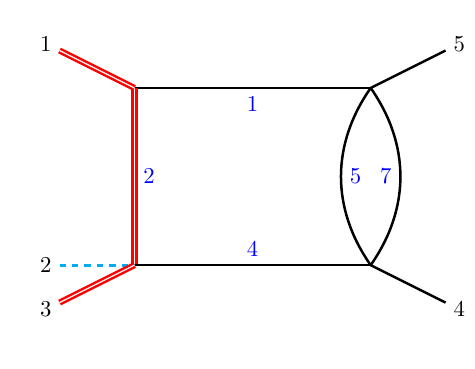}
\end{figure}
\end{minipage}
\begin{minipage}{0.65\textwidth}
\begin{align}
 \mathcal{N}^{(1)}_{41} &= \ep^3(1-2\ep)\sqrt{\gram{3}}\;, \label{eq:T1_N41} \\
 \mathcal{N}^{(1)}_{42} &= \ep^3 \vij{45}\vij{15}\frac{1}{\rho_5}\;. \label{eq:T1_N42}
\end{align}
\end{minipage}

\begin{minipage}{0.3\textwidth}
{\flushleft Sector: 93}\\
\begin{figure}[H]
 \centering
 \includegraphics[width=4.5cm]{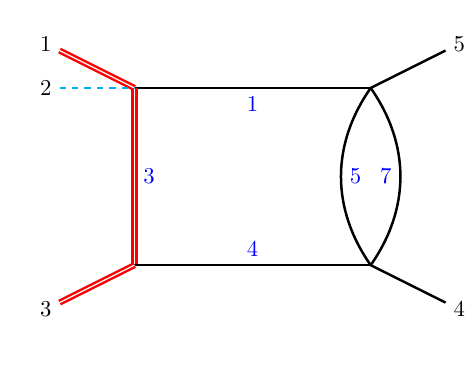}
\end{figure}
\end{minipage}
\begin{minipage}{0.65\textwidth}
\begin{align}
 \mathcal{N}^{(1)}_{43} &= \ep^3(1-2\ep)\sqrt{\gram{4}}\;, \label{eq:T1_N43} \\
 \mathcal{N}^{(1)}_{44} &= \ep^3 \vij{34}\vij{45}\frac{1}{\rho_5}\;. \label{eq:T1_N44}
\end{align}
\end{minipage}

\begin{minipage}{0.3\textwidth}
{\flushleft Sector: 94}\\
\begin{figure}[H]
 \centering
 \includegraphics[width=4.5cm]{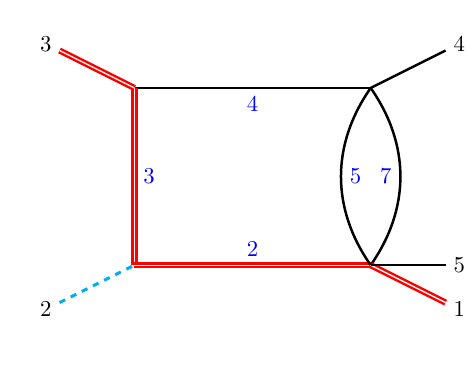}
\end{figure}
\end{minipage}
\begin{minipage}{0.65\textwidth}
\begin{align}
 \mathcal{N}^{(1)}_{45} &= \ep^3(1-2\ep)\sqrt{\gram{2}}\;, \label{eq:T1_N45} \\
 \mathcal{N}^{(1)}_{46} &= \ep^3 \vij{34}(\mHsq+\vij{12})\frac{1}{\rho_5}\;. \label{eq:T1_N46}
\end{align}
\end{minipage}

\begin{minipage}{0.3\textwidth}
{\flushleft Sector: 118}\\
\begin{figure}[H]
 \centering
 \includegraphics[width=4.5cm]{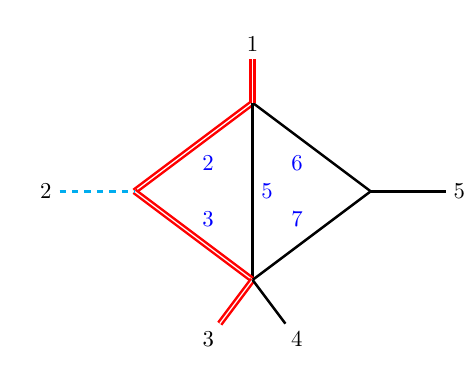}
\end{figure}
\end{minipage}
\begin{minipage}{0.65\textwidth}
\begin{align}
 \mathcal{N}^{(1)}_{47} &= \ep^4(\mHsq+\vij{12}+\vij{15}-\vij{34})\;, \label{eq:T1_N47} \\
 \mathcal{N}^{(1)}_{48} &= \ep^3\vij{15}(\mHsq+\vij{12})\frac{1}{\rho_5}\;. \label{eq:T1_N48}
\end{align}
\end{minipage}

\begin{minipage}{0.3\textwidth}
{\flushleft Sector: 122}\\
\begin{figure}[H]
 \centering
 \includegraphics[width=4.5cm]{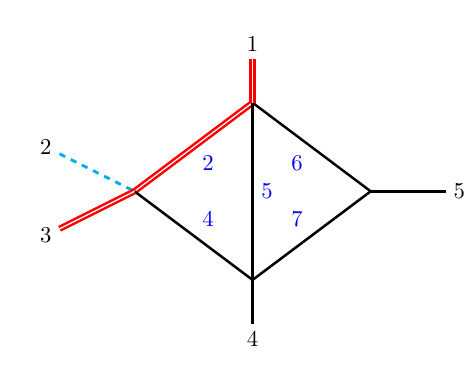}
\end{figure}
\end{minipage}
\begin{minipage}{0.65\textwidth}
\begin{align}
 \mathcal{N}^{(1)}_{49} &= \ep^4(\vij{45}+\vij{15})\;, \label{eq:T1_N49} \\
 \mathcal{N}^{(1)}_{50} &= \ep^3\vij{45}\vij{15}\frac{1}{\rho_5}\;. \label{eq:T1_N50}
\end{align}
\end{minipage}

\begin{minipage}{0.3\textwidth}
{\flushleft Sector: 124}\\
\begin{figure}[H]
 \centering
 \includegraphics[width=4.5cm]{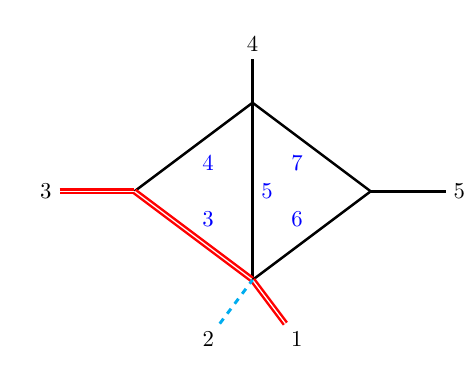}
\end{figure}
\end{minipage}
\begin{minipage}{0.65\textwidth}
\begin{align}
 \mathcal{N}^{(1)}_{51} &= \ep^4(\vij{34}+\vij{45}-\vij{12}-\mHsq)\;, \label{eq:T1_N51} \\
 \mathcal{N}^{(1)}_{52} &= \ep^3\vij{45}\vij{34}\frac{1}{\rho_5}\;. \label{eq:T1_N52}
\end{align}
\end{minipage}

\begin{minipage}{0.3\textwidth}
{\flushleft Sector: 151}\\
\begin{figure}[H]
 \centering
 \includegraphics[width=4.5cm]{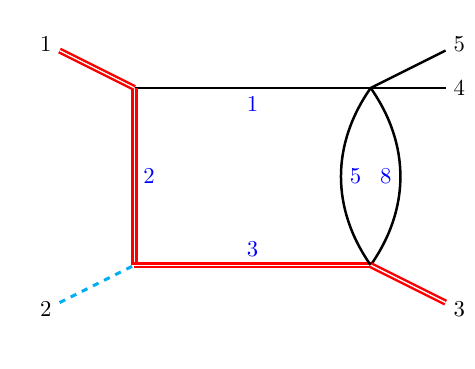}
\end{figure}
\end{minipage}
\begin{minipage}{0.65\textwidth}
\begin{align}
 \mathcal{N}^{(1)}_{53} &= \ep^3(1-2\ep)\sqrt{\gram{1}}\;, \label{eq:T1_N53} \\
 \mathcal{N}^{(1)}_{54} &= \ep^3\sqrt{\cayley{2}}\frac{1}{\rho_5}\;. \label{eq:T1_N54}
\end{align}
\end{minipage}

\begin{minipage}{0.3\textwidth}
{\flushleft Sector: 167}\\
\begin{figure}[H]
 \centering
 \includegraphics[width=4.5cm]{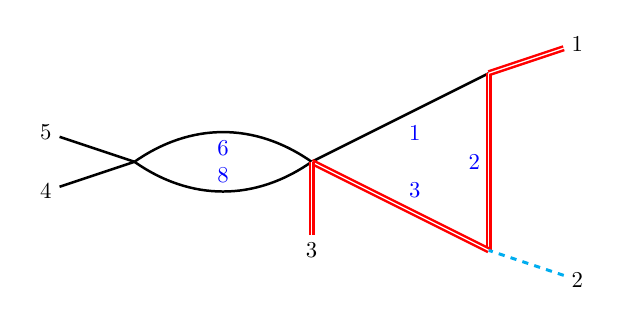}
\end{figure}
\end{minipage}
\begin{minipage}{0.65\textwidth}
\begin{align}
 \mathcal{N}^{(1)}_{55} &= \ep^3(1-2\ep)\sqrt{\gram{1}}\;. \label{eq:T1_N55}
\end{align}
\end{minipage}

\begin{minipage}{0.3\textwidth}
{\flushleft Sector: 171}\\
\begin{figure}[H]
 \centering
 \includegraphics[width=4.5cm]{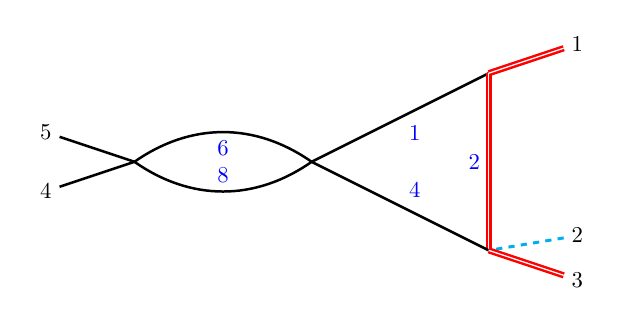}
\end{figure}
\end{minipage}
\begin{minipage}{0.65\textwidth}
\begin{align}
 \mathcal{N}^{(1)}_{56} &= \ep^3(1-2\ep)\sqrt{\gram{3}}\;. \label{eq:T1_N56}
\end{align}
\end{minipage}

\begin{minipage}{0.3\textwidth}
{\flushleft Sector: 173}\\
\begin{figure}[H]
 \centering
 \includegraphics[width=4.5cm]{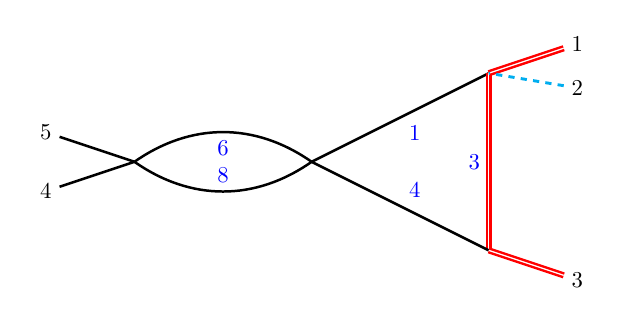}
\end{figure}
\end{minipage}
\begin{minipage}{0.65\textwidth}
\begin{align}
 \mathcal{N}^{(1)}_{57} &= \ep^3(1-2\ep)\sqrt{\gram{4}}\;. \label{eq:T1_N57}
\end{align}
\end{minipage}

\begin{minipage}{0.3\textwidth}
{\flushleft Sector: 174}\\
\begin{figure}[H]
 \centering
 \includegraphics[width=4.5cm]{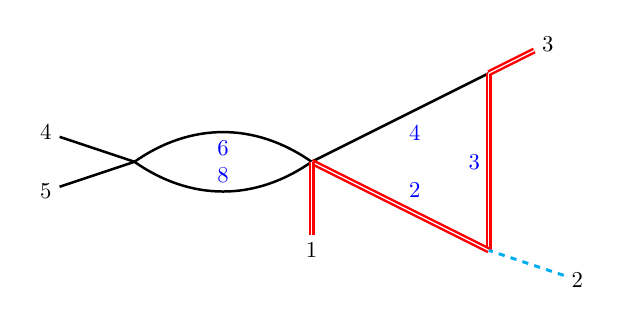}
\end{figure}
\end{minipage}
\begin{minipage}{0.65\textwidth}
\begin{align}
 \mathcal{N}^{(1)}_{58} &= \ep^3(1-2\ep)\sqrt{\gram{2}}\;. \label{eq:T1_N58}
\end{align}
\end{minipage}

\begin{minipage}{0.3\textwidth}
{\flushleft Sector: 181}\\
\begin{figure}[H]
 \centering
 \includegraphics[width=4.5cm]{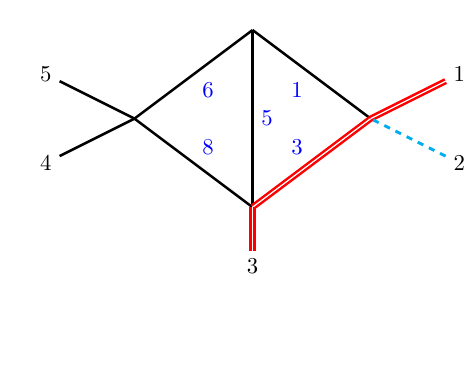}
\end{figure}
\end{minipage}
\begin{minipage}{0.65\textwidth}
\begin{align}
 \mathcal{N}^{(1)}_{59} &= \ep^4\sqrt{\gram{4}}\;. \label{eq:T1_N59}
\end{align}
\end{minipage}

\begin{minipage}{0.3\textwidth}
{\flushleft Sector: 182}\\
\begin{figure}[H]
 \centering
 \includegraphics[width=4.5cm]{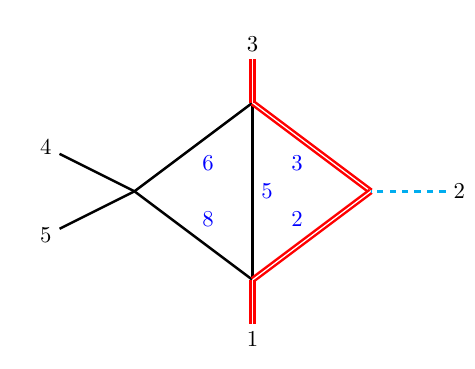}
\end{figure}
\end{minipage}
\begin{minipage}{0.65\textwidth}
\begin{align}
 \mathcal{N}^{(1)}_{60} &= \ep^4 \sqrt{r_1}\;, \label{eq:T1_N60} \\
 \mathcal{N}^{(1)}_{61} &= \ep^3 \vij{45} \sqrt{\gram{2}}\frac{1}{\rho_8}\;, \label{eq:T1_N61} \\
 \mathcal{N}^{(1)}_{62} &= \ep^3 \vij{45} \sqrt{\gram{1}}\frac{1}{\rho_6}\;, \label{eq:T1_N62} \\
 \mathcal{N}^{(1)}_{63} &= \ep^3 \sqrt{\cayley{2}}\frac{1}{\rho_5}\;, \label{eq:T1_N63} \\
 \mathcal{N}^{(1)}_{64} &= \ep^3 \left[\frac{\sqrt{N_+}}{2} \left(\frac{1}{\rho_3}-\frac{1}{\rho_2}\right) + \frac{\sqrt{\cayley{1}}\sqrt{N_-}}{2\mHsq} \left(\frac{1}{\rho_3}+\frac{1}{\rho_2}\right)\right]\;, \label{eq:T1_N64} \\
 \mathcal{N}^{(1)}_{65} &= \ep^3 \left[\frac{\sqrt{N_-}}{2} \left(\frac{1}{\rho_3}-\frac{1}{\rho_2}\right) + \frac{\sqrt{\cayley{1}}\sqrt{N_+}}{2\mHsq} \left(\frac{1}{\rho_3}+\frac{1}{\rho_2}\right)\right]\;, \label{eq:T1_N65} \\
 \mathcal{N}^{(1)}_{66} &= \ep^2 \frac{\mTsq \vij{45}(\mHsq+\vij{12})(\mHsq+\vij{23})}{2\mHsq+\vij{12}+\vij{23}}\left(\frac{1}{\rho_2\rho_6} + \frac{1}{\rho_3\rho_8}\right) \nonumber \\ 
&+ \ep^3\left(C_{66}^{(1)} \frac{1}{\rho_5} + C_{66}^{(2)} \frac{1}{\rho_6} + C_{66}^{(3)}\frac{1}{\rho_8} \right)\nonumber \\
&+ \ep^3\left( C_{66}^{(4)}\left(\frac{1}{\rho_3}+\frac{1}{\rho_2}\right) + C_{66}^{(5)}\left(\frac{1}{\rho_3}-\frac{1}{\rho_2}\right)\right) \label{eq:T1_N66}\\
&+ C_{66}^{(6)}\left[\rho_2~\mathcal{N}^{(1)}_{36}-\rho_3~\mathcal{N}^{(1)}_{33} + \rho_2\rho_8~\mathcal{N}^{(1)}_3\right] + C_{66}^{(7)}\left[\rho_5~\mathcal{N}^{(1)}_{28}\right] \nonumber \\
&+ C_{66}^{(8)}\left[\rho_6~\mathcal{N}^{(1)}_{26}-\rho_8~\mathcal{N}^{(1)}_{13}\right] + C_{66}^{(9)}\left[\rho_3\rho_6~\mathcal{N}^{(1)}_{10} + \rho_2\rho_8~\mathcal{N}^{(1)}_2\right] \nonumber \\
&+ C_{66}^{(10)}\left[\rho_3\rho_6~\mathcal{N}^{(1)}_9\right]\;. \nonumber
\end{align}
\end{minipage}

The coefficients for $\mathcal{N}^{(1)}_{66}$ read
\begin{align}
 C_{66}^{(1)} & = \frac{(\mHsq+\vij{12})(\mHsq+\vij{23})-(\mHsq-2\mTsq)\vij{45}}{2}\;,\\
 C_{66}^{(2)} & = \frac{(2\mTsq+\vij{12})(\mHsq+\vij{23})\vij{45}}{2\mHsq+\vij{12}+\vij{23}}\;,\\
 C_{66}^{(3)} & = \frac{(2\mTsq+\vij{23})(\mHsq+\vij{12})\vij{45}}{2\mHsq+\vij{12}+\vij{23}}\;,\\
 C_{66}^{(4)} & = \frac{1}{2}\left[\frac{N_b-\mTsq(\vij{12}-\vij{23})^2-2\mHsq\vij{45}^2}{2\mHsq+\vij{12}+\vij{23}} +(\mHsq+2\mTsq)\vij{45}\right]\;,\\
 C_{66}^{(5)} & = \frac{(\vij{12}-\vij{23})}{2}\left[(\mHsq-\mTsq)- \frac{\mHsq\vij{45}}{2\mHsq+\vij{12}+\vij{23}}\right]\;,\\
 C_{66}^{(6)} & = \frac{2(\mHsq+\vij{23})}{2\mHsq+\vij{12}+\vij{23}}\;,\\
 C_{66}^{(7)} & = \frac{\mHsq-2\mTsq}{\sqrt{\cayley{1}}}\;,\\
 C_{66}^{(8)} & = \frac{(\mHsq-2\mTsq)(\vij{12}-\vij{23})}{2(2\mHsq+\vij{12}+\vij{23})\sqrt{\cayley{1}}}\;,\\
 C_{66}^{(9)} & = \frac{\mHsq+\vij{12}}{2(2\mHsq+\vij{12}+\vij{23})}\;,\\
 C_{66}^{(10)} & = \frac{\vij{12}-\vij{23}}{2\mHsq+\vij{12}+\vij{23}}\;.
\end{align}

\begin{minipage}{0.3\textwidth}
{\flushleft Sector: 186}\\
\begin{figure}[H]
 \centering
 \includegraphics[width=4.5cm]{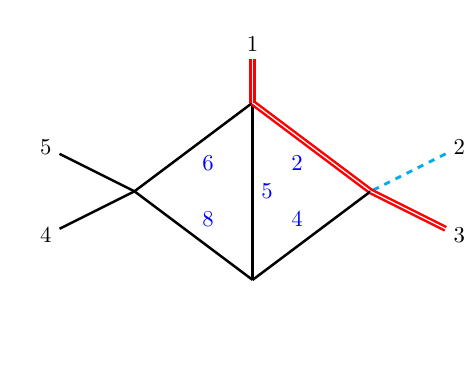}
\end{figure}
\end{minipage}
\begin{minipage}{0.65\textwidth}
\begin{align}
 \mathcal{N}^{(1)}_{67} &= \ep^4 \sqrt{\gram{3}}\;. \label{eq:T1_N67}
\end{align}
\end{minipage}

\begin{minipage}{0.3\textwidth}
{\flushleft Sector: 211}\\
\begin{figure}[H]
 \centering
 \includegraphics[width=4.5cm]{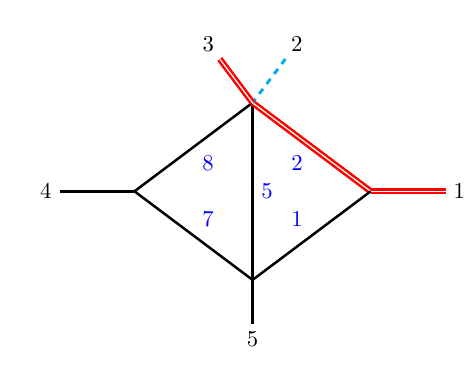}
\end{figure}
\end{minipage}
\begin{minipage}{0.65\textwidth}
\begin{align}
 \mathcal{N}^{(1)}_{68} &= \ep^4(\vij{15}+\vij{45}-\vij{23}-\mHsq)\;, \label{eq:T1_N68} \\
 \mathcal{N}^{(1)}_{69} &= \ep^3\vij{45}\vij{15}\frac{1}{\rho_5}\;. \label{eq:T1_N69}
\end{align}
\end{minipage}

\begin{minipage}{0.3\textwidth}
{\flushleft Sector: 213}\\
\begin{figure}[H]
 \centering
 \includegraphics[width=4.5cm]{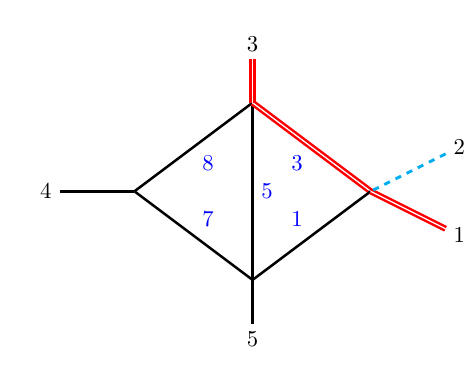}
\end{figure}
\end{minipage}
\begin{minipage}{0.65\textwidth}
\begin{align}
 \mathcal{N}^{(1)}_{70} &= \ep^4(\vij{34}+\vij{45})\;, \label{eq:T1_N70} \\
 \mathcal{N}^{(1)}_{71} &= \ep^3 \vij{34}\vij{45}\frac{1}{\rho_5}\;. \label{eq:T1_N71}
\end{align}
\end{minipage}

\begin{minipage}{0.3\textwidth}
{\flushleft Sector: 214}\\
\begin{figure}[H]
 \centering
 \includegraphics[width=4.5cm]{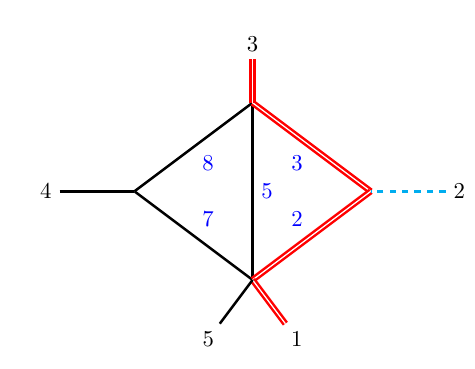}
\end{figure}
\end{minipage}
\begin{minipage}{0.65\textwidth}
\begin{align}
 \mathcal{N}^{(1)}_{72} &= \ep^4(\mHsq+\vij{23}+\vij{34}-\vij{15})\;, \label{eq:T1_N72} \\
 \mathcal{N}^{(1)}_{73} &= \ep^3 \vij{34}(\mHsq+\vij{23})\frac{1}{\rho_5}\;. \label{eq:T1_N73}
\end{align}
\end{minipage}

\begin{minipage}{0.3\textwidth}
{\flushleft Sector: 242}\\
\begin{figure}[H]
 \centering
 \includegraphics[width=4.5cm]{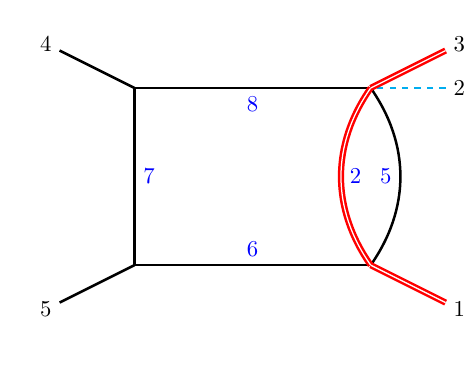}
\end{figure}
\end{minipage}
\begin{minipage}{0.65\textwidth}
\begin{align}
 \mathcal{N}^{(1)}_{74} &= \ep^3 \vij{45}(\mTsq+\vij{15})\frac{1}{\rho_2}\;, \label{eq:T1_N74} \\
 \mathcal{N}^{(1)}_{75} &= \ep^3 \vij{45}\left(\frac{\vij{15}}{\rho_5} - \frac{\mTsq}{\rho_2}\right)\;. \label{eq:T1_N75}
\end{align}
\end{minipage}

\begin{minipage}{0.3\textwidth}
{\flushleft Sector: 244}\\
\begin{figure}[H]
 \centering
 \includegraphics[width=4.5cm]{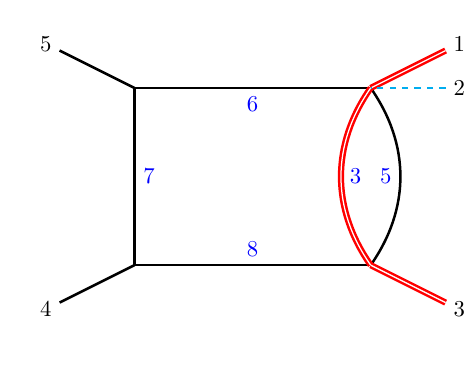}
\end{figure}
\end{minipage}
\begin{minipage}{0.65\textwidth}
\begin{align}
 \mathcal{N}^{(1)}_{76} &= \ep^3 \vij{45}(\mTsq+\vij{34})\frac{1}{\rho_3}\;, \label{eq:T1_N76} \\
 \mathcal{N}^{(1)}_{77} &= \ep^2 \vij{45}\left(\frac{\vij{34}}{\rho_5} - \frac{\mTsq}{\rho_3}\right)\;. \label{eq:T1_N77}
\end{align}
\end{minipage}
\subsection*{$\vec I_1$: 6 Propagator Integrals}
$\qquad$
\begin{minipage}{0.3\textwidth}
{\flushleft Sector: 95}\\
\begin{figure}[H]
 \centering
 \includegraphics[width=5cm]{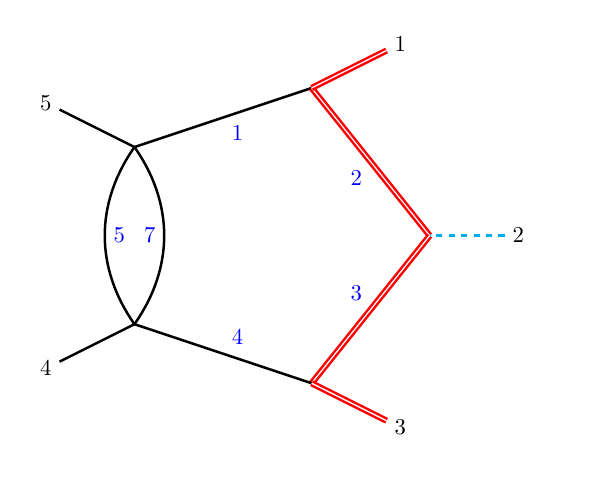}
\end{figure}
\end{minipage}
\begin{minipage}{0.65\textwidth}
\begin{align}
 \mathcal{N}^{(1)}_{78} &= \ep^3(1-2\ep)\sqrt{\cayley{2}}\;, \label{eq:T1_N78} \\
 \mathcal{N}^{(1)}_{79} &= \ep^3 \sqrt{\Delta_5} \frac{\mu_{11}}{\rho_5}\;. \label{eq:T1_N79}
\end{align}
\end{minipage}

\begin{minipage}{0.3\textwidth}
{\flushleft Sector: 126}\\
\begin{figure}[H]
 \centering
 \includegraphics[width=5cm]{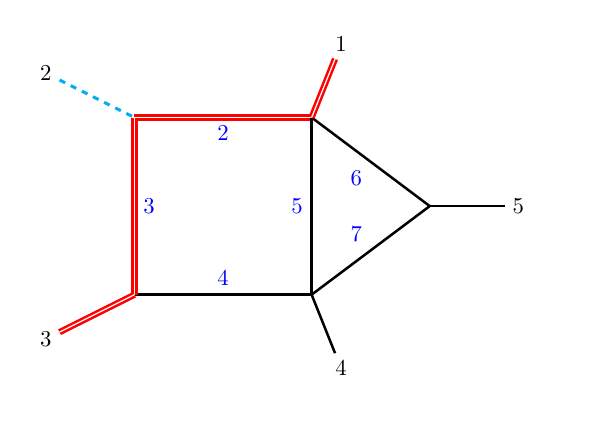}
\end{figure}
\end{minipage}
\begin{minipage}{0.65\textwidth}
\begin{align}
 \mathcal{N}^{(1)}_{80} &= \ep^4\sqrt{r_2}\;, \label{eq:T1_N80} \\
 \mathcal{N}^{(1)}_{81} &= \ep^3\sqrt{\Delta_5}\frac{\mu_{11}}{\rho_5}\;. \label{eq:T1_N81}
\end{align}
\end{minipage}

\begin{minipage}{0.3\textwidth}
{\flushleft Sector: 175}\\
\begin{figure}[H]
 \centering
 \includegraphics[width=5cm]{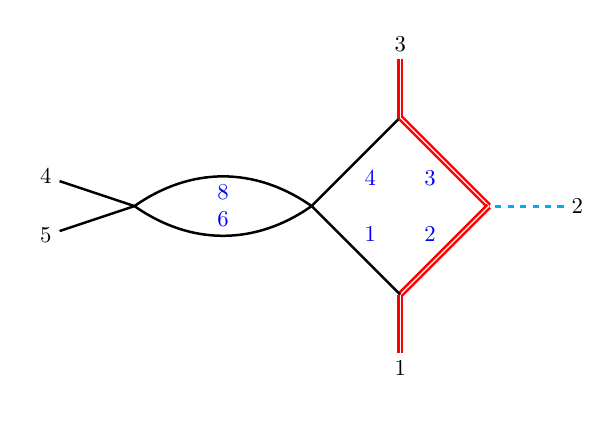}
\end{figure}
\end{minipage}
\begin{minipage}{0.65\textwidth}
\begin{align}
 \mathcal{N}^{(1)}_{82} &= \ep^3(1-2\ep)\sqrt{\cayley{2}}\;. \label{eq:T1_N82}
\end{align}
\end{minipage}

\begin{minipage}{0.3\textwidth}
{\flushleft Sector: 183}\\
\begin{figure}[H]
 \centering
 \includegraphics[width=5cm]{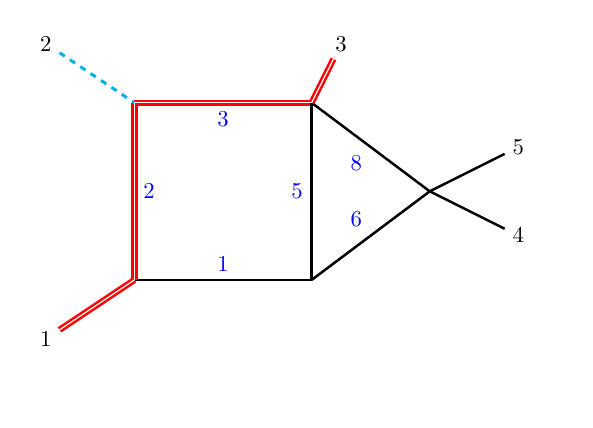}
\end{figure}
\end{minipage}
\begin{minipage}{0.65\textwidth}
\begin{align}
 \mathcal{N}^{(1)}_{83} &= \ep^4(\mHsq+\vij{12})\sqrt{\gram{3}}\;. \label{eq:T1_N83}
\end{align}
\end{minipage}

\begin{minipage}{0.3\textwidth}
{\flushleft Sector: 190}\\
\begin{figure}[H]
 \centering
 \includegraphics[width=5cm]{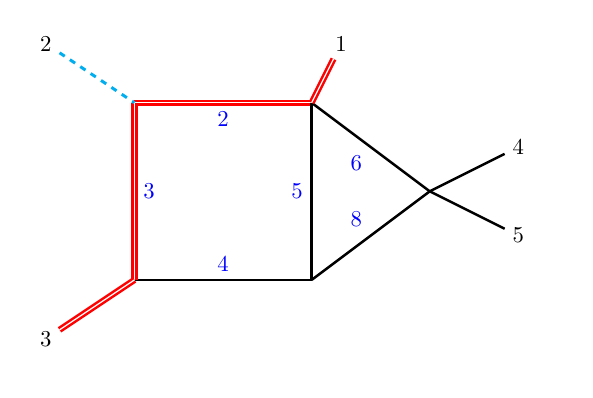}
\end{figure}
\end{minipage}
\begin{minipage}{0.65\textwidth}
\begin{align}
 \mathcal{N}^{(1)}_{84} &= \ep^4(\mHsq+\vij{23})\sqrt{\gram{4}}\;. \label{eq:T1_N84}
\end{align}
\end{minipage}

\begin{minipage}{0.3\textwidth}
{\flushleft Sector: 215}\\
\begin{figure}[H]
 \centering
 \includegraphics[width=5cm]{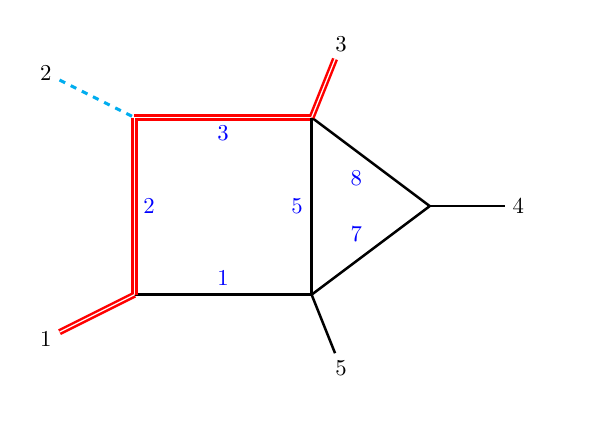}
\end{figure}
\end{minipage}
\begin{minipage}{0.65\textwidth}
\begin{align}
 \mathcal{N}^{(1)}_{85} &= \ep^4 \sqrt{r_3}\;, \label{eq:T1_N85} \\
 \mathcal{N}^{(1)}_{86} &= \ep^3\sqrt{\Delta_5} \frac{\mu_{11}}{\rho_5}\;. \label{eq:T1_N86}
\end{align}
\end{minipage}

\begin{minipage}{0.3\textwidth}
{\flushleft Sector: 245}\\
\begin{figure}[H]
 \centering
 \includegraphics[width=5cm]{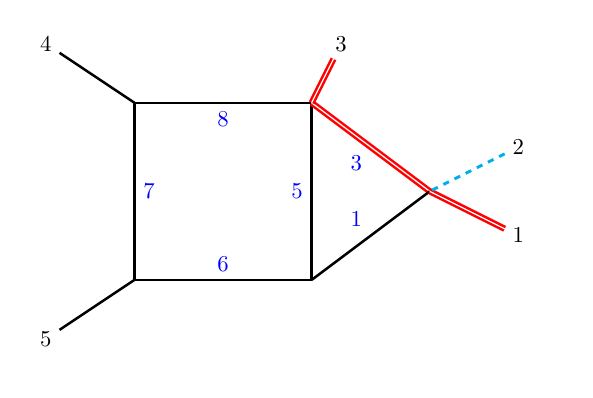}
\end{figure}
\end{minipage}
\begin{minipage}{0.65\textwidth}
\begin{align}
 \mathcal{N}^{(1)}_{87} &= \ep^4 \vij{45}(\vij{12}-\vij{34}+\mHsq)\;. \label{eq:T1_N87}
\end{align}
\end{minipage}

\begin{minipage}{0.3\textwidth}
{\flushleft Sector: 246}\\
\begin{figure}[H]
 \centering
 \includegraphics[width=5cm]{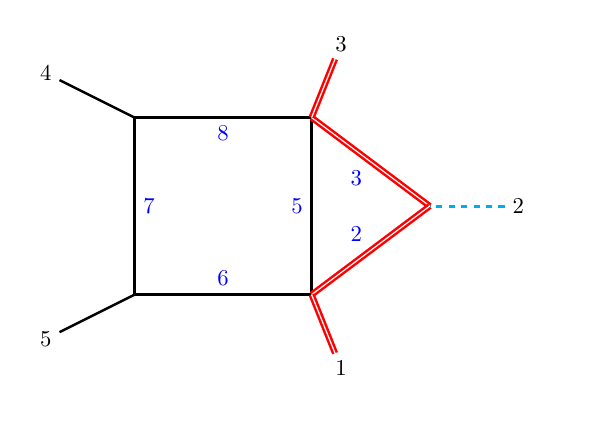}
\end{figure}
\end{minipage}
\begin{minipage}{0.65\textwidth}
\begin{align}
 \mathcal{N}^{(1)}_{88} &= \ep^4 \vij{45}\sqrt{\gram{5}}\;, \label{eq:T1_N88} \\
 \mathcal{N}^{(1)}_{89} &= \ep^3 \sqrt{\Delta_5} \frac{\mu_{12}}{\rho_5}\;, \label{eq:T1_N89} \\
 \mathcal{N}^{(1)}_{90} &= \ep^3 \sqrt{\Delta_5} \frac{\mu_{22}}{\rho_5}\;, \label{eq:T1_N90} \\
 \mathcal{N}^{(1)}_{91} &= \ep^3 \vij{45} \Bigg[\mHsq\rho_9\left(\frac{1}{\rho_2}-\frac{1}{\rho_3}\right) +2\ep(\vij{34}-\vij{15}) \nonumber \\
 &- (\mHsq+\vij{23})\frac{\rho_7}{\rho_8} +(\mHsq+\vij{12})\frac{\rho_7}{\rho_6} \Bigg]\;, \label{eq:T1_N91}\\
 \mathcal{N}^{(1)}_{92} &= \ep^3 \vij{45}\sqrt{\cayley{1}}\Bigg[\rho_9\left(\frac{1}{\rho_2}+\frac{1}{\rho_3}\right) + 2\ep \nonumber \\ 
 &\qquad \qquad \qquad\qquad\qquad+\rho_7\left(\frac{1}{\rho_8}+\frac{1}{\rho_6}\right)\Bigg]\;,\label{eq:T1_N92} \\
 \mathcal{N}^{(1)}_{93} &= \frac{\ep^3~\vij{45}}{\vij{15}-\vij{34}}\Bigg[ \ep\big( (\vij{15}-\vij{34})^2 - \mHsq(\vij{15}+\vij{34})\big) \nonumber \\
 & + (\mTsq(\vij{15}-\vij{34})^2 + \mHsq\vij{15}\vij{34})\left(\frac{1}{\rho_2} + \frac{1}{\rho_3}\right)\Bigg] \nonumber \\
& - \frac{\vij{15}+\vij{34}}{\vij{15}-\vij{34}}\frac{\mHsq}{2\sqrt{\cayley{1}}}\mathcal{N}^{(1)}_{92}\;.\label{eq:T1_N93}
\end{align}
\end{minipage}

\begin{minipage}{0.3\textwidth}
{\flushleft Sector: 250}\\
\begin{figure}[H]
 \centering
 \includegraphics[width=5cm]{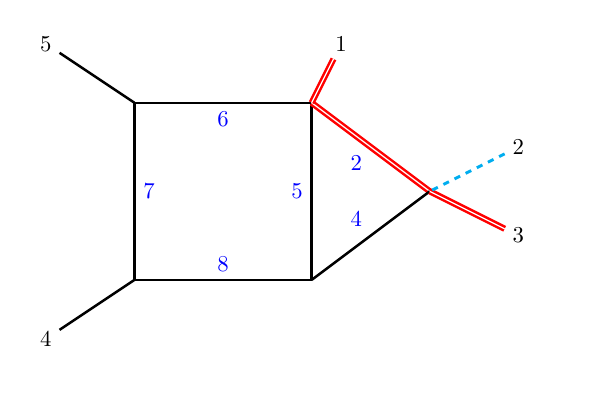}
\end{figure}
\end{minipage}
\begin{minipage}{0.65\textwidth}
\begin{align}
 \mathcal{N}^{(1)}_{94} &= \ep^4 \vij{45}(\vij{23}-\vij{15}+\mHsq)\;. \label{eq:T1_N94}
\end{align}
\end{minipage}
\subsection*{$\vec I_1$: 7 Propagator Integrals}
$\qquad$
\begin{minipage}{0.3\textwidth}
{\flushleft Sector: 247}\\
\begin{figure}[H]
 \centering
 \includegraphics[width=5cm]{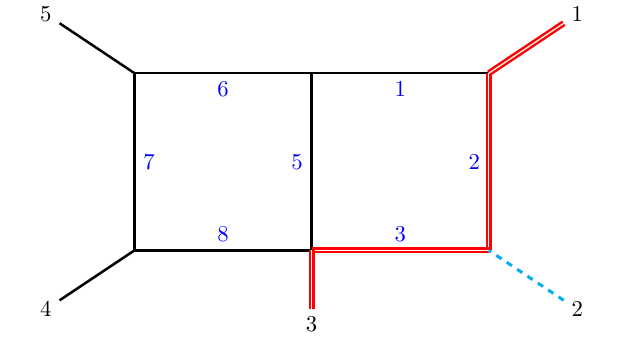}
\end{figure}
\end{minipage}
\begin{minipage}{0.7\textwidth}
\begin{align}
 \mathcal{N}^{(1)}_{95} &= \ep^4(\mHsq+\vij{12})\vij{15}\vij{45}\;, \label{eq:T1_N95} \\
 \mathcal{N}^{(1)}_{96} &= \ep^4\vij{45}\sqrt{\gram{1}}\rho_9\;, \label{eq:T1_N96} \\
 \mathcal{N}^{(1)}_{97} &= \ep^4(\mHsq+\vij{12})\Big[\vij{45}\rho_{11} - (\vij{23}-\vij{15}+\mHsq)\rho_6\Big]\;, \label{eq:T1_N97} \\
 \mathcal{N}^{(1)}_{98} &= \ep^4\sqrt{\Delta_5}\mu_{12}\;. \label{eq:T1_N98}
\end{align}
\end{minipage}

\begin{minipage}{0.3\textwidth}
{\flushleft Sector: 251}\\
\begin{figure}[H]
 \centering
 \includegraphics[width=5cm]{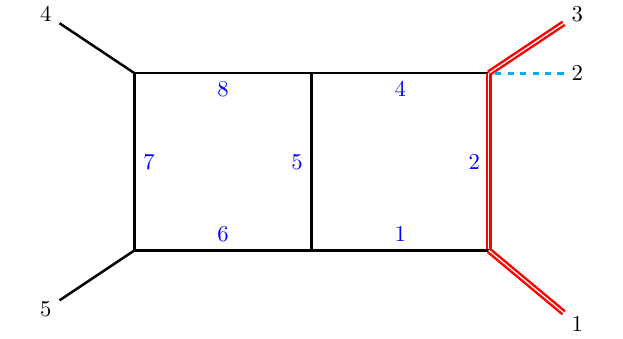}
\end{figure}
\end{minipage}
\begin{minipage}{0.7\textwidth}
\begin{align}
 \mathcal{N}^{(1)}_{99} &= \ep^4\vij{15}\vij{45}^2\;, \label{eq:T1_N99} \\
 \mathcal{N}^{(1)}_{100} &= \ep^4\vij{45}\sqrt{\gram{3}}\rho_9\;, \label{eq:T1_N100} \\
 \mathcal{N}^{(1)}_{101} &= \ep^4\vij{45}\left[ \vij{45}\rho_{11} + \frac{1}{\ep}(\mHsq+\vij{12})\frac{\rho_6}{\rho_5}\left(\rho_8 - \frac{\rho_4}{2}\right)\right]\;. \label{eq:T1_N101}
\end{align}
\end{minipage}

\begin{minipage}{0.3\textwidth}
{\flushleft Sector: 253}\\
\begin{figure}[H]
 \centering
 \includegraphics[width=5cm]{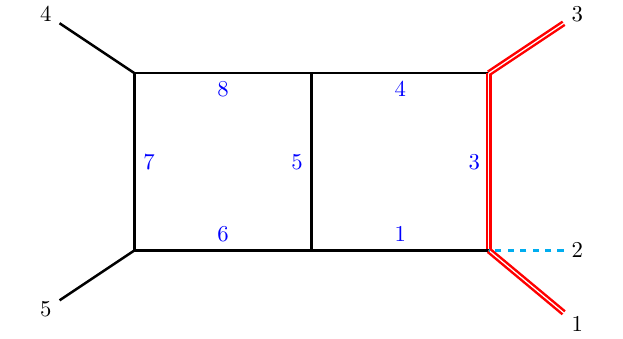}
\end{figure}
\end{minipage}
\begin{minipage}{0.7\textwidth}
\begin{align}
 \mathcal{N}^{(1)}_{102} &= \ep^4\vij{34}\vij{45}^2\;, \label{eq:T1_N102} \\
 \mathcal{N}^{(1)}_{103} &= \ep^4\vij{45}\sqrt{\gram{4}}\rho_9\;, \label{eq:T1_N103} \\
 \mathcal{N}^{(1)}_{104} &= \ep^4\vij{45}\left[ \vij{45}\rho_{10} + \frac{1}{\ep}(\mHsq+\vij{23})\frac{\rho_8}{\rho_5}\left(\rho_6 - \frac{\rho_1}{2}\right)\right]\;. \label{eq:T1_N104}
\end{align}
\end{minipage}

\begin{minipage}{0.3\textwidth}
{\flushleft Sector: 254}\\
\begin{figure}[H]
 \centering
 \includegraphics[width=5cm]{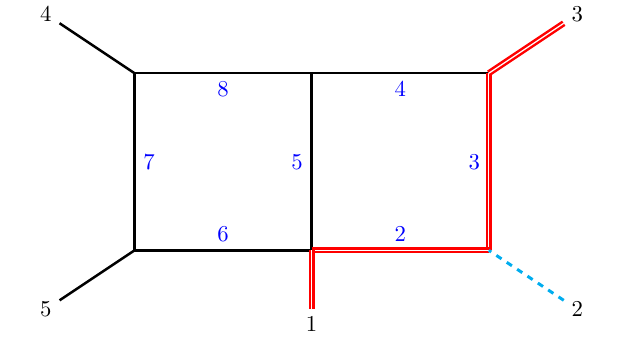}
\end{figure}
\end{minipage}
\begin{minipage}{0.7\textwidth}
\begin{align}
 \mathcal{N}^{(1)}_{105} &= \ep^4(\mHsq+\vij{23})\vij{45}\vij{34}\;, \label{eq:T1_N105} \\
 \mathcal{N}^{(1)}_{106} &= \ep^4\vij{45}\sqrt{\gram{2}}\rho_9\;, \label{eq:T1_N106} \\
 \mathcal{N}^{(1)}_{107} &= \ep^4(\mHsq+\vij{23})\Big[\vij{45}\rho_{10} - (\vij{12}-\vij{34}+\mHsq)\rho_8\Big]\;, \label{eq:T1_N107} \\
 \mathcal{N}^{(1)}_{108} &= \ep^4\sqrt{\Delta_5}\mu_{12}\;. \label{eq:T1_N108}
\end{align}
\end{minipage}
\subsection*{$\vec I_1$: 8 Propagator Integrals}
$\qquad$
\begin{minipage}{0.3\textwidth}
{\flushleft Sector: 255}\\
\begin{figure}[H]
 \centering
 \includegraphics[width=5cm]{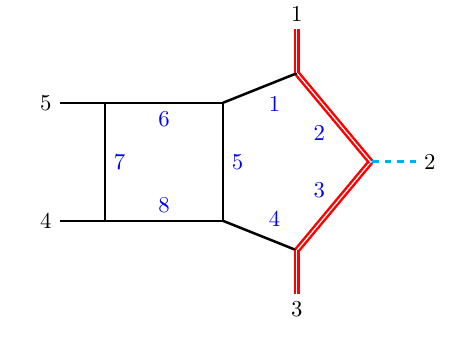}
\end{figure}
\end{minipage}
\begin{minipage}{0.65\textwidth}
\begin{align}
 \mathcal{N}^{(1)}_{109} &= \ep^4 \vij{45}\sqrt{\Delta_5} \mu_{11}\;, \label{eq:T1_N109} \\
 \mathcal{N}^{(1)}_{110} &= \ep^4 \vij{45}\sqrt{\Delta_5} \mu_{12}\;, \label{eq:T1_N110} \\
 \mathcal{N}^{(1)}_{111} &= \ep^4 \vij{45}\sqrt{\cayley{2}}\rho_9\;. \label{eq:T1_N111}
\end{align}
\end{minipage}

\section{Master Integral Basis for the \Tht{} Feynman Integral Family}
\label{app:2looppbub}

In this appendix we provide the definitions of all master integrals that we have
computed for the Feynman integral family \Tht{} which is shown in
equations~\eqref{eq:T2tfamily}, \eqref{eq:T2tprops}, and \eqref{eq:T2tisps}.
Notice that we do not include definitions of integrals in this family that
coincide with those presented in appendix~\ref{app:2looppbox} (see
section~\ref{sec:intfamilies} for details). We present the definitions in
subsections organized from the integrals with the least (4) to the
integrals with the most (6) propagators.

For each integral we provide information which exactly specifies it,
as explained in the introduction to Appendix~\ref{app:2looppbox}.
The generic integral belonging to this integral family is defined as:
\begin{equation*}
 \left(\vec I_{2}\right)_j = \int \frac{d^d\ell_1}{i\pi^{d/2}}\frac{d^d\ell_2}{i\pi^{d/2}}
 \frac{\mathcal{N}^{(2)}_j}{\rho_1^{\nu_1}\rho_2^{\nu_2}\rho_3^{\nu_3}\rho_4^{\nu_4}\rho_5^{\nu_5}\rho_6^{\nu_6}}\;,
\end{equation*}
where the superscript of the numerator insertion $\mathcal{N}^{(2)}_j$
indicates that this integrand belongs to the \Tht{} family, where the index
$j$ is an integer between 1 and 19, and where $\nu_i \in \{0,1\}$.
Notice that we employ kinematic invariants and
functions defined in section~\ref{sec:kinematics}.

For completeness, we also print the values of the integrals at weight 0 as used
in the discussion of iterated integral solutions of section~\ref{sec:iterints},
\begin{equation}
  \vec{I}_2^{\;(0)} = \mathcal{Z}\left(\vec{I}_2^{\;(0)}\right)  = 
  \Big\{-\frac{1}{2},0,-1,1,0,0,-\frac{1}{8},0,0,0,0,\frac{1}{2},0,0,-\frac{1}{8},0,0,\frac{1}{2},0\Big\}\;.
\end{equation}

\subsection*{$\vec I_2$: 4 Propagator Integrals}
$\qquad$
\begin{minipage}{0.3\textwidth}
{\flushleft Sector: 53}\\
\begin{figure}[H]
 \centering
 \includegraphics[width=4.5cm]{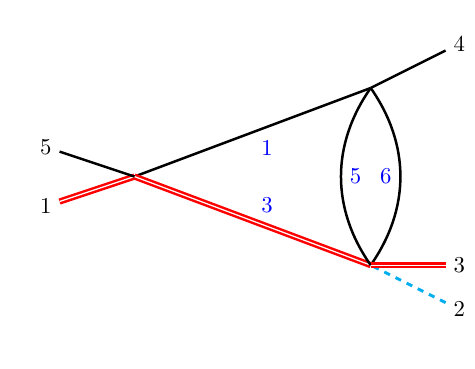}
\end{figure}
\end{minipage}
\begin{minipage}{0.65\textwidth}
\begin{align}
 \mathcal{N}^{(2)}_{5} &= \ep^3(\mHsq+\vij{23}-\vij{15})\frac{1}{\rho_5}\;. \label{eq:T2_N5}
\end{align}
\end{minipage}

\begin{minipage}{0.3\textwidth}
{\flushleft Sector: 57}\\
\begin{figure}[H]
 \centering
 \includegraphics[width=4.5cm]{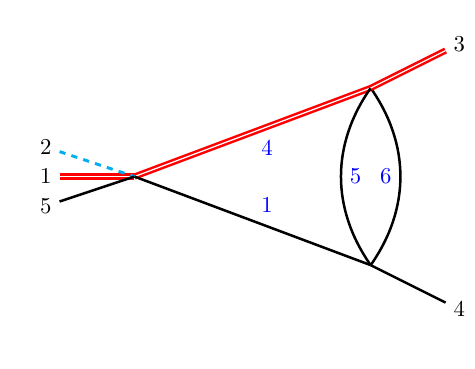}
\end{figure}
\end{minipage}
\begin{minipage}{0.65\textwidth}
\begin{align}
 \mathcal{N}^{(2)}_{7} &= \ep^3\vij{34}\frac{1}{\rho_5}\;. \label{eq:T2_N7}
\end{align}
\end{minipage}
\subsection*{$\vec I_2$: 5 Propagator Integrals}
$\qquad$
\begin{minipage}{0.3\textwidth}
{\flushleft Sector: 55}\\
\begin{figure}[H]
 \centering
 \includegraphics[width=4.5cm]{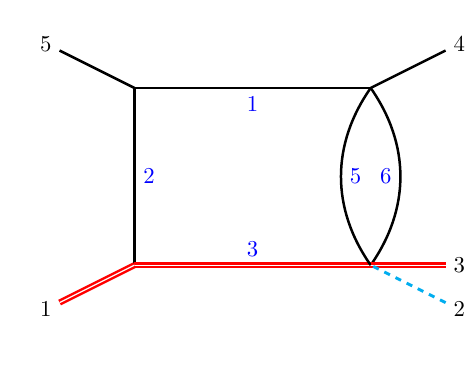}
\end{figure}
\end{minipage}
\begin{minipage}{0.65\textwidth}
\begin{align}
 \mathcal{N}^{(2)}_{12} &= \ep^3(1-2\ep)\vij{15}\;. \label{eq:T2_N12}
\end{align}
\end{minipage}

\begin{minipage}{0.3\textwidth}
{\flushleft Sector: 59}\\
\begin{figure}[H]
 \centering
 \includegraphics[width=4.5cm]{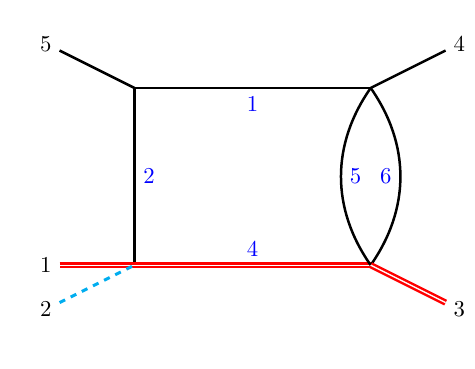}
\end{figure}
\end{minipage}
\begin{minipage}{0.65\textwidth}
\begin{align}
 \mathcal{N}^{(2)}_{13} &= \ep^3(1-2\ep)(\mHsq+\vij{12}-\vij{34})\;. \label{eq:T2_N13}
\end{align}
\end{minipage}

\begin{minipage}{0.3\textwidth}
{\flushleft Sector: 61}\\
\begin{figure}[H]
 \centering
 \includegraphics[width=4.5cm]{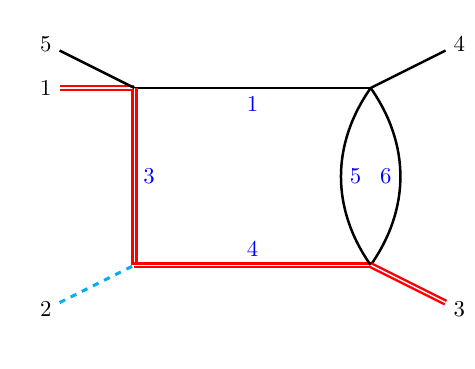}
\end{figure}
\end{minipage}
\begin{minipage}{0.65\textwidth}
\begin{align}
 \mathcal{N}^{(2)}_{14} &= \ep^3(1-2\ep)\sqrt{\gram{5}}\;, \label{eq:T2_N14} \\
 \mathcal{N}^{(2)}_{15} &= \ep^3 (\mHsq+\vij{23})\vij{34}\frac{1}{\rho_5}\;. \label{eq:T2_N15}
\end{align}
\end{minipage}
\subsection*{$\vec I_2$: 6 Propagator Integrals}
$\qquad$
\begin{minipage}{0.3\textwidth}
{\flushleft Sector: 63}\\
\begin{figure}[H]
 \centering
 \includegraphics[width=4.5cm]{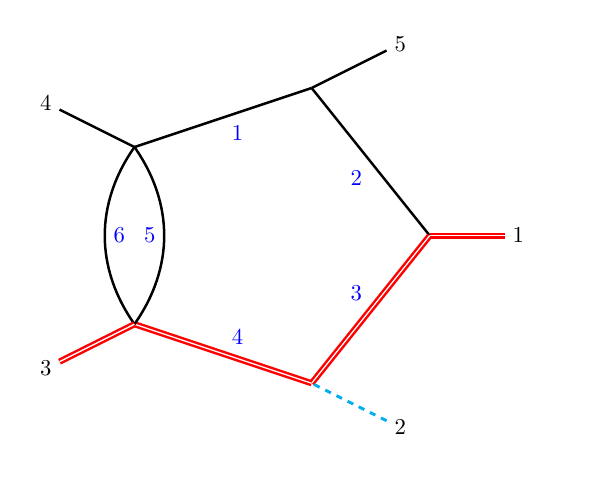}
\end{figure}
\end{minipage}
\begin{minipage}{0.65\textwidth}
\begin{align}
 \mathcal{N}^{(2)}_{18} &= \ep^3(1-2\ep)\vij{15}(\mHsq+\vij{12})\;, \label{eq:T2_N18} \\
 \mathcal{N}^{(2)}_{19} &= \ep^3\sqrt{\Delta_5}~\frac{\mu_{11}}{\rho_5}\;. \label{eq:T2_N19}
\end{align}
\end{minipage}

\section{Master Integral Basis for the $T_0$ Feynman Integral Family}
\label{app:1loop}

\begin{figure}[ht]
\begin{center}
 \includegraphics[scale=0.6]{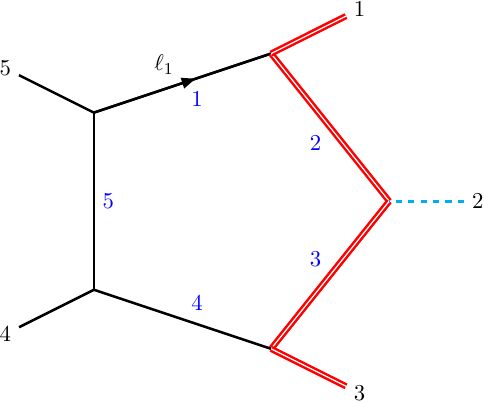}
\end{center}
\caption{The propagator structure asociated to the $T_0$ integral
  family, with the routing of loop momenta chosen in
  equation~(\ref{eq:T0family}). See caption of figure~\ref{fig:T1} for
  details on the notation.}
 \label{fig:1loopdiag}
\end{figure}

For completeness in this appendix we provide a basis of pure master
integrals for the one-loop Feynman integral family  related to the propagator
structure of the diagram in figure~\ref{fig:1loopdiag} and defined as
follows:
\begin{equation}
 T_0[\vec{\nu}] = \int \frac{d^d\ell_1}{i\pi^{d/2}}
 \frac{1}
 {\rho_1^{\nu_1}\rho_2^{\nu_2}\rho_3^{\nu_3}\rho_4^{\nu_4}\rho_5^{\nu_5}}\;,
 \label{eq:T0family}
\end{equation}
where the inverse propagators are defined by:
\begin{align}
 &\rho_1 = \ell_1^2\;, &&\rho_2 = (\ell_1+p_1)^2 - \mTsq\;, &&&\rho_3
=(\ell_1+p_{12})^2 - \mTsq\;,\nonumber \\
 &\rho_4 = (\ell_1+p_{123})^2\;, &&\rho_5 = (\ell_1-p_5)^2\;\ .
&&&\label{eq:T0props} 
\end{align}
This family has $\textrm{dim}\left(T_0\right) = 18$ and we present our
choice of pure master integral in the following subsections organized
from the integrals with the least (1) to the integrals with the most
(5) propagators.\footnote{We notice that a canonical basis for this integral
family has already been presented in Ref.~\cite{Chen:2022nxt}.}

For each integral we provide information which exactly specifies it,
as explained in the introduction to Appendix~\ref{app:2looppbox}.
The generic integral belonging to this integral family is defined as:
\begin{equation*}
 \left(\vec I_{0}\right)_j = \int \frac{d^d\ell_1}{i\pi^{d/2}}
 \frac{\mathcal{N}^{(0)}_j}{\rho_1^{\nu_1}\rho_2^{\nu_2}\rho_3^{\nu_3}\rho_4^{\nu_4}\rho_5^{\nu_5}}\;,
\end{equation*}
where the superscript of the numerator insertion $\mathcal{N}^{(0)}_j$
indicates that this integrand belongs to the $T_0$ family, where the index
$j$ is an integer between 1 and 18, and where $\nu_i \in \{0,1\}$.  Notice that
we employ kinematic invariants and
functions defined in section~\ref{sec:kinematics}.

For completeness, we also print the values of the integrals at weight 0 as used
in the discussion of iterated integral solutions of section~\ref{sec:iterints},
\begin{equation}
  \vec{I}_0^{\;(0)} =
  \Big\{1,0,0,1,0,0,0,0,0,0,0,0,0,\frac{1}{2},\frac{3}{2},\frac{3}{2},\frac{1}{2},0\Big\}\;.
\end{equation}

\subsection*{$\vec I_0$: 1 Propagator Integral}
$\qquad$
\begin{minipage}{0.3\textwidth}
{\flushleft Sector: 2}\\
\begin{figure}[H]
 \centering
 \includegraphics[width=2.5cm]{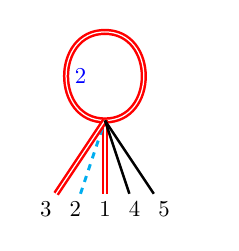}
\end{figure}
\end{minipage}
\begin{minipage}{0.65\textwidth}
\begin{equation}
 \mathcal{N}^{(0)}_{1} = \frac{\ep(1-\ep)}{\mTsq}\;. \label{eq:T0_N1}
\end{equation}
\end{minipage}

\subsection*{$\vec I_0$: 2 Propagator Integrals}
$\qquad$
\begin{minipage}{0.3\textwidth}
{\flushleft Sector: 5}\\
\begin{figure}[H]
 \centering
 \includegraphics[width=4.5cm]{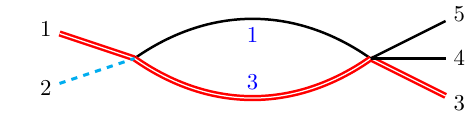}
\end{figure}
\end{minipage}
\begin{minipage}{0.65\textwidth}
\begin{equation}
 \mathcal{N}^{(0)}_{2} = \ep (\mHsq+\mTsq+\vij{12})\frac{1}{\rho_3}\;. \label{eq:T0_N2}
\end{equation}
\end{minipage}

\begin{minipage}{0.3\textwidth}
{\flushleft Sector: 6}\\
\begin{figure}[H]
 \centering
 \includegraphics[width=4.5cm]{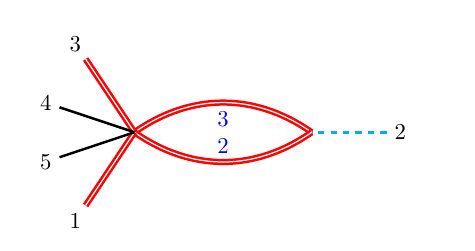}
\end{figure}
\end{minipage}
\begin{minipage}{0.65\textwidth}
\begin{equation}
 \mathcal{N}^{(0)}_{3} = \ep\sqrt{\cayley{1}}\frac{1}{\rho_2}\;. \label{eq:T0_N3}
\end{equation}
\end{minipage}

\begin{minipage}{0.3\textwidth}
{\flushleft Sector: 9}\\
\begin{figure}[H]
 \centering
 \includegraphics[width=4.5cm]{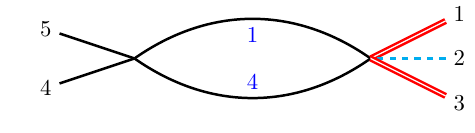}
\end{figure}
\end{minipage}
\begin{minipage}{0.65\textwidth}
\begin{equation}
 \mathcal{N}^{(0)}_{4} = \ep(1-\ep)\;. \label{eq:T0_N4}
\end{equation}
\end{minipage}

\begin{minipage}{0.3\textwidth}
{\flushleft Sector: 10}\\
\begin{figure}[H]
 \centering
 \includegraphics[width=4.5cm]{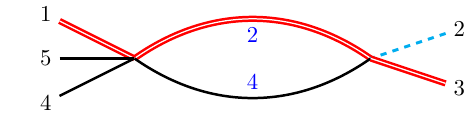}
\end{figure}
\end{minipage}
\begin{minipage}{0.65\textwidth}
\begin{equation}
 \mathcal{N}^{(0)}_{5} = \ep (\mHsq+\mTsq+\vij{23})\frac{1}{\rho_2}\;. \label{eq:T0_N5}
\end{equation}
\end{minipage}

\begin{minipage}{0.3\textwidth}
{\flushleft Sector: 18}\\
\begin{figure}[H]
 \centering
 \includegraphics[width=4.5cm]{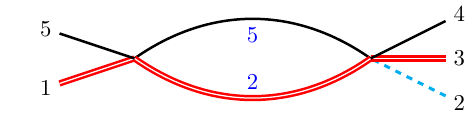}
\end{figure}
\end{minipage}
\begin{minipage}{0.65\textwidth}
\begin{equation}
 \mathcal{N}^{(0)}_{6} = \ep (\mTsq+\vij{15})\frac{1}{\rho_2}\;. \label{eq:T0_N6}
\end{equation}
\end{minipage}

\begin{minipage}{0.3\textwidth}
{\flushleft Sector: 20}\\
\begin{figure}[H]
 \centering
 \includegraphics[width=4.5cm]{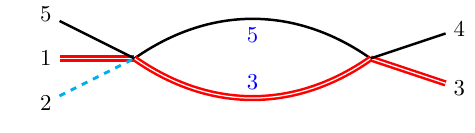}
\end{figure}
\end{minipage}
\begin{minipage}{0.65\textwidth}
\begin{equation}
 \mathcal{N}^{(0)}_{7} = \ep(\mTsq+\vij{34})\frac{1}{\rho_3}\;. \label{eq:T0_N7}
\end{equation}
\end{minipage}
\subsection*{$\vec I_0$: 3 Propagator Integrals}
$\qquad$
\begin{minipage}{0.3\textwidth}
{\flushleft Sector: 7}\\
\begin{figure}[H]
 \centering
 \includegraphics[width=4.5cm]{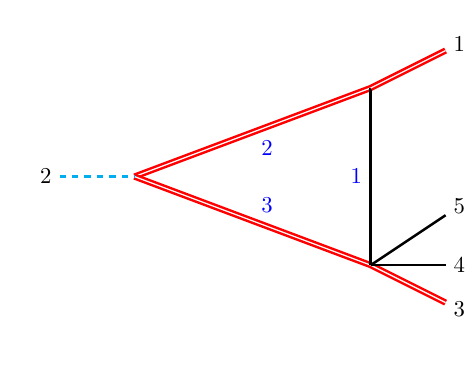}
\end{figure}
\end{minipage}
\begin{minipage}{0.65\textwidth}
\begin{equation}
 \mathcal{N}^{(0)}_{8} = \ep^2\sqrt{\gram{1}}\;. \label{eq:T0_N8}
\end{equation}
\end{minipage}

\begin{minipage}{0.3\textwidth}
{\flushleft Sector: 11}\\
\begin{figure}[H]
 \centering
 \includegraphics[width=4.5cm]{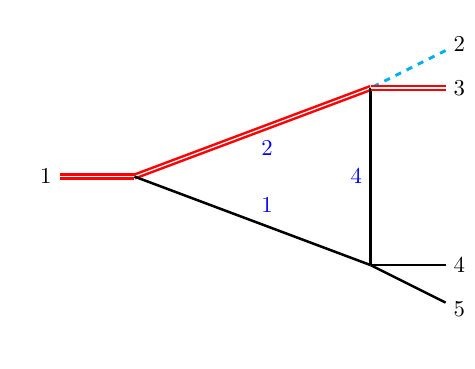}
\end{figure}
\end{minipage}
\begin{minipage}{0.65\textwidth}
\begin{equation}
 \mathcal{N}^{(0)}_{9} = \ep^2\sqrt{\gram{3}}\;. \label{eq:T0_N9}
\end{equation}
\end{minipage}

\begin{minipage}{0.3\textwidth}
{\flushleft Sector: 13}\\
\begin{figure}[H]
 \centering
 \includegraphics[width=4.5cm]{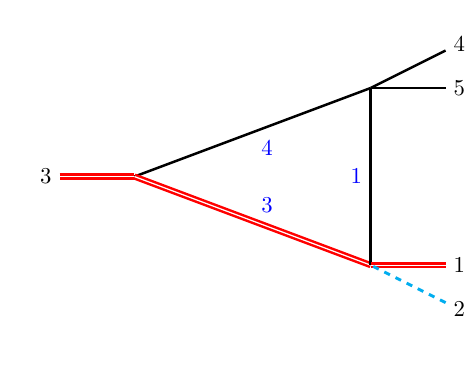}
\end{figure}
\end{minipage}
\begin{minipage}{0.65\textwidth}
\begin{equation}
 \mathcal{N}^{(0)}_{10} = \ep^2\sqrt{\gram{4}}\;. \label{eq:T0_N10}
\end{equation}
\end{minipage}

\begin{minipage}{0.3\textwidth}
{\flushleft Sector: 14}\\
\begin{figure}[H]
 \centering
 \includegraphics[width=4.5cm]{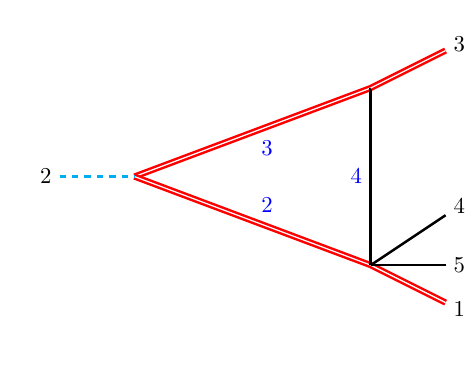}
\end{figure}
\end{minipage}
\begin{minipage}{0.65\textwidth}
\begin{equation}
 \mathcal{N}^{(0)}_{11} = \ep^2\sqrt{\gram{2}}\;. \label{eq:T0_N11}
\end{equation}
\end{minipage}

\begin{minipage}{0.3\textwidth}
{\flushleft Sector: 22}\\
\begin{figure}[H]
 \centering
 \includegraphics[width=4.5cm]{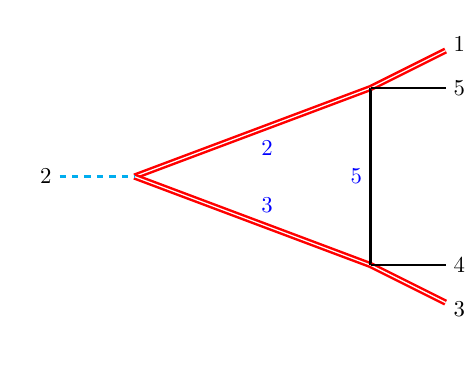}
\end{figure}
\end{minipage}
\begin{minipage}{0.65\textwidth}
\begin{equation}
 \mathcal{N}^{(0)}_{12} = \ep^2\sqrt{\gram{5}}\;. \label{eq:T0_N12}
\end{equation}
\end{minipage}
\subsection*{$\vec I_0$: 4 Propagator Integrals}
$\qquad$
\begin{minipage}{0.3\textwidth}
{\flushleft Sector: 15}\\
\begin{figure}[H]
 \centering
 \includegraphics[width=4.5cm]{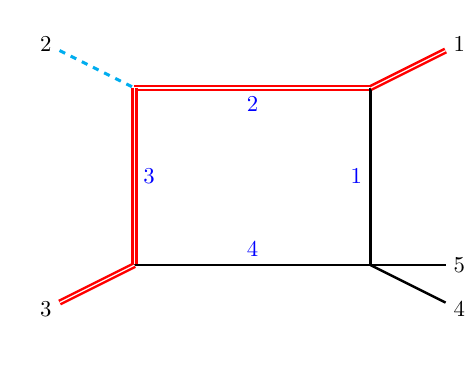}
\end{figure}
\end{minipage}
\begin{minipage}{0.65\textwidth}
\begin{equation}
 \mathcal{N}^{(0)}_{13} = \ep^2\sqrt{\cayley{2}}\;. \label{eq:T0_N13}
\end{equation}
\end{minipage}

\begin{minipage}{0.3\textwidth}
{\flushleft Sector: 23}\\
\begin{figure}[H]
 \centering
 \includegraphics[width=4.5cm]{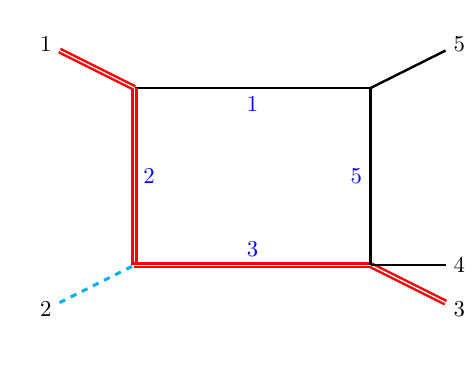}
\end{figure}
\end{minipage}
\begin{minipage}{0.65\textwidth}
\begin{equation}
 \mathcal{N}^{(0)}_{14} = \ep^2\vij{15}(\mHsq+\vij{12})\;. \label{eq:T0_N14}
\end{equation}
\end{minipage}

\begin{minipage}{0.3\textwidth}
{\flushleft Sector: 27}\\
\begin{figure}[H]
 \centering
 \includegraphics[width=4.5cm]{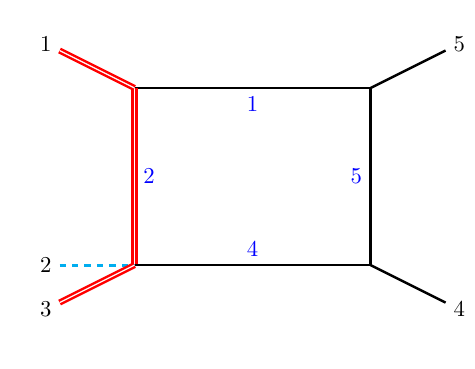}
\end{figure}
\end{minipage}
\begin{minipage}{0.65\textwidth}
\begin{equation}
 \mathcal{N}^{(0)}_{15} = \ep^2\vij{45}\vij{15}\;. \label{eq:T0_N15}
\end{equation}
\end{minipage}

\begin{minipage}{0.3\textwidth}
{\flushleft Sector: 29}\\
\begin{figure}[H]
 \centering
 \includegraphics[width=4.5cm]{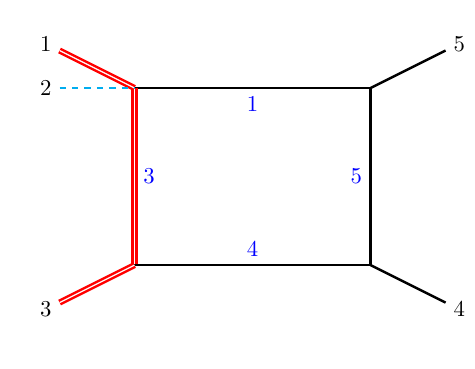}
\end{figure}
\end{minipage}
\begin{minipage}{0.65\textwidth}
\begin{equation}
 \mathcal{N}^{(0)}_{16} = \ep^2\vij{45}\vij{34}\;. \label{eq:T0_N16}
\end{equation}
\end{minipage}

\begin{minipage}{0.3\textwidth}
{\flushleft Sector: 30}\\
\begin{figure}[H]
 \centering
 \includegraphics[width=4.5cm]{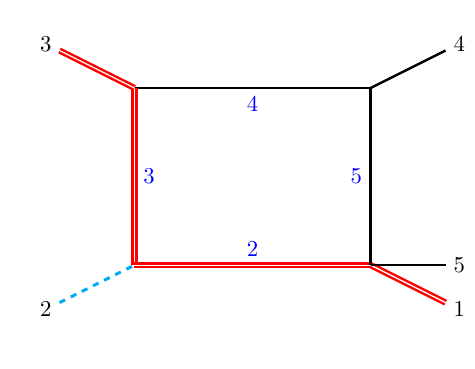}
\end{figure}
\end{minipage}
\begin{minipage}{0.65\textwidth}
\begin{equation}
 \mathcal{N}^{(0)}_{17} = \ep^2\vij{34}(\mHsq+\vij{23})\;. \label{eq:T0_N17}
\end{equation}
\end{minipage}
\subsection*{$\vec I_0$: 5 Propagator Integrals}
$\qquad$
\begin{minipage}{0.3\textwidth}
{\flushleft Sector: 31}\\
\begin{figure}[H]
 \centering
 \includegraphics[width=4.5cm]{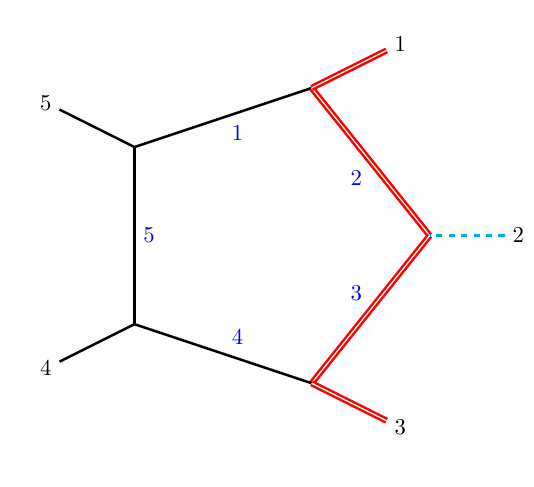}
\end{figure}
\end{minipage}
\begin{minipage}{0.65\textwidth}
\begin{equation}
 \mathcal{N}^{(0)}_{18} = \ep^2\sqrt{\Delta_5}~\mu_{11}\;. \label{eq:T0_N18}
\end{equation}
\end{minipage}

\bibliographystyle{JHEP}
\bibliography{ttHints}
\end{document}